\documentclass[referee]{aa}   
\usepackage{graphicx}
\usepackage{txfonts}
\begin{document}
%
%
%
%
\title{Shock waves in tidally compressed stars by massive black holes}
  
\author{M. Brassart \and J.-P. Luminet}

\offprints{M. Brassart}

\institute{Laboratoire Univers et Th\'eories (LUTH), 
           Observatoire de Paris,
	   CNRS, 
	   Universit\'e Paris Diderot;
	   5 place Jules Janssen, 
	   92190 Meudon, France \\
           \email{matthieu.brassart@obspm.fr, jean-pierre.luminet@obspm.fr}
           }
	     
\date{Received date; accepted date}

\abstract
%
%
{}
%
%
{We interest in the case of a main-sequence star deeply penetrating within the tidal radius of a massive black hole. 
We focus on the compression phase leading to a so-called pancake configuration of the star at the instant of maximal compression. 
The aim is to study the tidal compression process paying particular attention to the development of shock waves;
to deduce reliable estimates of the thermodynamical quantities involved in the pancake star;
and to solve a controversy about whether or not thermonuclear reactions can be triggered in the core of a tidally compressed star.}
%
%
{We have set up a one-dimensional hydrodynamical model well-adapted to the geometry of the problem.
Based on the high-resolution shock-capturing Godunov-type approach, it allows to study the compression phase undergone by the star in the direction orthogonal to its orbital plane.}
%
%
{We show the existence of two regimes depending on whether shock waves develop before or after the instant of maximal core compression. 
In both cases we confirm high compression and heating factors in the stellar core able to trigger a thermonuclear explosion. 
Moreover, we show that the shock waves carry outwards a brief but very high peak of temperature from the centre to the surface of the star. 
We tentatively conclude that the phenomenon could give rise to hard electromagnetic radiation, to be compared to some X-ray flares already observed in some galactic nuclei harbouring massive black holes. 
Finally, we estimate that the rate of pancake stars should be about $10^{-5}$ per galaxy per year. If generated in hard X- or $\gamma$-ray band, several events of this kind per year should be detectable within the full observable universe.}
%
%
{}


\authorrunning{M. Brassart \and J.-P. Luminet}
  
\titlerunning{Shock waves in tidally compressed stars}

\maketitle
%
%
%
%
\section{Introduction} \label{Intro}
A massive black hole (BH), of mass $10^{4} M_{\odot} \lesssim M_{\bullet} \lesssim 10^{7} M_{\odot}$, can tidally disrupt (small) main-sequence stars penetrating within the so-called \textit{tidal radius}, defined in order of magnitude by  
\begin{equation}
R_{\mathrm{T}} \equiv R_{*} \left( \frac{M_{\bullet}}{M_{*}} \right)^{1/3} \label{P01Eq01}
\end{equation}
with $M_{*}$ and $R_{*}$ respectively the mass and radius of the star.
With an estimated frequency of one stellar disruption every $10^{4}$ years (Magorrian \& Tremaine \cite{Mag99}), or every $10^{2}$ years in the presence of a self-gravitating accretion disc around the massive BH (Karas \& \v Subr \cite{Kar07}), such an astrophysical event is expected to provide gas for feeding a moderately active galactic nucleus, and for bringing back to active life an otherwise quiescent galactic nucleus such as that of the Galaxy. 
Some observations of UV flares (Renzini et al. \cite{Ren95}) and X-ray flares (see Komossa \cite{Kom02} for a review) from the core of non-active galaxies have already been tentatively interpreted in this sense.

Several analytical and hydrodynamical models have been set up to study particular aspects of the tidal disruption process of a main-sequence star by a massive BH. 
When the star \textit{slightly} penetrates within the tidal radius, the dynamics of disruption has been followed e.g. by Nolthenius \& Katz (\cite{Nol82}, \cite{Nol83}); Regev \& Portnoy (\cite{Reg87}); Evans \& Kochanek (\cite{Eva89}); Laguna et al. (\cite{Lag93b}); Khokhlov et al. (\cite{Kho93}); Fulbright (\cite{Fulthesis95}); Diener et al. (\cite{Die97}); Ayal et al. (\cite{Aya00}); Ivanov \& Novikov (\cite{Iva01}); Ivanov et al. (\cite{Iva03}); Ivanov \& Chernyakova (\cite{Iva06}). 
The distribution and subsequent evolution of the stellar debris have been followed e.g. by Evans \& Kochanek (\cite{Eva89}); Laguna et al. (\cite{Lag93b}); Kochanek (\cite{Koc94}); Lee \& Kang (\cite{Lee96}); Fulbright (\cite{Fulthesis95}); Kim et al. (\cite{Kim99}); Ayal et al. (\cite{Aya00}).  
A quite complete and coherent picture of the tidal disruption process can be drawn from these different studies in this range of moderate star-BH encounters.

In this paper, we interest in the case when the star \textit{deeply} penetrates within the tidal radius. 
In such strong encounters, it was noticed on the basis of geometrical and qualitative arguments (Carter \& Luminet \cite{Car82}) that before being fully disrupted, the star should pass through a short-lived strong compression phase,
at the end of which it should adopt a \textit{pancake}-shape configuration highly flattened in its orbital plane.
As a consequence, it was suggested that thermonuclear reactions could be triggered within the stellar core leading to an explosive disruption.
The astrophysical importance of such an event would reside in the fact that the additional thermonuclear energy release should produce a luminous flare from the core of the galaxy, hence providing an observational signature of a pancake star disruption; that it could be sufficient to give the stellar gas a velocity greater than the escape velocity from the BH, hence challenging the subsequent accretion scenario (see e.g. Rees \cite{Ree88}); finally that some heavy isotopes could be formed through nucleosynthesis processes which would be ultimately ejected in the interstellar medium.

We shortly review in the following section the main studies on strong star-BH encounters before explaining the motivation of the present paper.  
\section{The affine star model versus hydrodynamical models} \label{AMvsHydro}
The first investigations on pancake stars began more than twenty years ago within the framework of the semi-analytical affine star model (AM), both in Newtonian dynamics (Carter \& Luminet \cite{Car83}; Luminet \& Carter \cite{Lum86}) and in general relativity with Schwarzschild BHs (Luminet \& Marck \cite{Lum85}).
The mechanical properties of the AM are described in detail in Carter \& Luminet (\cite{Car85}).

The main result of the AM was to explicitly show that along a plunging orbit within the tidal radius, the tidal field of the BH increases sufficiently rapidly to transitorily surpass the pressure and self-gravity fields of the star (Carter \& Luminet \cite{Car83}).  
The stellar matter then lands up in \textit{free fall} in the external gravitational field, and compresses in the direction orthogonal to the orbital plane in direct response to tidal effects.
Finally near the periastron of the orbit, the pressure field suddenly increases within the stellar core to counteract the compressive contribution.
The stellar matter \textit{bounces} within the core, and quickly \textit{expands} in an almost symmetric way to the free fall while the star is still moving within the tidal radius. 
In the relativistic case, the star actually undergoes several successive compression phases when its orbit crosses itself inside the tidal radius.

At the instant of bounce, the density and temperature of the stellar core, initially equal to the equilibrium values $\rho_{*}$ and $T_{*}$, were found to follow the respective power laws 
\begin{eqnarray}
\frac{\rho_{*}^{\mathrm{m}}}{\rho_{*}} & \sim & \beta^{{\frac{2}{\gamma - 1}}}, \label{P01Eq02} \\
\frac{T_{*}^{\mathrm{m}}}{T_{*}}       & \sim & \beta^{2},                      \label{P01Eq03}
\end{eqnarray}
considering a polytropic gas of adiabatic index $\gamma$, and defining the \textit{penetration factor} within the tidal radius
\begin{equation}
\beta \equiv \frac{R_{\mathrm{T}}}{R_{\mathrm{p}}}, \label{P01Eq04}
\end{equation}
with $R_{\mathrm{p}}$ the star-BH distance at periastron. 
The study of thermonuclear reactions that could be triggered during the tidal compression was taken into account in great detail in Pichon (\cite{Pic85}) and Luminet \& Pichon (\cite{Lum89}).

To go beyond the AM in dealing with the behaviour of the tidally compressed star in a more realistic way, three-dimensional hydrodynamical models based on smoothed particle hydrodynamics (SPH) were developed by Bicknell \& Gingold (\cite{Bic83}), Laguna et al. (\cite{Lag93b}), and Fulbright (\cite{Fulthesis95}).
These models interested in polytropic stars $\gamma=5/3$ of solar mass and radius, but did not take into account the possible thermonuclear energy generation.
The tidal gravitational field was handled in Newtonian dynamics, except in Laguna et al. (\cite{Lag93b}) where it was described in general relativity for Schwarzschild BHs.

The SPH simulations qualitatively confirmed the tidal compression process proposed by the AM as a free fall phase followed by a bounce-expansion phase. 
They nevertheless reported that the stellar core could be less compressed than 
initially predicted, which could possibly impact the explosive disruption 
scenario. 
In particular, for star-BH encounters with $\beta \le 10$, Bicknell \& Gingold (\cite{Bic83}) found that the maximal density of the compressed stellar core scaled as $\beta^{1.5}$. Laguna et al. (\cite{Lag93b}) confirmed this result, whereas Fulbright (\cite{Fulthesis95}) found a $\beta^{2.5}$ dependence.
For $\beta > 10$, Bicknell \& Gingold (\cite{Bic83}) found that the maximal core density decreased as $\beta$ increased.

The authors of the SPH studies mentioned that the ellipsoidal assumption on which is based the AM could partly be responsible for the discrepancy. 
It is true that the SPH simulations clearly showed that the \textit{global} configuration of the star within the tidal radius was not ellipsoidal.
At the instant of maximal core compression, the star had more a crescent-like shape within the orbital plane, while it was not uniformly flattened in the orthogonal direction. 
However, it must be pointed out that this configuration essentially concerned the stellar envelope and hardly the stellar core.
Moreover, since the velocity of the star near the periastron is high enough, of the order of 
$(G M_{\mathrm{BH}}/R_{\mathrm{p}})^{1/2}$ and comparable to the speed of light when the periastron is close to the BH horizon, it must also be pointed out that the non-uniform flattening is fully negligible.
It is therefore quite unlikely that the ellipsoidal assumption of the stellar core postulated in the AM could be rightly questioned.

Noticing an increase of the stellar gas entropy in their simulations, Bicknell \& Gingold (\cite{Bic83}) and Fulbright (\cite{Fulthesis95}) also mentioned that shock waves developed during the compression.
On the contrary, the AM assumed that the compression process occurs without shock wave development at least until the instant of bounce, so that the kinetic energy of compressive motion fully converts into internal energy of stellar gas.
One could then imagine that the shock wave development prematurely halts the free fall phase, and thus leads 
to an effective decrease of the compression and heating factors (\ref{P01Eq02})-(\ref{P01Eq03}) as suggested by the SPH results.
However the true effect of shock waves was not studied in depth and remains up to now still unclear.
One could inversely argue that the lack of reliability of the SPH models could explain the discrepancy with the AM.

On the one hand, the simulations had a poor spatial resolution due to the excessively weak number of particles used ($\approx 500$ for Bicknell \& Gingold \cite{Bic83}, and $\approx 7000$ for Laguna et al. \cite{Lag93b} and Fulbright \cite{Fulthesis95}), but also due to an undesirable effect of the smoothing kernels.
In the SPH approach, the kernels behave as interpolating functions whose size is continually adjusted to keep a constant number of particles as interpolation points.
In these standard conditions, Fulbright et al. (\cite{Ful95}) showed that the usual spherically symmetric kernels were likely to artificially freeze the spatial resolution along the direction of compression of the star.
To improve this problem, Fulbright (\cite{Fulthesis95}) resorted to anisotropic spheroidal kernels.
Bicknell \& Gingold (\cite{Bic83}) also used anisotropic ellipsoidal kernels but with too few particles for the kernels geometry to be effective, whereas Laguna et al. (\cite{Lag93b}) used standard spherical kernels.

On the other hand, the artificial viscosity could produce spurious entropy during the tidal compression.
As is well known, the artificial viscosity is used to numerically deal with shock waves, in order to correctly reproduce the shock front without excessive oscillations and smearing, but also in order to produce entropy during the propagation of the shock wave.
As Fulbright et al. (\cite{Ful95}) also clearly showed, the usual artificial viscosity was likely to produce entropy during the free fall phase even in the physical absence of shock wave development.
Fulbright (\cite{Fulthesis95}) implemented a more adapted artificial viscosity to reduce this excessive dissipation, but Bicknell \& Gingold (\cite{Bic83}) and Laguna et al. (\cite{Lag93b}) used standard forms.
Luminet \& Carter (\cite{Lum86}) already glimpsed this issue where including viscosity in the AM, they showed that they could fully reproduce the results of Bicknell \& Gingold (\cite{Bic83}). 
The excessively high value of viscosity nevertheless suggested that too much dissipation really occured during the SPH simulations.

Apart from SPH, no other numerical strategy has been set up to follow the three-dimensional evolution of a tidally compressed star.
Khokhlov et al. (\cite{Kho93}) solved the hydrodynamical equations using the Eulerian finite-difference flux corrected transport technique, but they exclusively interested in weak penetration factors which did not lead to the compression of the stellar core, and did not investigate deeper encounters due to the lack of spatial resolution of the stationary numerical grid.
Marck et al. (\cite{Mar96}) designed a well-adapted formalism combining symmetry considerations, an adaptive moving grid, and pseudo-spectral methods, but they exemplified their model to the tidal disruption process for a weak penetration factor $\beta = 1.5$, and unfortunately did not performed any extensive study of deeper encounters.

Recently Kobayashi et al. (\cite{Kob04}) considered a one-dimensional hydrodynamical model based on a Lagrangian Godunov-type method (see appendix of their paper) to simulate the direction of compression of the star orthogonal to the orbital plane.
They clearly showed in the single simulated case $\beta = 10$, through 
the density and pressure profiles, that a shock wave formed after the pressure at the centre of the star reached its maximum.
Actually Fulbright (\cite{Fulthesis95}) noticed a quite similar behaviour 
for $\beta=5$ when the increase of the gas entropy in the stellar envelope occurred after the instant of maximal core compression.
Both results then indicate that shock waves are likely to develop until shortly after the bounce phase as long as $\beta \lesssim 10$. 
Kobayashi et al. (\cite{Kob04}) also underlined that the propagation of the shock wave within the star could produce a X-ray flare in the keV energy, thus leading to interesting observational consequences.

In this context, the main motivation of the present investigation is to solve the still remaining controversy between calculations based on the simplified AM and on the SPH models regarding the influence of shock wave development on the compression of the stellar core.
We have performed a detailed hydrodynamical study of the tidal compression process for penetration factors $3 \le \beta \le 15$ covering all the realistic cases of strong encounters between a main-sequence star and a massive BH.    
We have restricted to follow the evolution of the star from the tidal radius until the neighbourhood of the periastron, i.e. during the only free fall and bounce-expansion phases of the stellar matter.
As justified in the next section, this restriction allows to deal with the tidal compression process in a one-dimensional way, considering only the direction of compression orthogonal to the orbital plane and neglecting the motion of the stellar matter along others directions. 
In a similar way of Kobayashi et al. (\cite{Kob04}), the hydrodynamical equations have been solved through a Godunov-type approach, highly competitive in the numerical treatment of shock waves.
Because such an approach does not rely on the artificial viscosity technique, the dissipation issue which was encountered by the SPH models during the compression of the star cannot arise.
Estimates of the compression and heating factors at the instant of bounce have been calculated in order to investigate the possibility of the thermonuclear explosion of the star.

In this preliminary study, we have considered minimal assumptions focusing on a Newtonian description of the BH gravitational field, and a main-sequence star modeled by a polytropic equation of state as was already the case in previous hydrodynamical simulations. 
The extension to general relativistic backgrounds and to a more elaborate equation of state describing the stellar structure is postponed to a forthcoming investigation.

The article is organised as follows.
In Sect.~3, we describe the hydrodynamical model and the Godunov-type numerical method for dealing with shock waves.
In Sect.~4, we present our results and show the existence of two regimes depending on whether the shock wave development occurs before or after the instant of maximal compression of the stellar core.  
The critical value of the penetration factor separating the two regimes sensibly depends on the assumed equation of state.
In Sect.~5, we discuss some astrophysical perspectives.
%
%
%
%
\section{Hydrodynamical model}
\subsection{One-dimensional approximation}
If the full tidal disruption process of the star obviously requires three-dimensional calculations, it is however possible to reduce the dynamics of the \textit{only} tidal compression process to one-dimensional calculations.

As previously mentioned, as soon as the star substantially penetrates (say with $\beta \gtrsim 3$) within the tidal radius, the BH tidal field quickly dominates the internal pressure and self-gravity fields.
The stellar matter gets into a transitional free fall phase in the external gravitational field which fully controls its own dynamics.

It is well known (see appendix~\ref{A02}) that the tidal field produces \textit{stretching} along one of the tidal principal directions parallel to the orbital plane, and \textit{compression} at once along the second tidal principal direction parallel to the orbital plane and along the tidal principal direction orthogonal to the orbital plane (hereafter called the \textit{vertical} direction).
The compressive vertical direction is fixed, whereas both compressive and stretching directions within the orbital plane continually change as the star moves (see Fig.~\ref{A02Fig01} of appendix~\ref{A02}).
Moreover, by symmetry relative to the orbital plane, the induced vertical motion is fully decoupled from the induced motion in the orbital plane so that they can be considered as independent.

Therefore, moving within the tidal radius until the neighbourhood of the periastron, the star is continually compressed along the vertical direction, while in the orbital plane the stretching and compression partly cancel out following the rotation through a nearly right angle of both tidal principal directions. 
The AM (Carter \& Luminet \cite{Car83}) indeed showed that the sections of the star in the orbital plane do not significantly change by a factor more than two, which was further confirmed by the SPH simulations (Fulbright \cite{Fulthesis95}).
For these reasons, the dynamics within the orbital plane can be quite neglected relative to the dynamics along the vertical direction in a first approximation.
It has to be noted that during the subsequent bounce phase, when the stellar matter vertically expands, the deformations within the orbital plane also remain negligible due to the rapidity of the process.    
\subsection{Governing equations}
We assume that the motion of the star in the external gravitational field is equivalent to the motion of a particle of mass $M_{*}$, since the size of the star is small enough relative to its distance from the BH so that the deviation effect on the motion of its centre of mass is weak.
Moreover, the asymptotic velocity $v_{\infty}$ of the star far from the BH being typically of the order of a few hundred $\mathrm{km} \, \mathrm{s}^{-1}$, the specific kinetic energy $v_{\infty}^{2}/2$ is much inferior to (the absolute value of) the BH gravitational potential at the tidal radius $G M_{\bullet}/R_{\mathrm{T}}$ so that
the orbit of the star is assumed parabolic (for a strictly parabolic orbit $v_{\infty}=0$). 
Note that being only interested in the behaviour of the star inside the tidal radius, our results would be applicable to sufficiently elliptic or hyperbolic orbits as well.

\begin{figure}[h!]
\centering
\includegraphics[width=8cm]{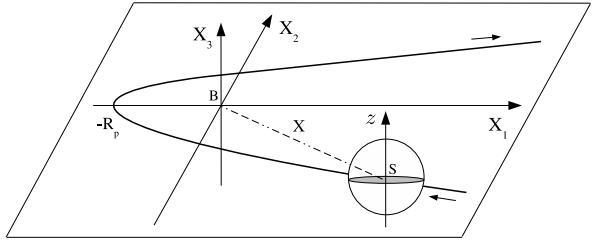}
\caption{Newtonian parabolic orbit of the star centre of mass $S$ relative to the point-mass BH $B$. 
The BH is at the origin of the Cartesian coordinate system $(X_{1},X_{2},X_{3})$ used to follow the star orbit, and oriented such that the motion is contained in the $(X_{1},X_{2})$ plane. 
The distance between the star and the BH is $X$, and the periastron of the orbit is at distance $R_{\mathrm{p}}$ from the BH. 
The small arrows indicate the direction of the star motion as given by (\ref{P02Eq01})-(\ref{P02Eq02}). 
The $z$ axis is attached to the star centre of mass and is at each time orthogonal to the orbital plane $(X_{1},X_{2})$.} 
\label{P02Fig01}
\end{figure}

Considering a Cartesian coordinate system $(X_{1},X_{2},X_{3})$ with origin at the BH and oriented such that the motion of the star is contained in the $(X_{1},X_{2})$ plane (Fig.~\ref{P02Fig01}), the Newtonian parabolic orbital law writes 
\begin{eqnarray}
X_{2}^{3}(t) + 12 \, R_{\mathrm{p}}^{2} \, X_{2}(t) & = & 
12 \, R_{\mathrm{p}} \, (2 G M_{\bullet} R_{\mathrm{p}})^{1/2} \, t, 
\label{P02Eq01} \\
X_{1}(t)                                            & = & 
\frac{X_{2}^{2}(t)}{4 R_{\mathrm{p}}} - R_{\mathrm{p}}, 
\label{P02Eq02} \\ 
X_{3}(t)                                            & = & 
0, 
\label{P02Eq03}
\end{eqnarray}
taking the origin of time $t$ at the periastron of the orbit, and the star-BH distance is
\begin{equation}
X(t) = \frac{X_{2}^{2}(t)}{4 R_{\mathrm{p}}} + R_{\mathrm{p}}. \label{P02Eq04}
\end{equation}

The hydrodynamics is described in the reference frame of the star centre of mass. 
To simulate the compression orthogonal to the orbital plane of the star, we consider the vertical direction $z$ passing through the centre of mass and joining the poles of the star.

The contribution to internal forces is due to the pressure field. 
The simulation of the tidal compression process starts at the tidal radius so that the self-gravity of the star can be neglected relative to the tidal contribution. 
The effects of viscosity, heat conduction, mixing and convection within the stellar matter are neglected because their timescales are long enough compared to the duration of the compression.

The contribution to external forces is due to the BH tidal field.
The general expression of the tidal gravitational acceleration is recalled in appendix \ref{A01}.
From (\ref{A01Eq16}) and (\ref{A02Eq01}), the tidal acceleration at position $z$, when the star is at distance $X(t)$ from the BH, is given by 
\begin{equation}
g(z,t) = -\frac{G M_{\bullet}}{X^{3}(t)} \, z. \label{P02Eq05} 
\end{equation}

The hydrodynamics is governed by the one-dimensional Euler equations for the conservation of mass, momentum, and energy (see e.g. Landau \& Lifshitz \cite{Land63}):
\begin{eqnarray}
&\partial_{t} \, \rho        +  \partial_{z} \, ( \rho v )           = 0,&        \label{P02Eq06} \\
&\partial_{t} \, ( \rho v )  +  \partial_{z} \, ( P + \rho v^{2} )   = \rho g,&   \label{P02Eq07} \\
&\partial_{t} \, ( \rho e )  +  \partial_{z} \, ( ( P + \rho e ) v ) = \rho v g,& \label{P02Eq08}
\end{eqnarray}
where $\rho$ is the density, $P$ the pressure, $v$ the velocity, and $e=\varepsilon + v^{2}/2$ the specific total energy which includes the specific internal energy $\varepsilon$ and the specific kinetic energy.

The stellar matter is considered as a polytropic (ideal) gas of adiabatic index $\gamma$:
\begin{eqnarray}
P & = & \mathcal{R} \, \rho \, T  \label{P02Eq10} \\
  & = & \rho \, \varepsilon \, (\gamma - 1), \label{P02Eq11}
\end{eqnarray}
where $\mathcal{R}$ is the specific gas constant and $T$ the temperature. The (local) speed of sound is given by 
\begin{equation}
a = \sqrt{\frac{\gamma \, P}{\rho}}. \label{P02Eq12}
\end{equation}

The Euler equations (\ref{P02Eq06})-(\ref{P02Eq08}) are expressed in conservation form  
\begin{equation}
\partial_{t} \, \vec{U} + \partial_{z} \, \vec{F}(\vec{U}) = \vec{S}(\vec{U},z,t), \label{P02Eq16}
\end{equation}
defining
\begin{eqnarray}
\vec{U}                 & \equiv & \left[ \rho,   \rho v,         \rho e           \right]^{T}, \label{P02Eq13} \\
\vec{F} ( \vec{U} )     & \equiv & \left[ \rho v, P + \rho v^{2}, ( P + \rho e ) v \right]^{T}, \label{P02Eq14} \\
\vec{S} ( \vec{U},z,t ) & \equiv & \left[ 0,      \rho g,         \rho v g         \right]^{T}, \label{P02Eq15}
\end{eqnarray}
respectively the vector of conserved variables, of fluxes, and of sources.
\subsection{Numerical method}
The Euler hyperbolic system of conservation laws (\ref{P02Eq16}) is solved by a Godunov-type high-resolution shock-capturing finite-volume method (see e.g. LeVeque \cite{Lev02}; Toro \cite{Tor99} for textbooks).

Dividing the spatial domain into cells of right (resp. left) boundary $z_{i \pm 1/2}$, centre $z_{i}=(z_{i-1/2}+z_{i+1/2})/2$, and width $\Delta z_{i}=z_{i+1/2} - z_{i-1/2}$,
\begin{equation}
\vec{\bar{U}}(z_{i},t) \equiv \frac{1}{\Delta z_{i}} \int_{z_{i-1/2}}^{z_{i+1/2}} \vec{U}(z,t) \, dz 
\label{P02Eq17}
\end{equation}
is defined as the average value over the $i^{\mathrm{th}}$ cell of the conserved variables (\ref{P02Eq13}), and 
\begin{equation}
\vec{\bar{U}}_{i}^{n} \approx \vec{\bar{U}}(z_{i},t^{n}) 
\label{P02Eq18}
\end{equation}
as its numerical approximation at time level $t^{n}$.

A source splitting approach is used to deal with the source terms, splitting (\ref{P02Eq16}) into the (homogeneous) system of conservation laws
\begin{equation}
\partial_{t} \, \vec{U} + \partial_{z} \, \vec{F}(\vec{U}) = \vec{0}, \label{P02Eq19}
\end{equation}
and the system of ordinary differential equations  
\begin{equation}
\partial_{t} \, \vec{U} = \vec{S}(\vec{U},z,t). \label{P02Eq20}
\end{equation}
We consider the Strang splitting where the set of numerical solutions $\left\{ \vec{\bar{U}}_{i}^{n} \right\}$ are advanced from time level $t^{n}$ to time level $t^{n+1} = t^{n} + \Delta t^{n}$ according to
\begin{equation} 
\vec{\bar{U}}_{i}^{n+1} = 
\mathcal{L}_{1} \left( \Delta t^{n}/2 \right) \,
\mathcal{L}_{2} \left( \Delta t^{n} \right) \,
\mathcal{L}_{1} \left( \Delta t^{n}/2 \right) \,
\left\{ \vec{\bar{U}}_{i}^{n} \right\},
\label{P02Eq21}
\end{equation}
with $\Delta t^{n}$ the time step constrained by the CFL condition, $\mathcal{L}_{1}$ and $\mathcal{L}_{2}$ the independent solution operators for (\ref{P02Eq19}) and (\ref{P02Eq20}) respectively.
For clarity, both solution operators are next expressed over the time step $\Delta t^{n}$ from time level $t^{n}$ to time level $t^{n+1}$. 
\\
\\
\textbf{Numerical scheme for the system of conservation laws}

From the integration of (\ref{P02Eq19}) over the space-time domain $[z_{i-1/2},z_{i+1/2}] \times [t^{n},t^{n+1}]$ and using (\ref{P02Eq17}), 
\begin{eqnarray}
\vec{\bar{U}}(z_{i},t^{n+1}) = \vec{\bar{U}}(z_{i},t^{n}) 
& + & \frac{1}{\Delta z_{i}} \Bigg( \int_{t^{n}}^{t^{n+1}} \vec{F}(\vec{U}(z_{i-1/2},t)) \, dt \nonumber \\ 
& - & \int_{t^{n}}^{t^{n+1}} \vec{F}(\vec{U}(z_{i+1/2},t)) \, dt \Bigg). 
\label{P02Eq22}
\end{eqnarray}
The temporal evolution of the numerical solution (\ref{P02Eq18}) is then computed by the explicit scheme in conservation form 
\begin{equation}
\vec{\bar{U}}_{i}^{n+1} = 
\vec{\bar{U}}_{i}^{n} 
+ \frac{\Delta t^{n}}{\Delta z_{i}} 
\left( \vec{\hat{F}}_{i-1/2} - \vec{\hat{F}}_{i+1/2} \right), 
\label{P02Eq23}
\end{equation}
where the numerical fluxes
\begin{equation}
\vec{\hat{F}}_{i + 1/2} \approx 
\frac{1}{\Delta t^{n}} \int_{t^{n}}^{t^{n+1}} \vec{F}(\vec{U}(z_{i + 1/2},t)) \, dt 
\label{P02Eq24}
\end{equation}
are defined as the approximation of the average value over the time interval $[t^{n},t^{n+1}]$ of the fluxes (\ref{P02Eq14}) at the right boundary of the $i^{\mathrm{th}}$ cell.

We have considered the conservation form scheme implemented in the Relativistic Piecewise Parabolic Method (RPPM) of Mart\'\i~\& M\"uller (\cite{Mar96}).
The RPPM actually solves the one-dimensional homogeneous Euler equations in special relativity. 
However, since in our present application of the tidal compression the velocity of the stellar gas is much smaller than the speed of light, the relativistic Euler equations reduces to the Newtonian version.
Following the Godunov-type approach, the numerical fluxes $\vec{\hat{F}}_{i + 1/2}$ are computed in the RPPM by solving at time level $t^{n}$ a Riemann problem at the cell boundary $z_{i+1/2}$:
\begin{displaymath}
\partial_{t} \, \vec{U} + \partial_{z} \, \vec{F}(\vec{U}) = \vec{0} 
\end{displaymath}
\begin{equation}
\vec{U}(z,t^{n}) = 
\left \{ 
\begin{array}{ll}
\vec{U}_{i+1/2}^{\mathrm{L}} & \quad z < z_{i+1/2} \\
\vec{U}_{i+1/2}^{\mathrm{R}} & \quad z > z_{i+1/2}
\end{array} 
\right. 
\label{P02Eq25}
\end{equation}
with constant states $\vec{U}_{i+1/2}^{\mathrm{L}}$ and $\vec{U}_{i+1/2}^{\mathrm{R}}$.
The RPPM implements an exact solution of the Riemann problem (\ref{P02Eq25}). 
For practical reasons, we have instead switched to an approximate solution through the use of a Roe-type Riemann solver where the numerical fluxes are given by
\begin{eqnarray}
\vec{\hat{F}}_{i + 1/2}
=
\frac{1}{2} \,
\Bigg( 
\vec{F}(\vec{U}_{i+1/2}^{\mathrm{L}}) 
&+& \vec{F}(\vec{U}_{i+1/2}^{\mathrm{R}}) \nonumber \\ 
&-& \sum_{k = 0, \pm} \vert \tilde{\lambda}_{k} \vert \,
\vec{\tilde{l}}_{k} \cdot ( \vec{U}_{i+1/2}^{\mathrm{R}} - \vec{U}_{i+1/2}^{\mathrm{L}} ) \,
\vec{\tilde{r}}_{k}
\Bigg)
\label{P02Eq26}
\end{eqnarray}
with $\tilde{\lambda}_{k}$, $\vec{\tilde{r}}_{k}$, and $\vec{\tilde{l}}_{k}$ respectively the eigenvalues, the right and left eigenvectors (see e.g. Toro \cite{Tor99}) of the Jacobian matrix 
\begin{equation}
\frac{d \vec{F}(\vec{U})}{d \vec{U}}  \label{P02Eq27}
\end{equation}
evaluated at the average value $\vec{U} = \frac{1}{2} \left( \vec{U}_{i+1/2}^{\mathrm{L}} + \vec{U}_{i+1/2}^{\mathrm{R}} \right)$.
\\
\\
\textbf{Numerical scheme for the system of ordinary differential equations}

From the integration of (\ref{P02Eq20}) over the space domain $[z_{i-1/2},z_{i+1/2}]$ and using (\ref{P02Eq17}),
\begin{equation}
\frac{d}{dt} \vec{\bar{U}}(z_{i},t) = \vec{\bar{S}}(z_{i},t) \label{P02Eq28}
\end{equation} 
where 
\begin{equation}
\vec{\bar{S}}(z_{i},t) \equiv \frac{1}{\Delta z_{i}} 
\int_{z_{i-1/2}}^{z_{i+1/2}} \vec{S}(\vec{U}(z,t),z,t) \, dz
\label{P02Eq29}
\end{equation}
is defined as the average value over the $i^{\mathrm{th}}$ cell of the sources (\ref{P02Eq15}).

It is possible to explicitly express $\vec{\bar{S}}$ as function of $\vec{\bar{U}}$.
Let us call $V_{j}$ the $j^{\mathrm{th}}$ component of any vector $\vec{V}$. 
From (\ref{P02Eq15}) and (\ref{P02Eq13}), (\ref{P02Eq29}) writes
\begin{eqnarray}
\vec{\bar{S}}(z_{i},t) = 
\left[
0,  
\frac{1}{\Delta z_{i}} \int_{z_{i-1/2}}^{z_{i+1/2}} U_{1}(z,t) \, g(z,t) \, dz, 
\right. \nonumber \\
\left. 
\frac{1}{\Delta z_{i}} \int_{z_{i-1/2}}^{z_{i+1/2}} U_{2}(z,t) \, g(z,t) \, dz
\right]^{T}.
\label{P02Eq30}
\end{eqnarray}
Since the cell width $\Delta z_{i}$ is small, 
\begin{eqnarray}
&         & \frac{1}{\Delta z_{i}} \int_{z_{i-1/2}}^{z_{i+1/2}} U_{j}(z,t) \, g(z,t) \, dz \nonumber \\
& \approx & 
            \frac{1}{\Delta z_{i}} \int_{z_{i-1/2}}^{z_{i+1/2}} U_{j}(z,t) \, dz
            \times
            \frac{1}{\Delta z_{i}} \int_{z_{i-1/2}}^{z_{i+1/2}} g(z,t) \, dz 
\label{P02Eq31}
\end{eqnarray}
so that 
\begin{equation}
\vec{\bar{S}}(z_{i},t) = 
\left[
0,\bar{U}_{1}(z_{i},t) \, \bar{g}(z_{i},t),\bar{U}_{2}(z_{i},t) \, \bar{g}(z_{i},t)
\right]^{T},
\label{P02Eq32}
\end{equation}
defining
\begin{equation}
\bar{g}(z_{i},t) \equiv \frac{1}{\Delta z_{i}} \int_{z_{i-1/2}}^{z_{i+1/2}} g(z,t) \, dz
\label{P02Eq33}
\end{equation}
the average value over the $i^{\mathrm{th}}$ cell of the tidal acceleration (\ref{P02Eq05}) which writes 
\begin{equation}
\bar{g}(z_{i},t) = -\frac{G M_{\bullet}}{X^{3}(t)} \, z_{i}, \label{P02Eq35}
\end{equation}
with the star-BH distance $X(t)$ evolving with time according to the orbital law (\ref{P02Eq01})-(\ref{P02Eq04}).

Therefore by (\ref{P02Eq28}) and (\ref{P02Eq32}), the temporal evolution of the numerical solution 
(\ref{P02Eq18}) is computed from the first-order system of ordinary differential equations
\begin{eqnarray}
\frac{d}{dt} \vec{\mathcal{U}}(t) 
& = & 
\vec{f}(\vec{\mathcal{U}}(t),t) \nonumber \\
& = &
\left[ 0, \mathcal{U}_{1}(t) \, \bar{g}(z_{i},t), \mathcal{U}_{2}(t) \, \bar{g}(z_{i},t) \right]^{T},
\end{eqnarray}
which is solved from time level $t^{n}$ where $\vec{\mathcal{U}}(t^{n})=\vec{\bar{U}}_{i}^{n}$ to time level $t^{n+1}$ where $\vec{\mathcal{U}}(t^{n+1})=\vec{\bar{U}}_{i}^{n+1}$.
We have performed the temporal integration with the explicit fourth-order accurate four-stages Runge-Kutta method (see e.g. Lambert \cite{Lam91}).           
%
%
%
%
\section{Results}
\subsection{Initial and boundary conditions}
In the following, we consider a BH of mass $M_{\bullet}=10^{6} M_{\odot}$, and a star of mass $M_{*}=M_{\odot}$ and radius $R_{*}=R_{\odot}$ initially modeled as a polytrope of polytropic index $3/2$.
The profiles of density and pressure along the vertical direction $z$ are obtained from the resolution of the Lane-Emden equation (see e.g. Chandrasekhar \cite{Cha67}).
The profile of temperature deduces from the ideal gas law (\ref{P02Eq10}), whereas the stellar gas is uniformly at rest.
Since the vertical compression of the star is symmetric relative to the orbital plane, the spatial domain is restricted to the positive values $0 \le z \le R_{*}$, where the origin corresponds to the centre of the star. 
We note $\rho_{*}$ the initial density, $P_{*}$ the initial pressure, and $T_{*}$ the initial temperature at the centre of the star, with $\rho_{*}=8.43 \times 10^{3} \, \mathrm{kg} \, \mathrm{m}^{-3}$, $P_{*}=8.65 \times 10^{14} \, \mathrm{kg} \, \mathrm{m}^{-1} \, \mathrm{s}^{-2}$, and $T_{*}=7.53 \times 10^{6} \, \mathrm{K}$. 
The speed of light in vacuum is noted $c$. 
As said before, in order to be able to neglect the self-gravity of the star during the tidal compression process, and to study the motion of the stellar matter in a one-dimensional way, the simulations start when the star is at the tidal radius (\ref{P01Eq01}).

Like in previous SPH calculations, the evolution of the stellar matter obeys the polytropic gas equation of state (\ref{P02Eq11}) of adiabatic index $\gamma = 5/3$, well-adapted to the structure of small main-sequence stars.
However we have also interested in cases with $\gamma = 4/3$ in order to see the influence of the adiabatic index on the shock wave development.

We consider penetration factors (\ref{P01Eq04}) of the star within the tidal radius $3 \le \beta \le 15$.
Let us mention that the full disruption of a polytropic star $\gamma=5/3$ on a parabolic orbit roughly occurs for $\beta \gtrsim 1$.
In particular, the star is disrupted for $\beta \ge 0.7$ in the AM (Luminet \& Carter \cite{Lum86}), whereas it is disrupted for $\beta \ge 0.9$ in the improved AM of Ivanov \& Novikov (\cite{Iva01}), or for $\beta \ge 1.5$ in the SPH model of Fulbright (\cite{Fulthesis95}).
Note that the parameter $\eta \equiv \left( R_{\mathrm{p}} / R_{*} \right)^{3/2}  \left( M_{*} / M_{\bullet} \right)^{1/2}$ is sometimes used to quantify the strength of the star-BH encounter instead of the penetration factor, with the relation $\eta = \beta^{-3/2}$, so that the disruption occurs for $\eta \lesssim 1$.

From a numerical point of view, the spatial domain $0 \le z \le R_{*}$ is divided in sub-intervals of unequal length uniformly discretized, and has been covered by $\approx 8000$ cells in the different simulations.  
During the compression of the stellar gas, the more exterior cells of the numerical domain become empty, i.e. the density in these cells decreases, the more interior cells fill. 
A cell is considered empty when the density becomes inferior to the critical value $10^{-5} \rho_{*}$, and is taken off the numerical domain by imposing vacuum values (finite small values for numerical reasons) to the hydrodynamical variables.
This condition allows to follow the decrease of the star radius with time. 
Also, ghost cells are added to compute the numerical fluxes (\ref{P02Eq26}) for the cells situated at the extremities of the numerical domain.
At each time step, reflecting boundary conditions are imposed to ghost cells to the left of the numerical domain due to the symmetry relative to the orbital plane, whereas vacuum conditions are imposed to ghost cells to the right.
\subsection{Case $\gamma = 5/3$}
\subsubsection{Description of the tidal compression process}
The hydrodynamical simulations performed for star-BH encounters in the range $3 \le \beta \le 15$ have underlined that the evolution of the tidal compression process actually depends on the penetration factor with respect to a critical value $\approx 12$ for a polytropic gas $\gamma=5/3$.  
\\
\\
\textbf{Encounters \mbox{\boldmath $3 \le \beta < 12$}}

\begin{figure}[h!]
\includegraphics[width=8.5cm]{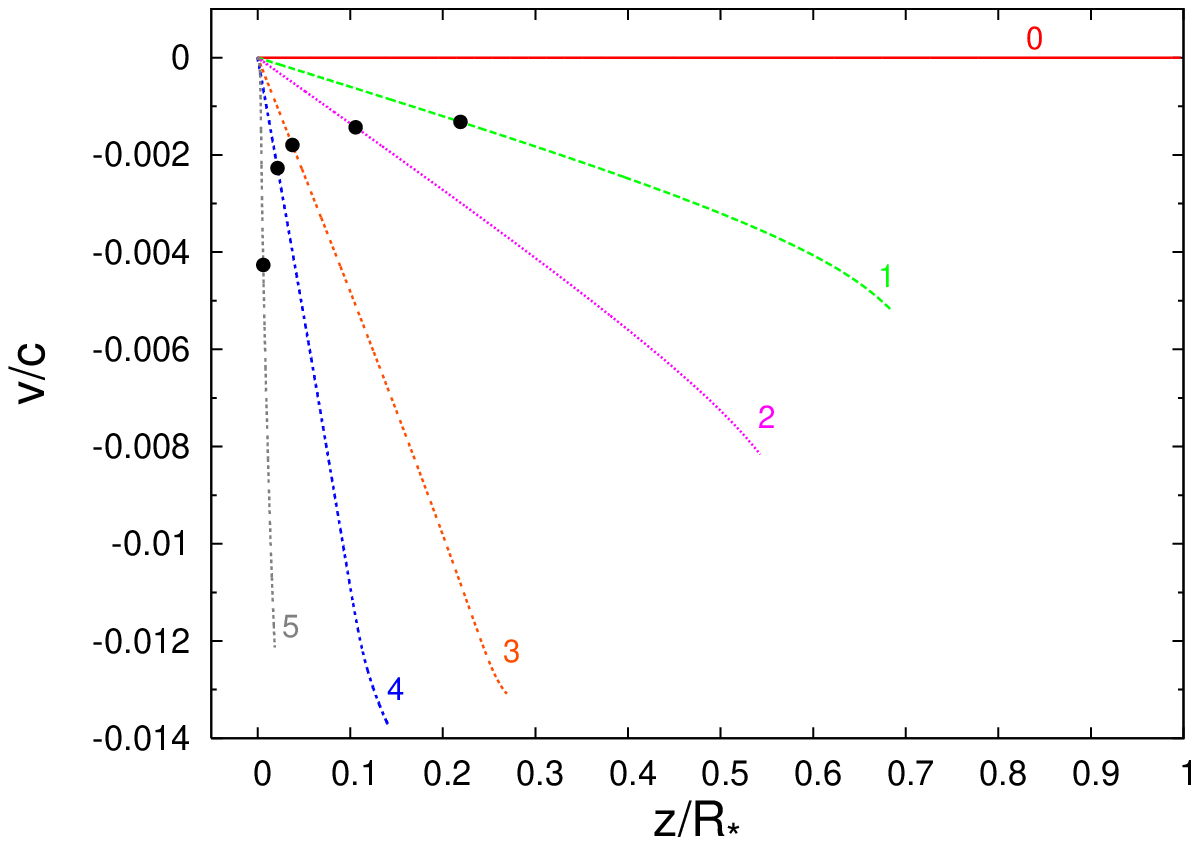} \\ \\
\includegraphics[width=8.5cm]{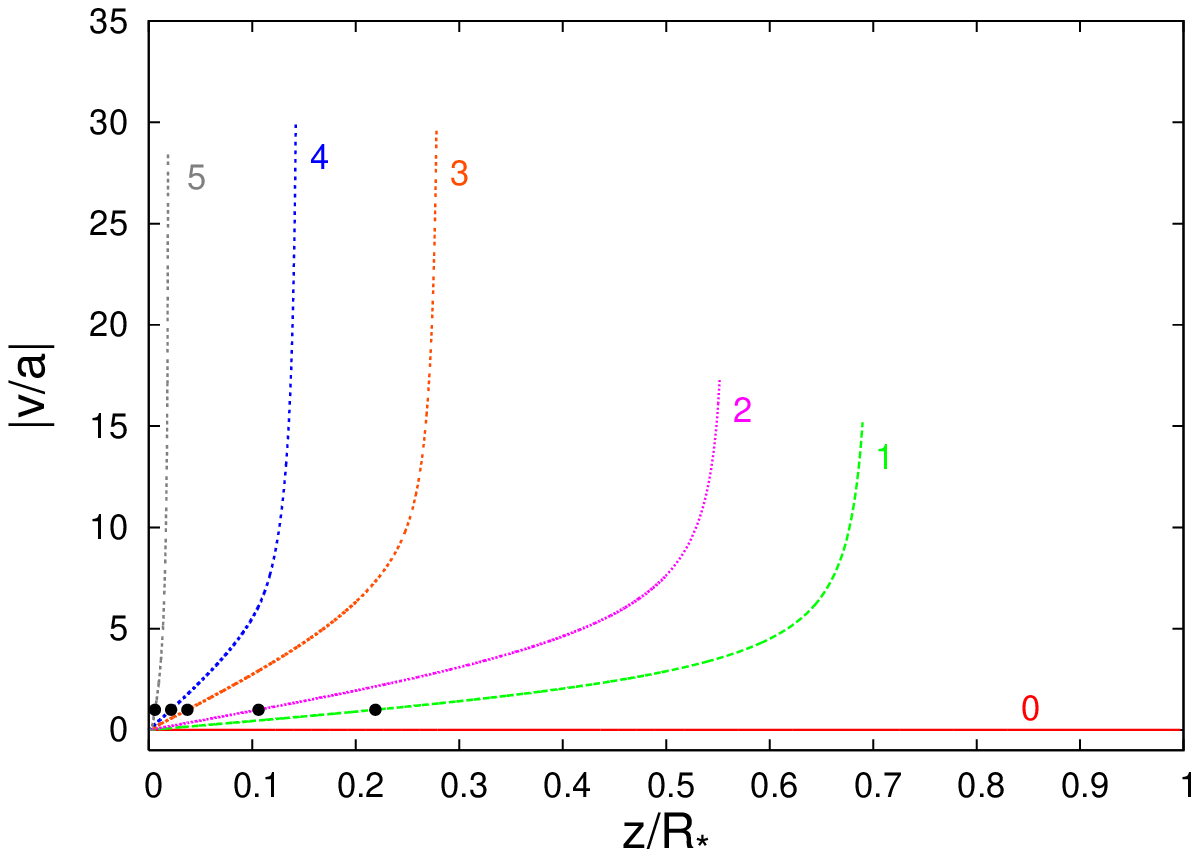}
\caption{ 
Velocity and Mach number profiles along the positive vertical direction $z$ at different times $t$ during the free fall phase for $\beta=7$ and $\gamma=5/3$. 
Labels stand for the following values of $t$ $[\mathrm{s}]$: (0)~$-894.56$, (1)~$-105.02$, (2)~$-58.57$, (3)~$6.30 \times 10^{-3}$, (4)~23.19, (5)~44.97.
The star is at the tidal radius at $t=-894.56$ and at the periastron at $t=0$. 
The Mach number is defined by $\vert v(z,t)/a(z,t) \vert$, with $a(z,t)$ the local speed of sound given by Eq.~(\ref{P02Eq12}).
The position $z_{\mathrm{s}}$ of the sonic point where $\vert v(z_{\mathrm{s}},t) \vert = a(z_{\mathrm{s}},t)$ is indicated by the black points.
At left (resp. right) to the sonic point, the flow is subsonic (resp. supersonic) with $\vert v(z<z_{\mathrm{s}},t) \vert < a(z<z_{\mathrm{s}},t)$ (resp. $\vert v(z>z_{\mathrm{s}},t) \vert > a(z>z_{\mathrm{s}},t)$).
The homologous velocity profile, i.e. $v(z,t) \sim z$, shows that the whole stellar matter collapses in free fall in the BH gravitational field.}
\label{P03Fig01}
\end{figure}

The evolution of the velocity profile for an encounter with $\beta=7$ is reproduced on Fig.~\ref{P03Fig01}, between the instant when the star is initially at the tidal radius and the instant when the compression at the centre of the star is maximum.
The \textit{homologous} velocity distribution, i.e. $v(z,t) \sim z$, at the different times indicates that the contribution of the internal pressure field is fully negligible relative to the contribution of the BH tidal field (see Bicknell \& Gingold \cite{Bic83}).
Therefore, the whole stellar matter vertically \textit{collapses} in \textit{free fall} in the external gravitational field, actually with both subsonic and supersonic velocities. 
The free fall velocities increase as the star approaches the periastron, and leads in the present case to high Mach numbers up to $\approx 25$ (Fig.~\ref{P03Fig01} bottom).

\begin{figure}[h!]
\resizebox{\hsize}{!}{\includegraphics{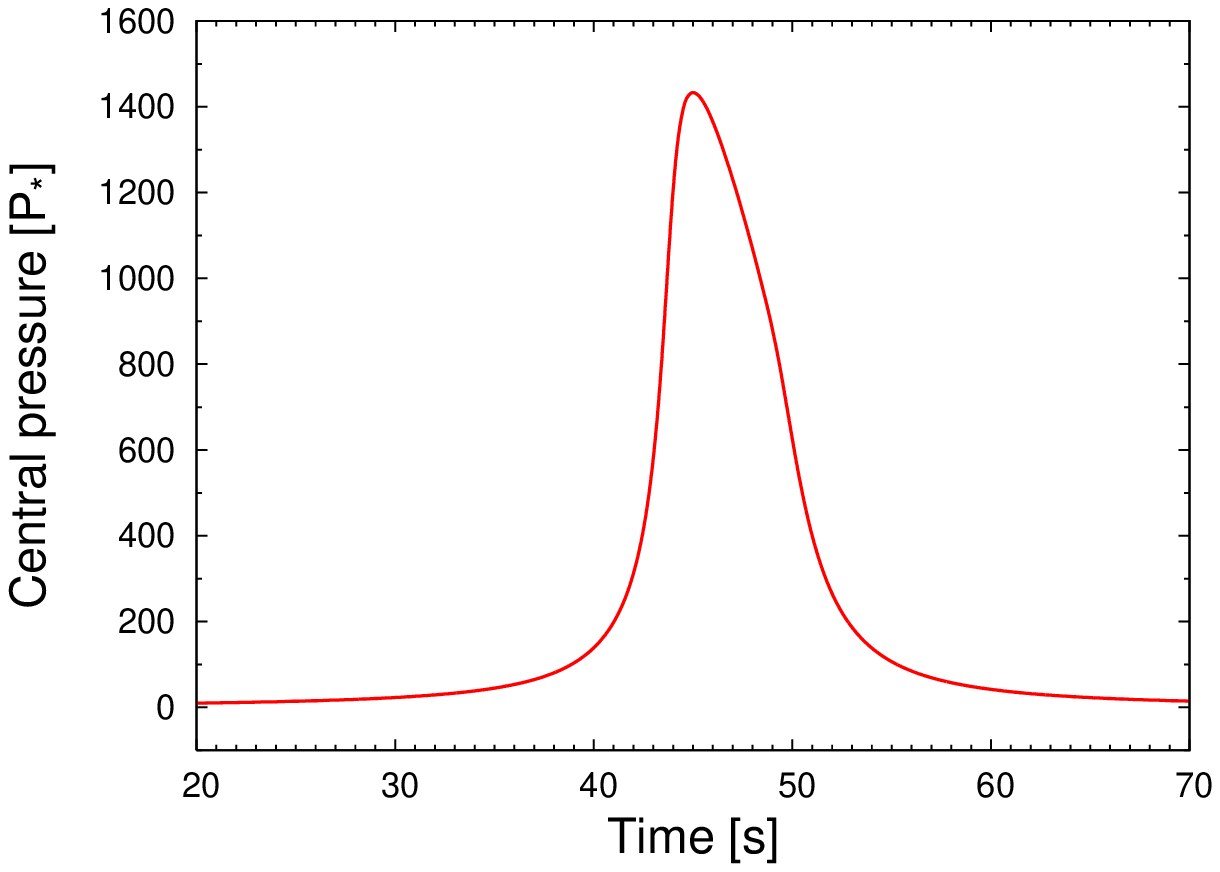}}
\caption{
Evolution of the central pressure as function of time $t$ for $\beta=7$ and $\gamma=5/3$.
After the passage of the star through the periastron at $t=0$, the stellar matter suddenly compresses at the centre of the star until the pressure reaches the sufficient magnitude to counteract the compressive contribution of the BH tidal field. 
The central pressure is maximal at $t \approx 44.97$.}
\label{P03Fig03}
\end{figure}

\begin{figure}[h!]
\includegraphics[width=8.5cm]{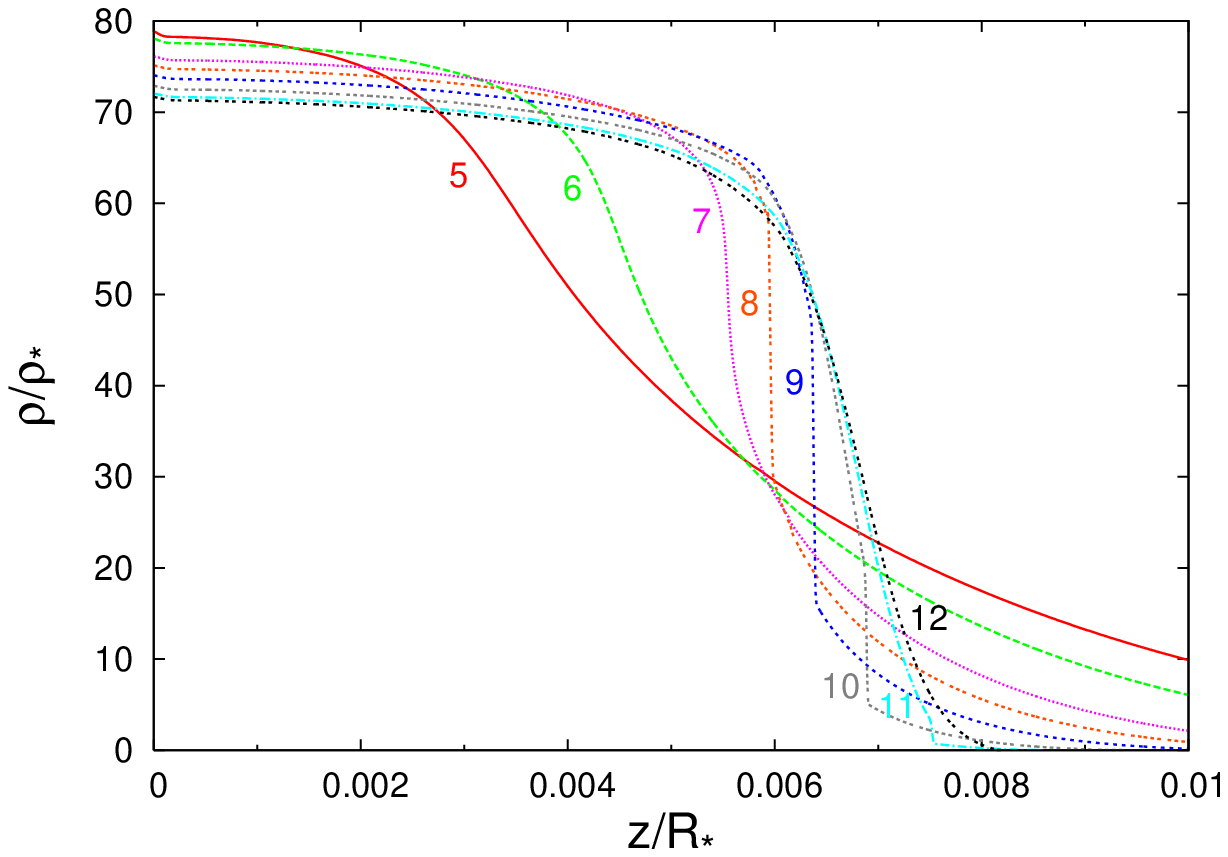} \\ \\
\includegraphics[width=8.5cm]{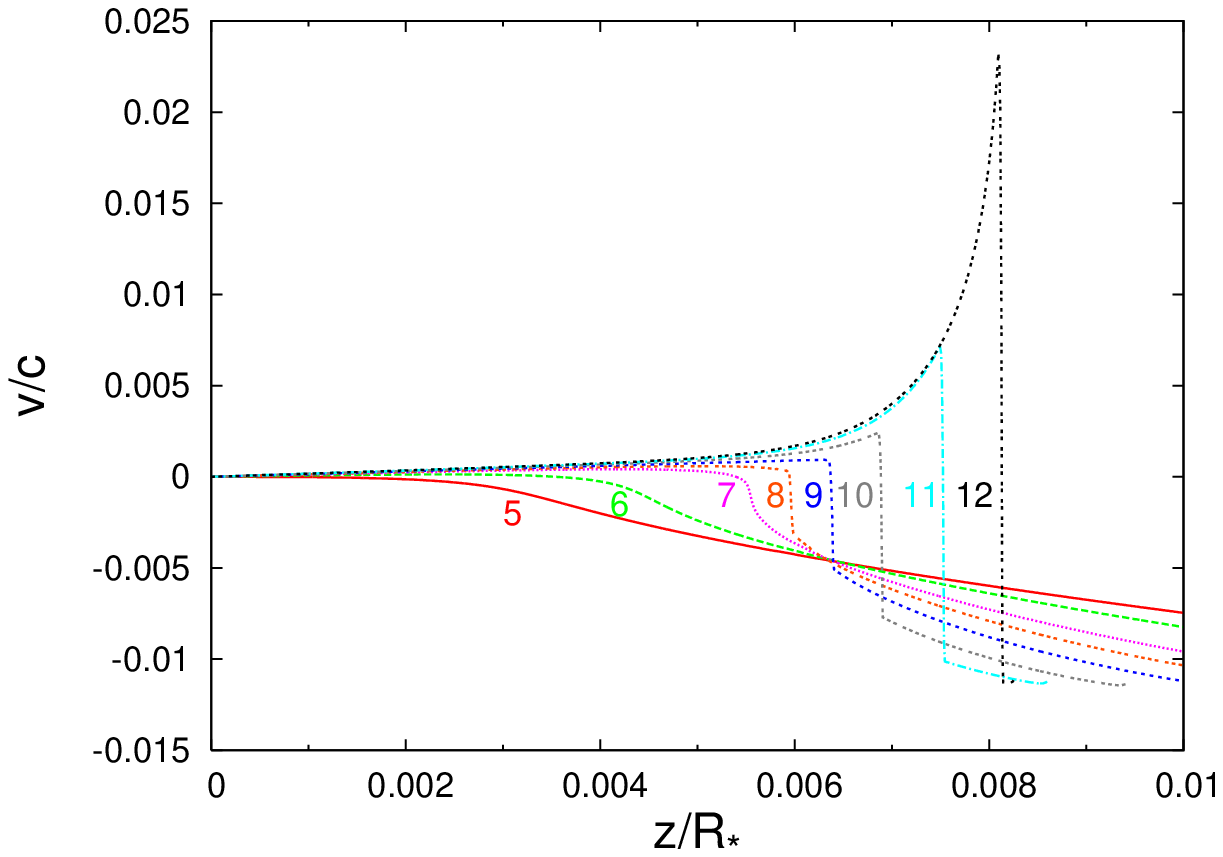}
\caption{
Density and velocity profiles along the positive vertical direction $z$ at different times $t$ during the bounce-expansion phase for $\beta=7$ and $\gamma=5/3$.
Labels stand for the following values of $t$ $[\mathrm{s}]$: (5)~44.97, (6)~45.51, (7)~46.08, (8)~46.32, (9)~46.56, (10)~46.79, (11)~46.96, (12)~47.03. 
The star is at the periastron at $t=0$. 
The central compression is maximal at $t \approx 44.97$. 
The stellar matter expands from the centre of the star while it collapses elsewhere. 
A shock wave forms in front of the expansion and propagates outwards. 
The collapse stops at $t \approx 47.04$.}
\label{P03Fig05_A}
\end{figure}

\begin{figure}[h!]
\includegraphics[width=8.5cm]{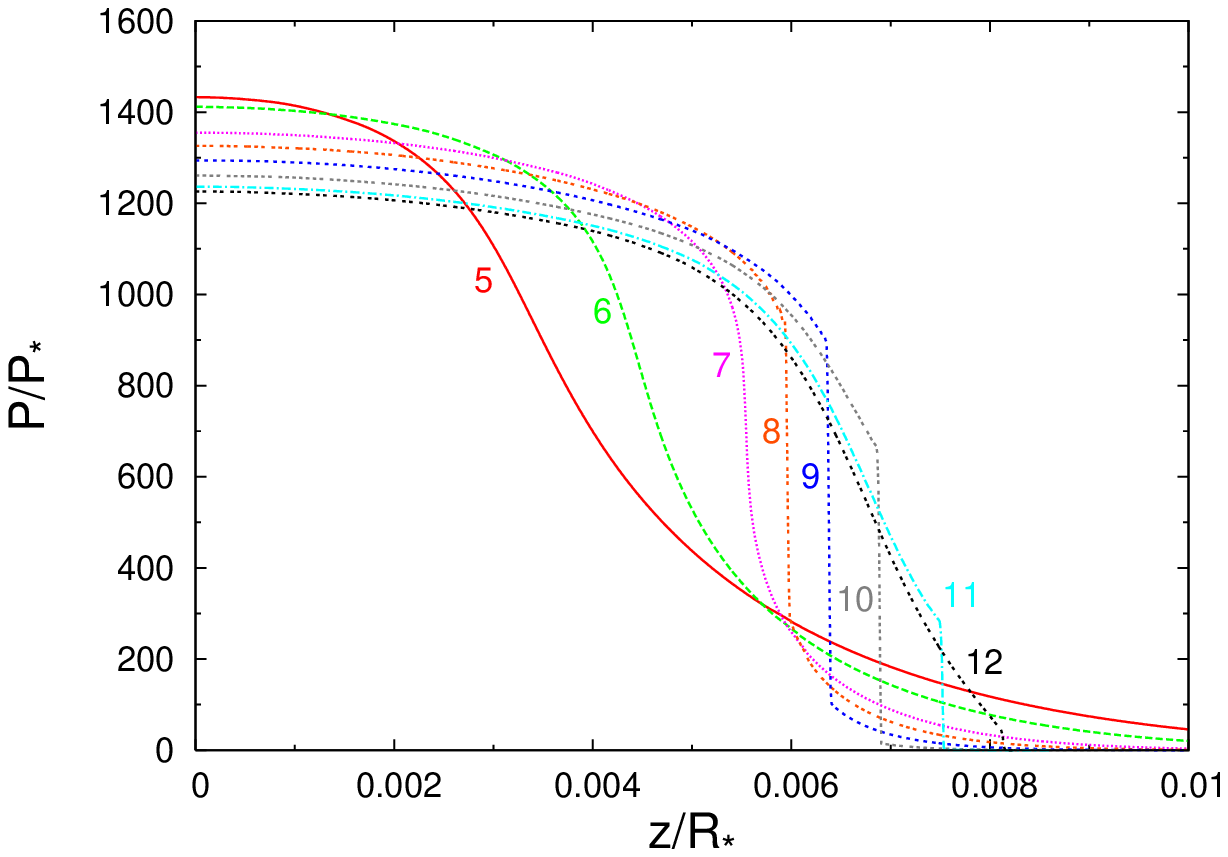} \\ \\
\includegraphics[width=8.5cm]{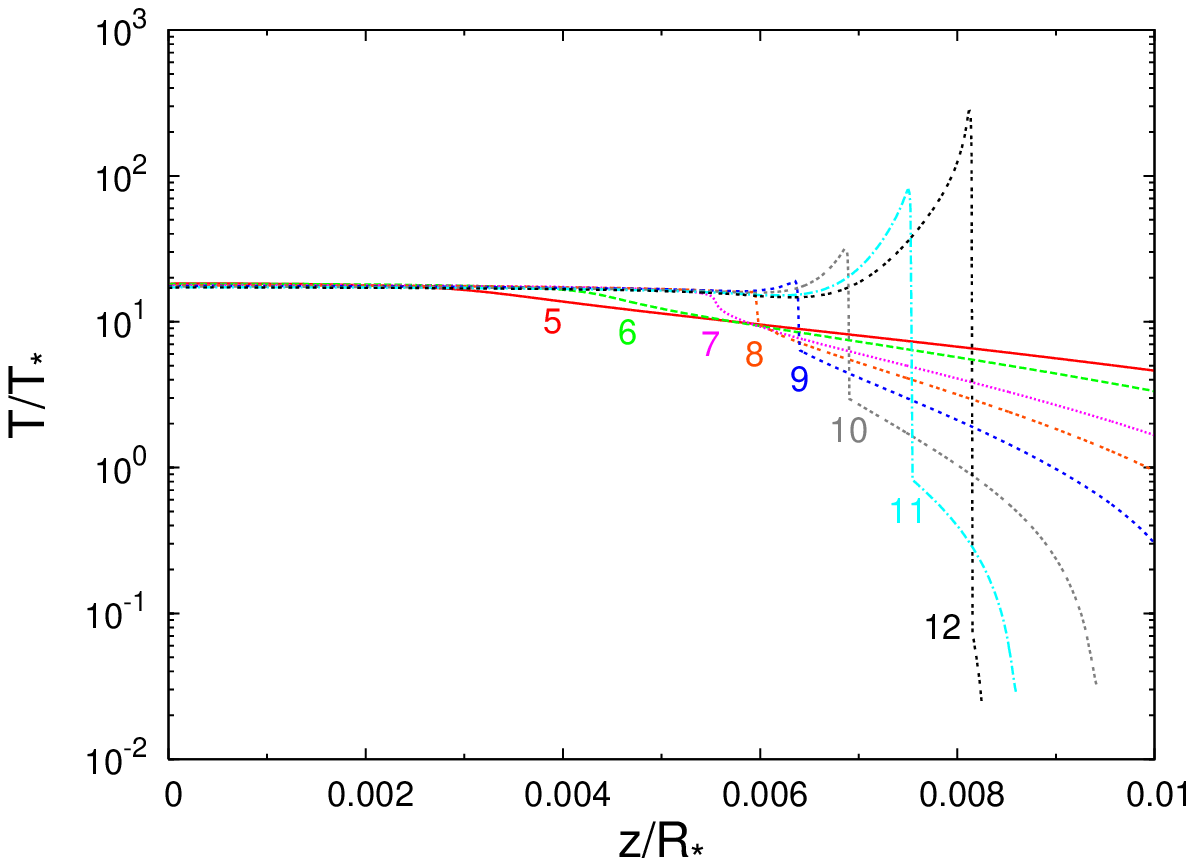}
\caption{ 
Pressure and temperature profiles along the positive vertical direction $z$ at different times $t$ during the bounce-expansion phase for $\beta=7$ and $\gamma=5/3$.
Same labels as in Fig.~\ref{P03Fig05_A}.}
\label{P03Fig05_B}
\end{figure}

The free fall phase finally breaks up following the sudden increase of the compression at the centre of the star (Fig.~\ref{P03Fig03}), just after the passage through the periastron of the orbit. 
Indeed the tidal field decreases after the periastron, but continues to compress the stellar matter until the central pressure suddenly builds up to fully counteract the compressive contribution.
The stellar matter then \textit{bounces} and \textit{expands} with subsonic velocities from the centre, while it continues to collapse elsewhere with subsonic and supersonic velocities.  
The evolution of the hydrodynamical variables can be followed on Figs.~\ref{P03Fig05_A}-\ref{P03Fig05_B}.

\begin{figure}[h!]
\resizebox{\hsize}{!}{\includegraphics{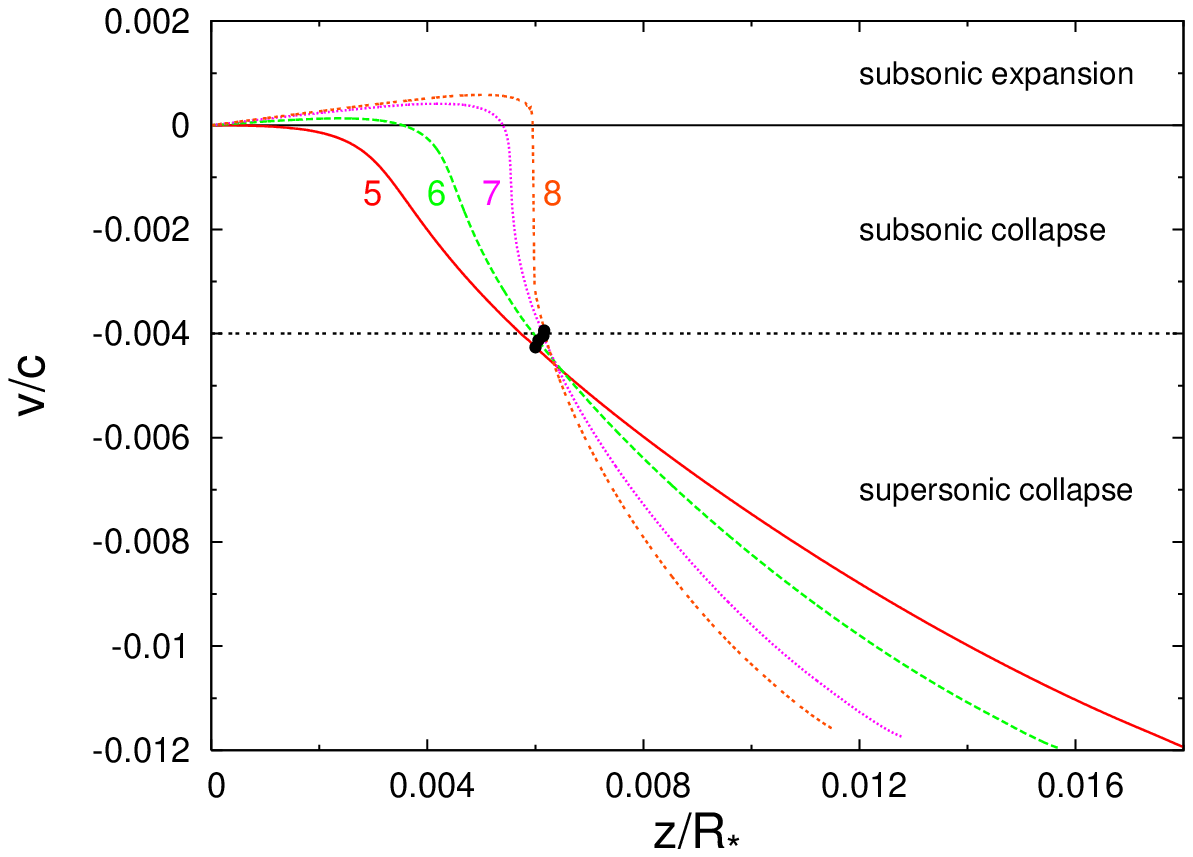}}
\caption{
Velocity profiles along the positive vertical direction $z$ at different times $t$ during the bounce-expansion phase for $\beta=7$ and $\gamma=5/3$ (enlargement of Fig.~\ref{P03Fig05_A} bottom). 
The stellar matter subsonically expands from the centre of the star while it both subsonically and supersonically collapses elsewhere. 
Both opposite motions produce pressure waves which propagate outwards with the speed of sound, and which accumulate near the sonic point (black points) where they steepen into a shock wave.}
\label{P03Fig04}
\end{figure}

During that phase, pressure waves are produced and accumulate near the sonic point, where they finally steepen into a shock wave (Fig.~\ref{P03Fig04}).
Powered by the expanding motion, the shock wave propagates outwards through the collapsing matter, imparts momentum in its own direction (Fig.~\ref{P03Fig05_A} bottom) and heats the medium (Fig.~\ref{P03Fig05_B} bottom), until it reaches the star radius.
The whole stellar matter then continues to expand, but we did not further follow the evolution because the one-dimensional approximation becomes thereafter less and less valid.

\begin{figure}[h!]
\includegraphics[width=8.5cm]{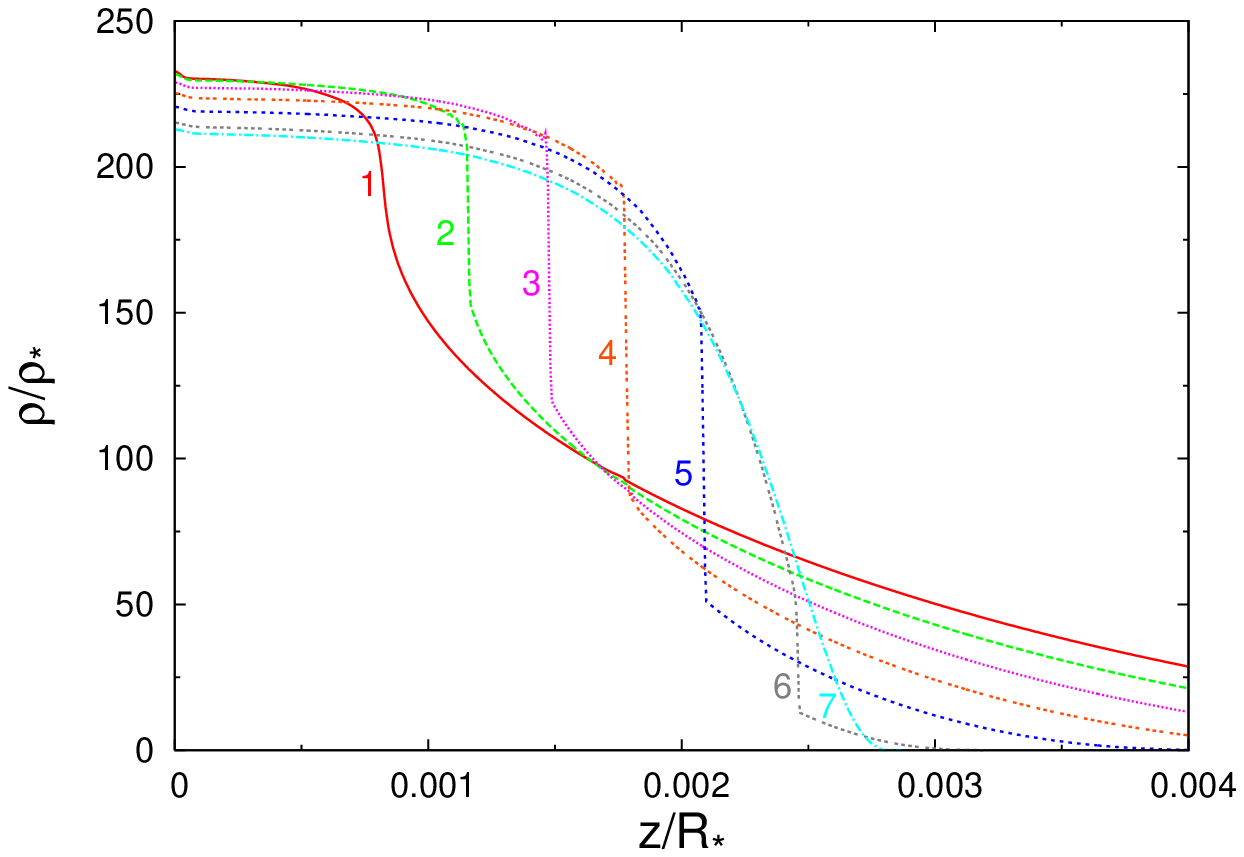} \\ \\
\includegraphics[width=8.5cm]{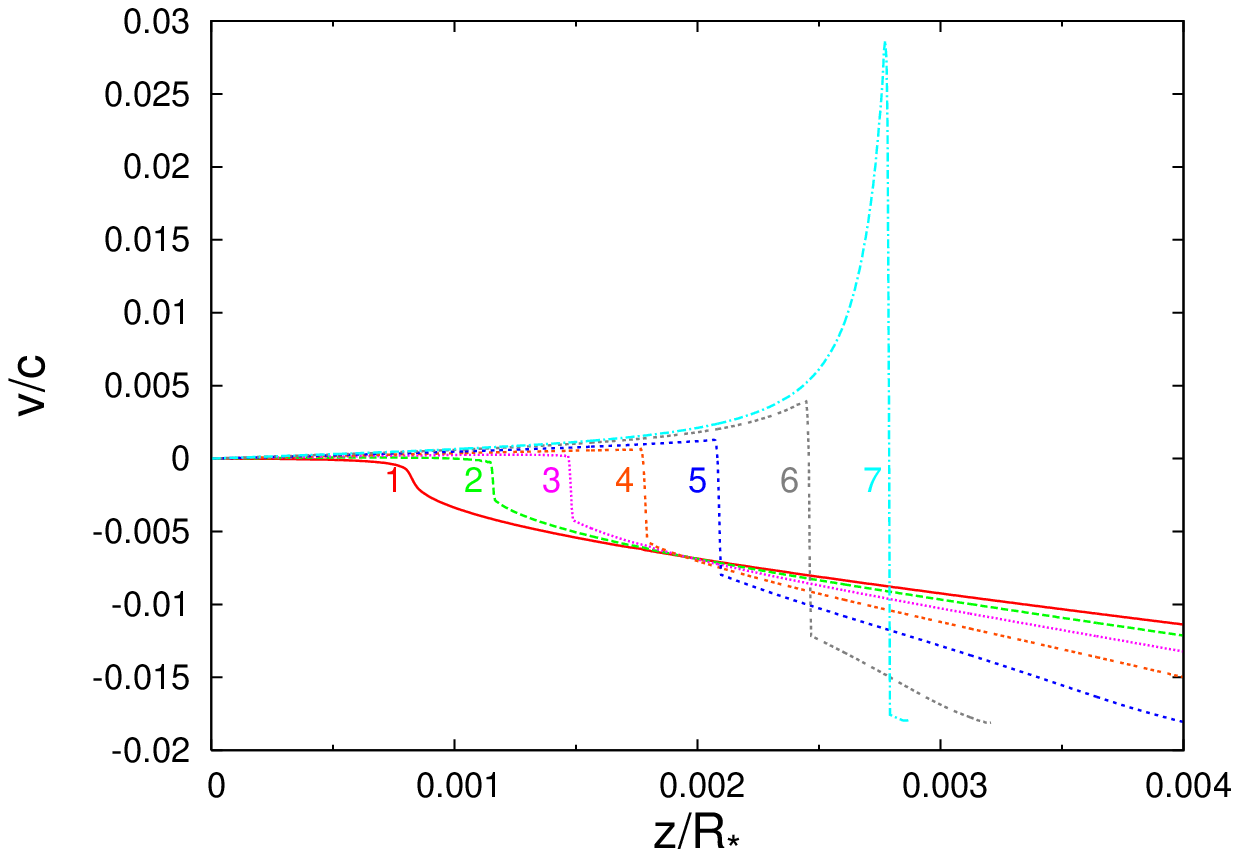}
\caption{ 
Density and velocity profiles along the positive vertical direction $z$ at different times $t$ during the bounce-expansion phase for $\beta=10$ and $\gamma=5/3$.
Labels stand for the following values of $t$ $[\mathrm{s}]$: (1)~20.80, (2)~20.91, (3)~21.02, (4)~21.14, (5)~21.25, (6)~21.37, (7)~21.41. 
The star is at the periastron at $t=0$. 
The central compression is maximal at $t \approx 20.80$.
The stellar matter expands from the centre of the star while it collapses elsewhere. 
A shock wave forms in front of the expansion and propagates outwards. 
The collapse stops at $t \approx 21.42$.}
\label{P03Fig07}
\end{figure}

As summarized in Table \ref{P03Tab01}, the formation properties of the shock wave during the bounce-expansion phase of the stellar matter actually depend on the penetration factor.
\begin{table}[h!]
\caption{
Characteristic quantities when the shock wave forms for penetration factors $3 < \beta < 12$ in the case $\gamma=5/3$.
Column 2: time $t$ $[ \mathrm{s} ]$ when the shock wave forms (the star is at the periastron at $t=0$).
Column 3: position of the sonic point $z_{\mathrm{s}}$ $[ 10^{-2} \, R_{*} ]$ at time $t$.
Column 4: star radius $R$ $[ 10^{-2} \, R_{*} ]$ at time $t$.
Column 5: position of the sonic point relative to the star radius.
Column 6: time interval $\Delta t_{1}$ $[ \mathrm{s} ]$ between the maximum of central pressure and the formation of the shock wave. 
Column 7: time interval $\Delta t_{2}$ $[ \mathrm{s} ]$ between the formation of the shock wave and the end of the collapse of the whole stellar matter.
}
\label{P03Tab01}
\centering
\begin{tabular}{ccccccc}
\hline \hline
$\beta$ & $t$ & $z_{\mathrm{s}}$ & $R$ & $z_{\mathrm{s}}/R$ & $\Delta t_{1}$ & $\Delta t_{2}$ \\
\hline
4  & 173.00 & 4.19 & 4.68 & 0.89 & 16.82     & 1.50 \\
5  & 100.49 & 1.93 & 2.61 & 0.74 & 6.08      & 1.45 \\
6  & 65.58  & 1.04 & 1.74 & 0.60 & 2.49      & 1.15 \\
7  & 46.08  & 0.61 & 1.29 & 0.47 & 1.11      & 0.96 \\
8  & 34.21  & 0.39 & 0.98 & 0.40 & 0.58      & 0.71 \\
9  & 26.30  & 0.26 & 0.85 & 0.30 & 0.23      & 0.64 \\
10 & 20.90  & 0.18 & 0.81 & 0.22 & 0.10      & 0.52 \\
11 & 16.98  & 0.13 & 0.61 & 0.21 & $10^{-2}$ & 0.45 \\
\hline
\end{tabular}
\end{table}
In particular, as the penetration factor increases, the shock wave forms closer to the centre, as indicated by the position of the sonic point, and faster after the instant of maximal central pressure.
The shock wave then propagates within the collapsing matter during a shorter time before reaching the star radius, since the free fall velocities increase with the penetration factor.
The evolution of the density and velocity profiles in the case $\beta=10$ is reproduced on Fig.~\ref{P03Fig07}. 
Let us mention that no shock wave forms for $\beta=3$.

The propagation velocity of the shock wave and the compression ratios of the shocked matter are given in Table \ref{P03Tab02}.
The shock wave propagates within the stellar matter with velocities $\approx 10^{3} \, \mathrm{km} \, \mathrm{s}^{-1}$, and produces density jumps tending towards the strong shock limit $(\gamma + 1) / (\gamma - 1) = 4$ (see e.g. Landau \& Lifshitz \cite{Land63}).
\begin{table}[h!]
\caption{
Characteristic quantities of the shock wave for penetration factors $\beta=7$ and $10$ in the case $\gamma=5/3$.
Column 2: compression ratio $\sigma_{\mathrm{shock}}$ (defined as the ratio of density just upstream of the shock front to density just downstream of the shock front). 
Column 3: velocity of the shock front $v_{\mathrm{shock}}$ $[ 10^{3} \, \mathrm{km} \, \mathrm{s}^{-1} ]$.
The values of time $t$ $[ \mathrm{s} ]$ correspond to those of Fig.~\ref{P03Fig05_A} for $\beta=7$, and of Fig.~\ref{P03Fig07} for $\beta=10$.}
\label{P03Tab02}
\centering
\begin{tabular}{cccc}
\hline \hline
 & $t$ & $\sigma_{\mathrm{shock}}$ & $v_{\mathrm{shock}}$ \\
\hline
\hline
            & 46.32 & 1.9 & 1.2 \\
$\beta = 7$ & 46.56 & 2.9 & 1.5 \\
            & 46.79 & 3.7 & 1.5 \\
            & 46.96 & 3.9 & 2.5 \\
\hline
             & 21.02 & 1.7 & 1.8 \\
$\beta = 10$ & 21.14 & 2.2 & 1.8 \\
             & 21.25 & 3.1 & 1.8 \\
             & 21.37 & 3.7 & 2.4 \\
\hline
\end{tabular}
\end{table}

Such an evolution of the tidal compression process until $\beta < 12$ is in full agreement with the qualitative results of Fulbright (\cite{Fulthesis95}) and Kobayashi et al. (\cite{Kob04}) mentioned in Sect.~\ref{AMvsHydro} in the respective cases $\beta=5$ and $\beta=10$. 
It also confirms for these star-BH encounters the assumption of the AM where no shock wave forms before the instant of maximal compression of the stellar core.
\\
\\
\textbf{Encounters \mbox{\boldmath $12 \le \beta \le 15$}}

\begin{figure}[h!]
\includegraphics[width=8.5cm]{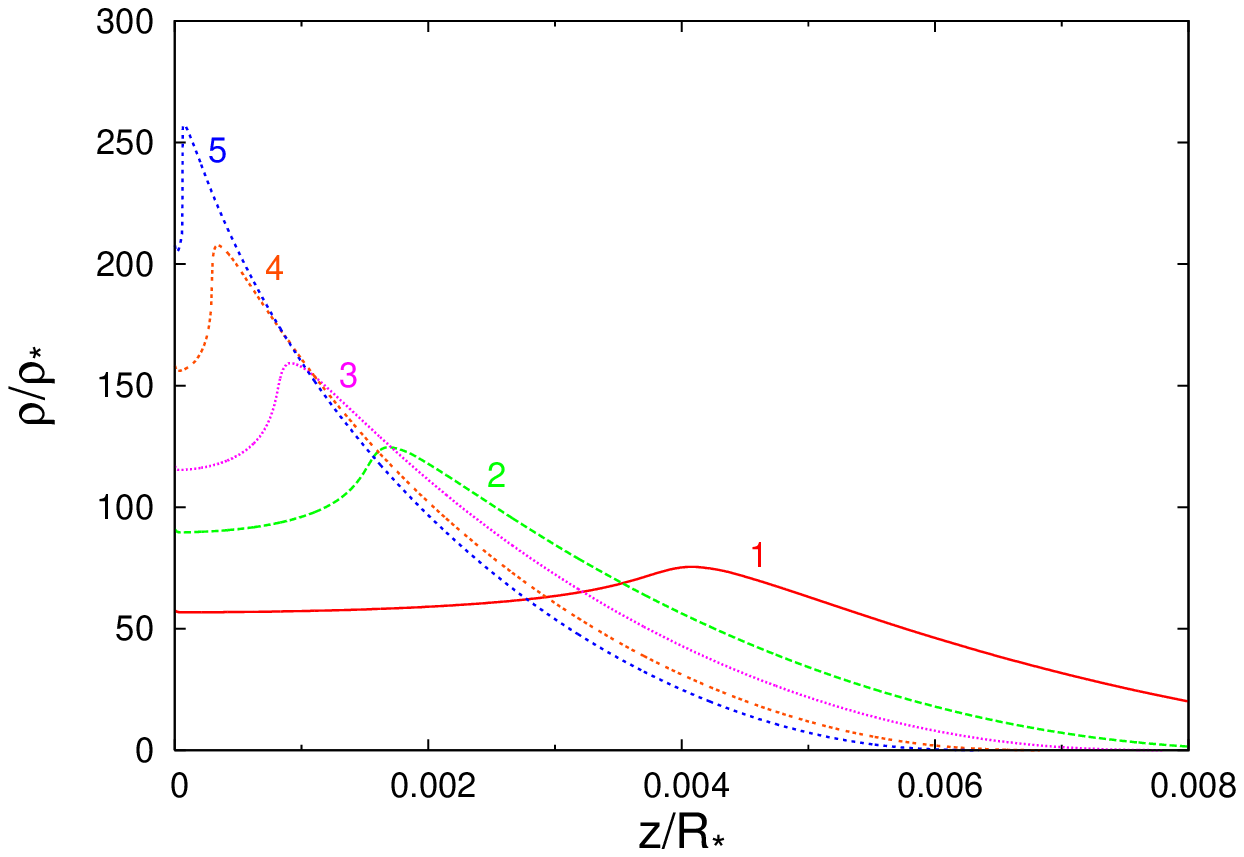} \\ \\
\includegraphics[width=8.5cm]{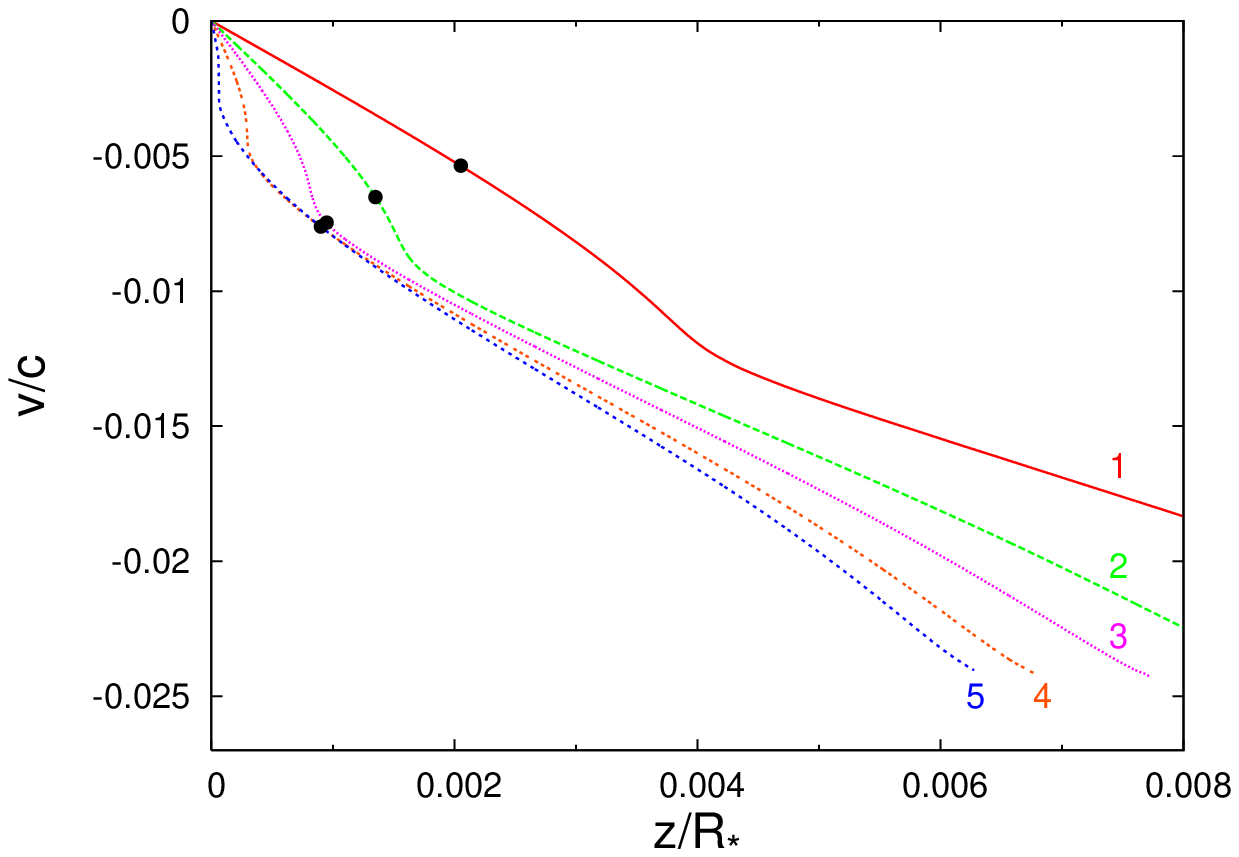}
\caption{ 
Density and velocity profiles along the positive vertical direction $z$ at different times $t$ during the free fall phase for $\beta=12$ and $\gamma=5/3$.
Labels stand for the following values of $t$ $[\mathrm{s}]$: (1)~13.40, (2)~13.73, (3)~13.85, (4)~13.94, (5)~13.99. 
The star is at the periastron at $t=0$. 
The position $z_{\mathrm{s}}$ of the sonic point where $\vert v(z_{\mathrm{s}},t) \vert = a(z_{\mathrm{s}},t)$, with $a(z,t)$ the local speed of sound given by Eq.~(\ref{P02Eq12}), is indicated by the black points. 
At left (resp. right) to the sonic point, the flow is subsonic (resp. supersonic) with $\vert v(z<z_{\mathrm{s}},t) \vert < a(z<z_{\mathrm{s}},t)$ (resp. $\vert v(z>z_{\mathrm{s}},t) \vert > a(z>z_{\mathrm{s}},t)$).
As the whole stellar matter collapses in free fall in the BH gravitational field, a shock wave forms and propagates up to the centre of the star.}
\label{P03Fig08_A}
\end{figure}

\begin{figure}[h!]
\includegraphics[width=8.5cm]{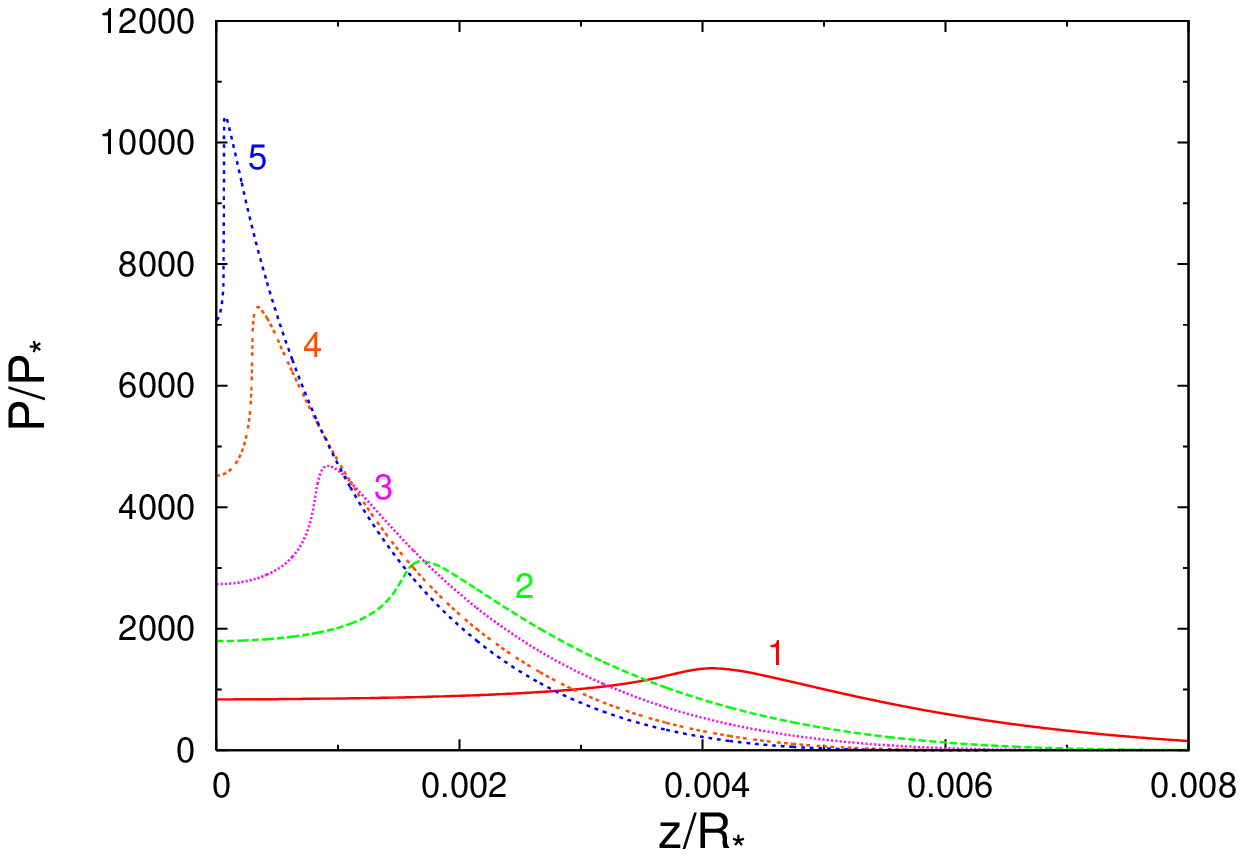} \\ \\
\includegraphics[width=8.5cm]{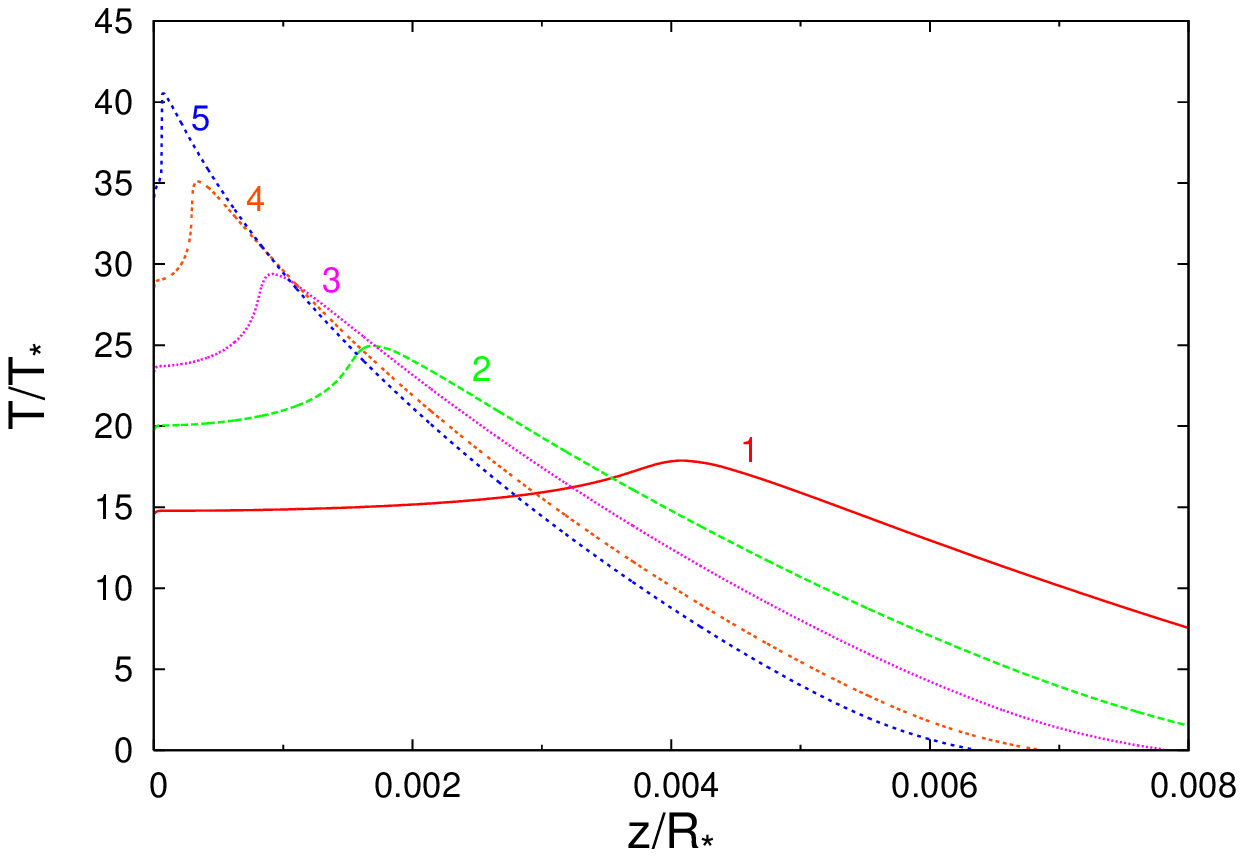}
\caption{
Pressure and temperature profiles along the positive vertical direction $z$ at different times $t$ during the free fall phase for $\beta=12$ and $\gamma=5/3$.
Same labels as in Fig.~\ref{P03Fig08_A}.}
\label{P03Fig08_B}
\end{figure}

\begin{figure}[h!]
\resizebox{\hsize}{!}{\includegraphics{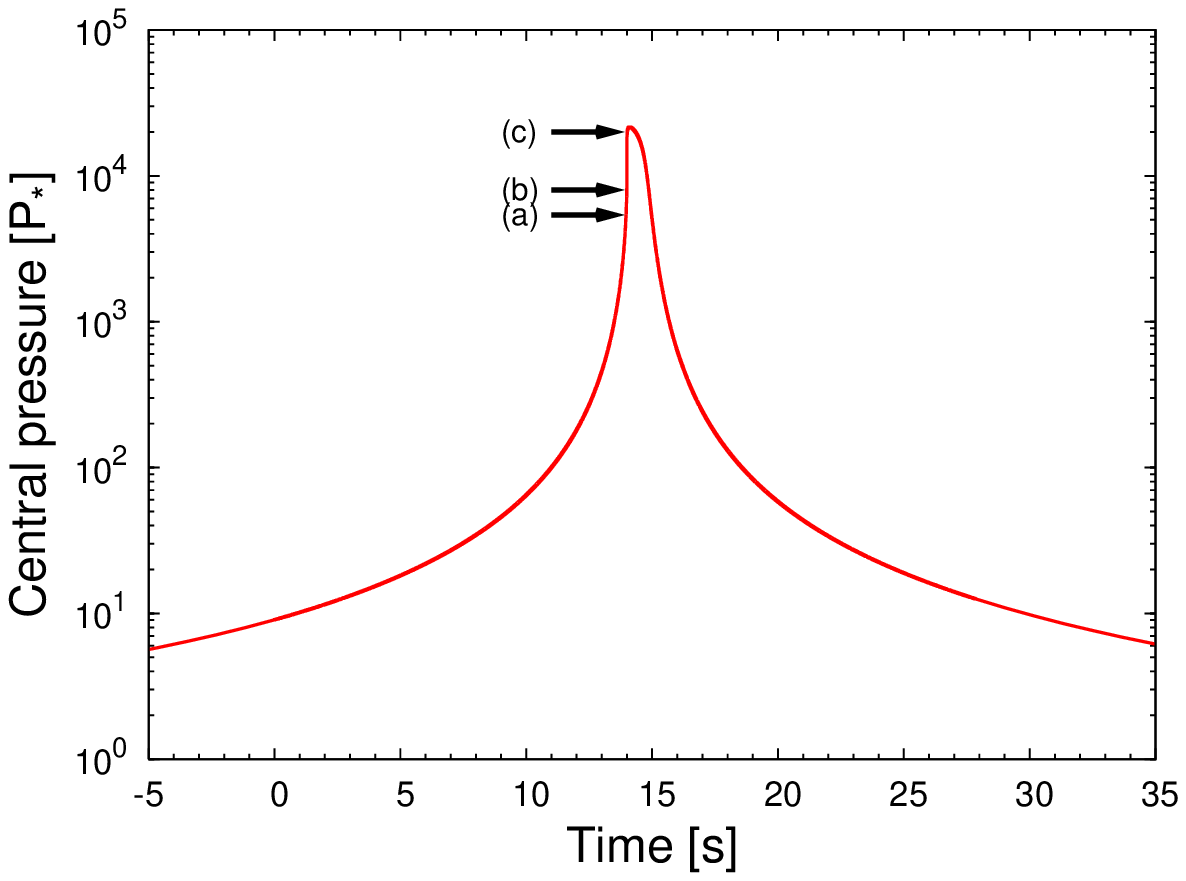}}
\caption{
Evolution of the central pressure as function of time $t$ for $\beta=12$ and $\gamma=5/3$.
After the passage of the star through the periastron at $t=0$, the stellar matter suddenly compresses at the centre of the star. 
During that phase, a shock wave forms at $t \approx 13.96$ (a) on both sides of the orbital plane (see Figs.~\ref{P03Fig08_A}-\ref{P03Fig08_B}).
Both symmetric shock waves propagate inwards until they collide at the centre of the star at $t \approx 13.98$ (b), which produces an additional (instantaneous) compression of the central matter (b) $\to$ (c).
The shock waves are then reflected outwards which stops the central compression at $t \approx 14.10$ (c).}
\label{P03Fig09}
\end{figure}

When the star penetrates within the tidal radius with a penetration factor $\beta \ge 12$, the hydrodynamical simulations have shown that a shock wave could indeed form during the free fall phase. 
The evolution of the hydrodynamical variables is given on Figs.~\ref{P03Fig08_A}-\ref{P03Fig08_B} for an encounter with $\beta=12$.
As the whole stellar matter collapses in the BH gravitational field, in contrast with weaker encounters the profiles progressively steepen until they finally break into a shock front.
The shock wave then propagates up to the centre where it collides with the symmetric shock wave propagating on the other side of the orbital plane, which produces an additional compression of the central matter.
It is interesting to notice that the shock waves precisely form during the sudden increase of the compression at the centre of the star (Fig.~\ref{P03Fig09}) induced by the tidal field after the passage through the periastron. 

\begin{figure}[h!]
\includegraphics[width=8.5cm]{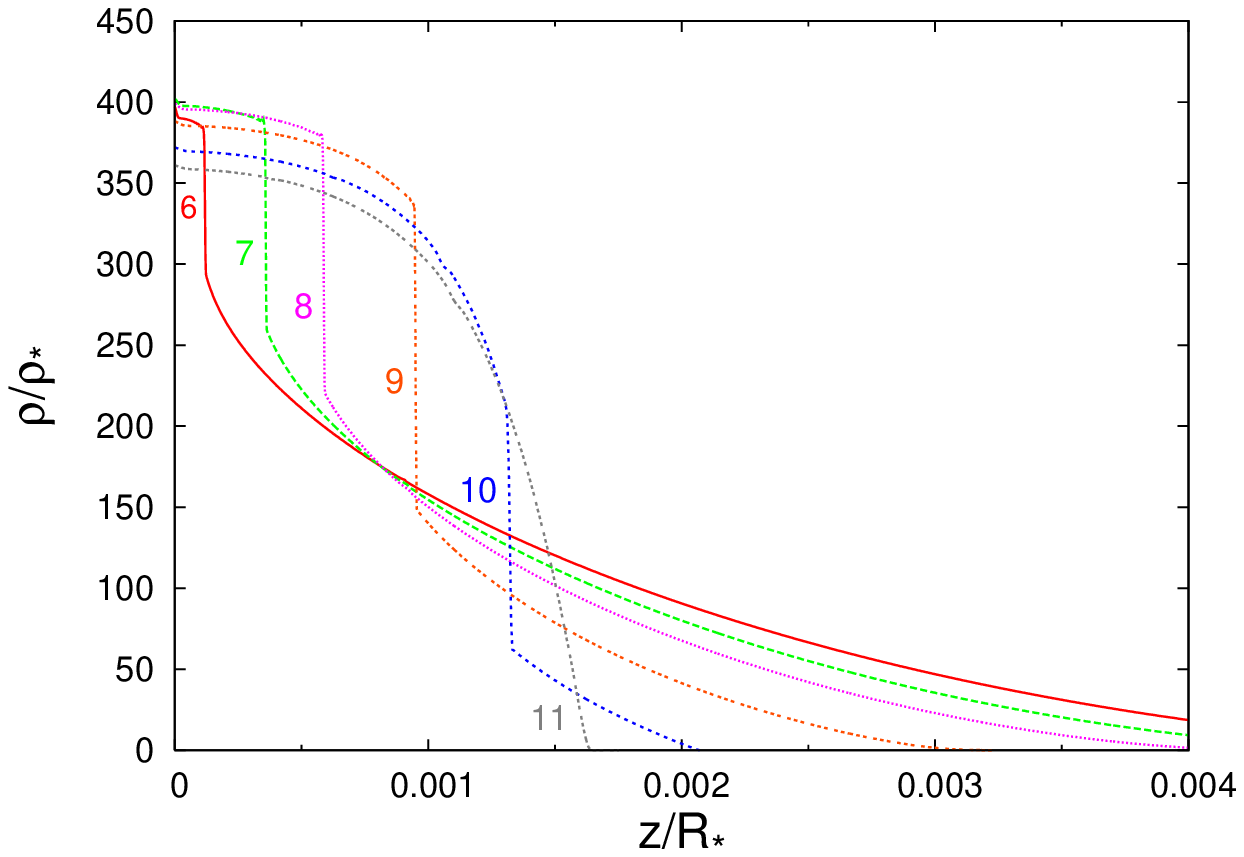} \\ \\
\includegraphics[width=8.5cm]{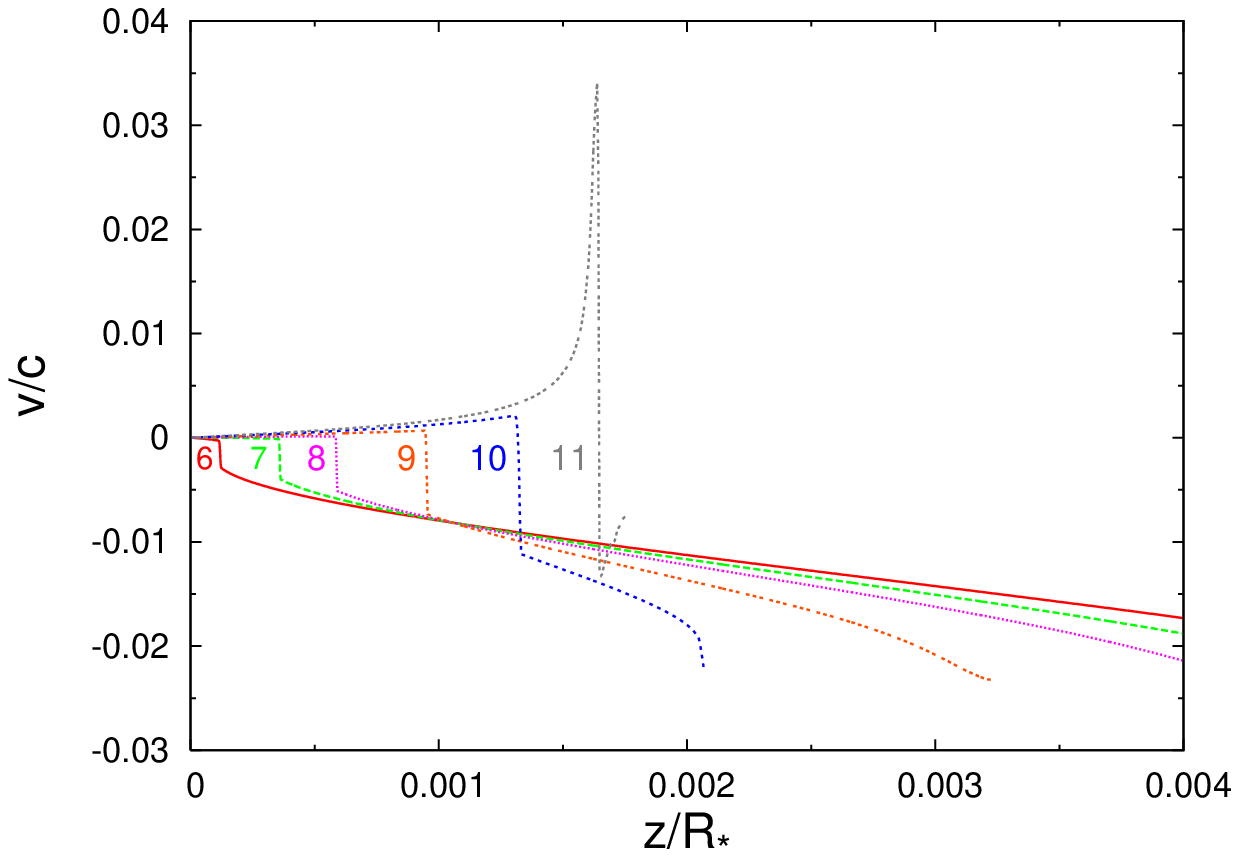}
\caption{ 
Density and velocity profiles along the positive vertical direction $z$ at different times $t$ during the bounce-expansion phase for $\beta=12$ and $\gamma=5/3$.
Labels stand for the following values of $t$ $[\mathrm{s}]$: (6)~14.04, (7)~14.11, (8)~14.18, (9)~14.29, (10)~14.41, (11)~14.46. 
The star is at the periastron at $t=0$. 
The central compression is maximal at $t \approx 14.10$.
The shock wave formed during the free fall phase, after reflexion at the centre of the star at $t \approx 13.98$, propagates outwards which reverts the collapse.
Behind the shock wave, the stellar matter expands from the centre under the effect of the central pressure.
The collapse stops at $t \approx 14.47$.}
\label{P03Fig10_A}
\end{figure}

\begin{figure}[h!]
\includegraphics[width=8.5cm]{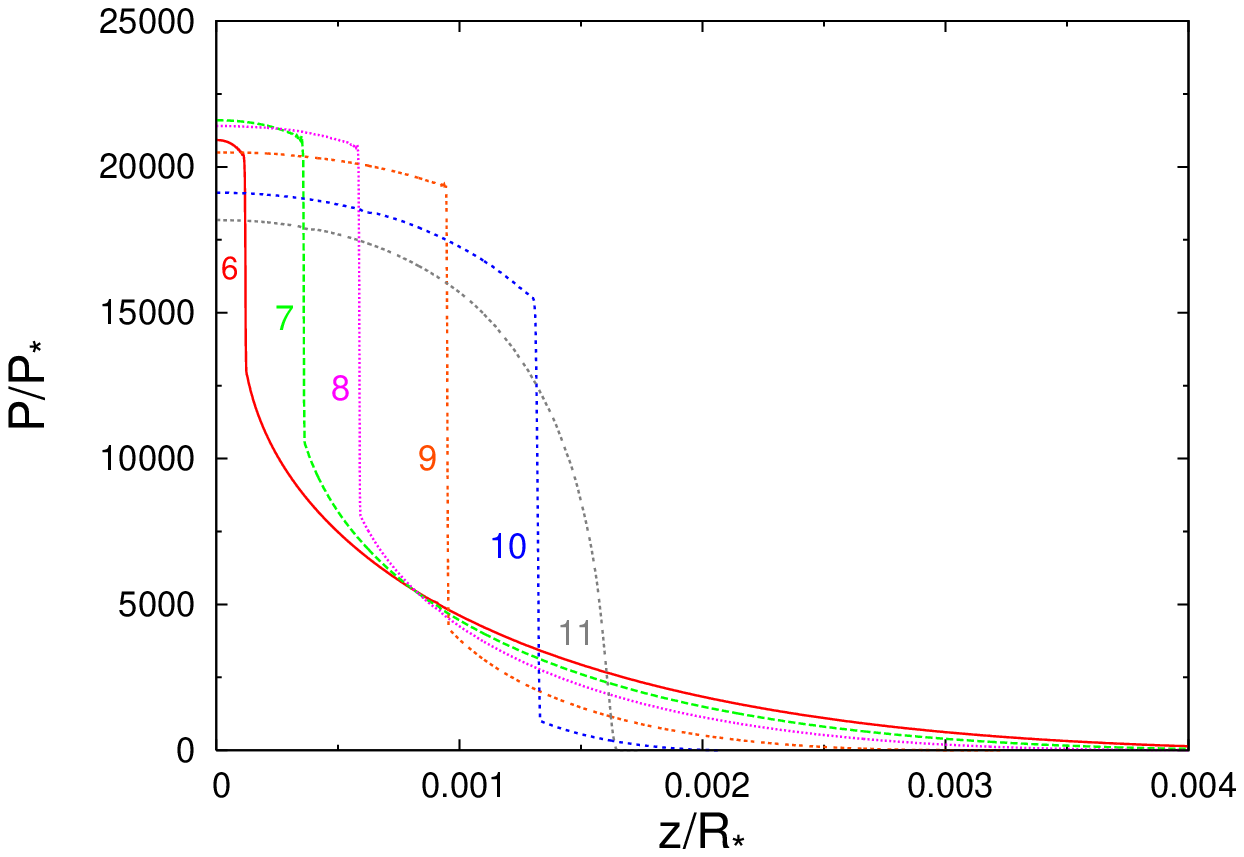} \\ \\
\includegraphics[width=8.5cm]{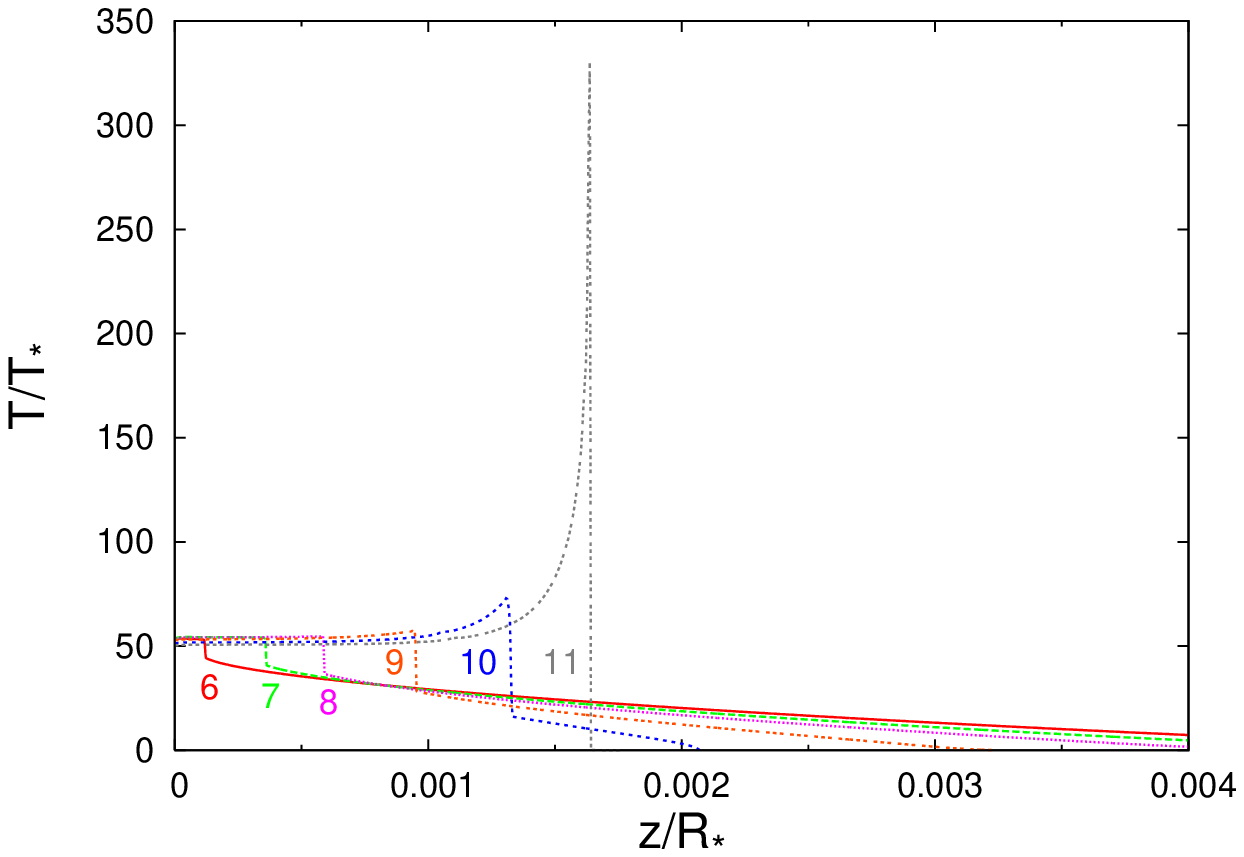}
\caption{ 
Pressure and temperature profiles along the positive vertical direction $z$ at different times $t$ during the bounce-expansion phase for $\beta=12$ and $\gamma=5/3$.
Same labels as in Fig.~\ref{P03Fig10_A}.}
\label{P03Fig10_B}
\end{figure}

After collision, both shock waves are reflected outwards which indeed stops the central compression. 
The evolution of the hydrodynamical variables during the bounce-expansion phase can be followed on Figs.~\ref{P03Fig10_A}-\ref{P03Fig10_B}.
The propagation of the shock wave reverts the collapse, while the stellar matter expands from the centre under the effect of the pressure.
The collapse definitely stops a few time after the instant of reflexion when the shock front reaches the star radius.
As in the previous description, the shock wave speeds up the expansion in the less dense regions (Fig.~\ref{P03Fig10_A} bottom) and heats the collapsing matter (Fig.~\ref{P03Fig10_B} bottom).

\begin{figure}[h!]
\includegraphics[width=8.5cm]{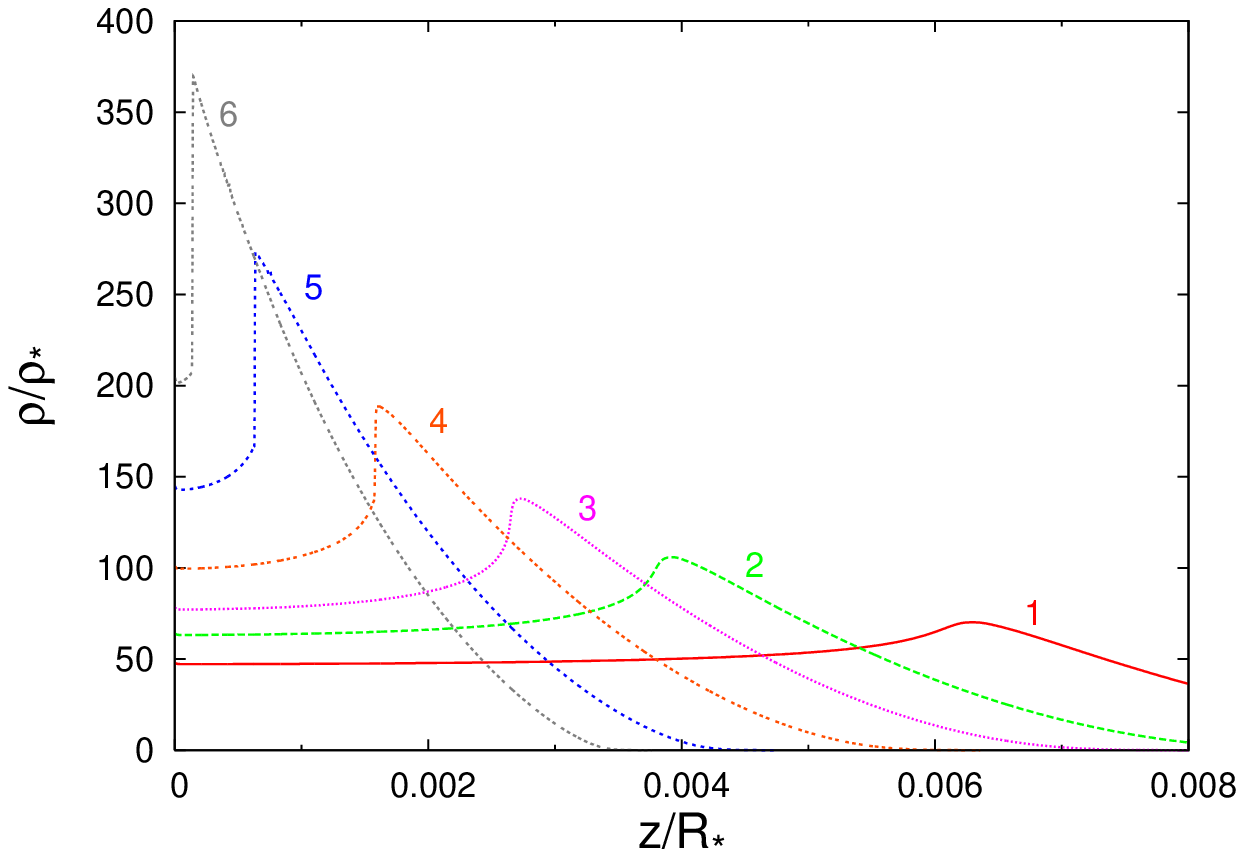} \\ \\
\includegraphics[width=8.5cm]{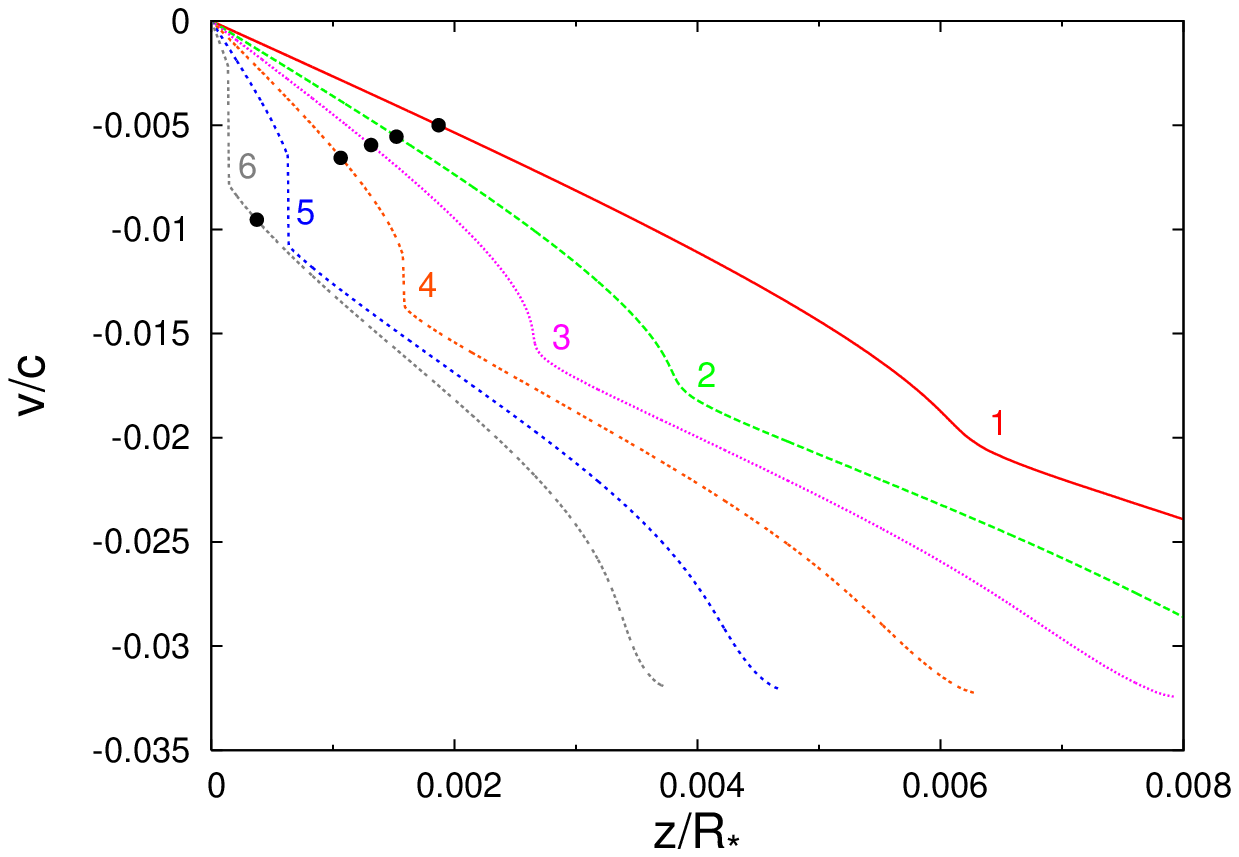}
\caption{ 
Density and velocity profiles along the positive vertical direction $z$ at different times $t$ during the free fall phase for $\beta=15$ and $\gamma=5/3$.
Labels stand for the following values of $t$ $[\mathrm{s}]$: (1)~8.13, (2)~8.36, (3)~8.47, (4)~8.59, (5)~8.71, (6)~8.77. 
The star is at the periastron at $t=0$.
The position $z_{\mathrm{s}}$ of the sonic point where $\vert v(z_{\mathrm{s}},t) \vert = a(z_{\mathrm{s}},t)$, with $a(z,t)$ the local speed of sound given by Eq.~(\ref{P02Eq12}), is indicated by the black points. 
At left (resp. right) to the sonic point, the flow is subsonic (resp. supersonic) with $\vert v(z<z_{\mathrm{s}},t) \vert < a(z<z_{\mathrm{s}},t)$ (resp. $\vert v(z>z_{\mathrm{s}},t) \vert > a(z>z_{\mathrm{s}},t)$).
As the whole stellar matter collapses in free fall in the BH gravitational field, a shock wave forms and propagates up to the centre of the star.}
\label{P03Fig11_A}
\end{figure}

\begin{figure}[h!]
\includegraphics[width=8.5cm]{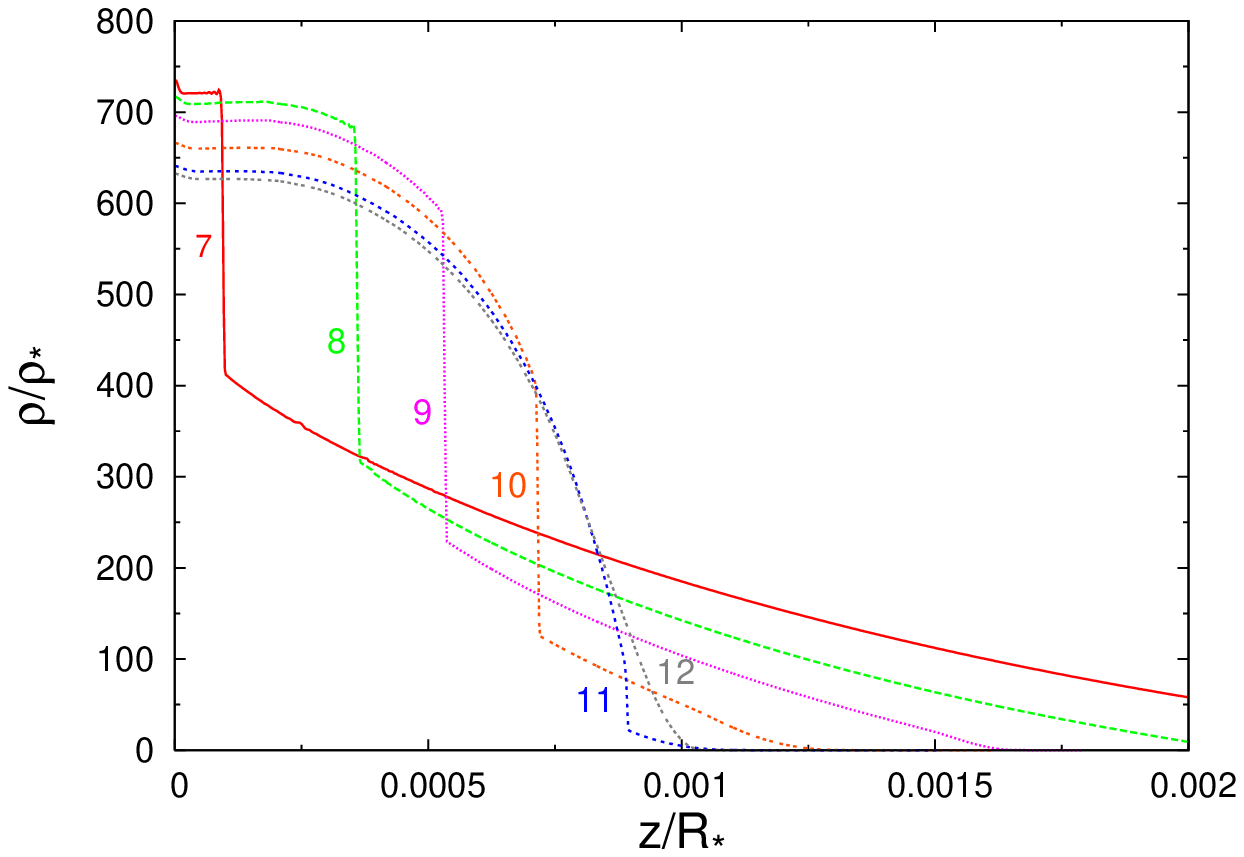} \\ \\
\includegraphics[width=8.5cm]{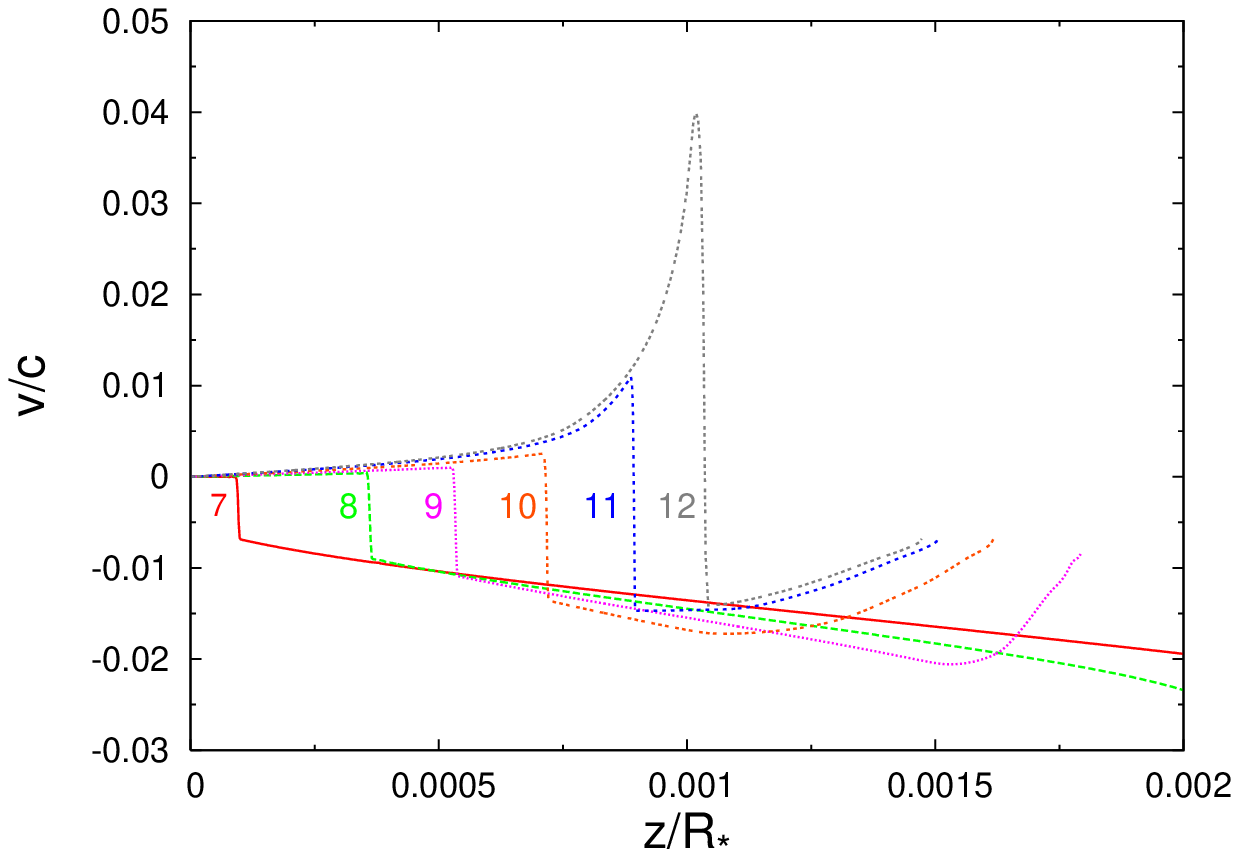}
\caption{ 
Density and velocity profiles along the positive vertical direction $z$ at different times $t$ during the bounce-expansion phase for $\beta=15$ and $\gamma=5/3$.
Labels stand for the following values of $t$ $[\mathrm{s}]$: (7)~8.82, (8)~8.89, (9)~8.94, (10)~8.98, (11)~9.02, (12)~9.03. 
The star is at the periastron at $t=0$. 
The central compression is maximal at $t \approx 8.81$.
The shock wave formed during the free fall phase, after reflexion at the centre of the star at $t \approx 8.80$, propagates outwards which reverts the collapse.
Behind the shock wave, the stellar matter expands from the centre under the effect of the central pressure.
The collapse stops at $t \approx 9.04$.}
\label{P03Fig11_B}
\end{figure}

The formation properties of the shock wave as function of the penetration factor are summarized in Table~\ref{P03Tab03}.
\begin{table}[h!]
\caption{
Characteristic quantities when the shock wave forms for penetration factors $12 \le \beta \le 15$ in the case $\gamma=5/3$.
Column 2: time $t$ $[ \mathrm{s} ]$ when the shock wave forms (the star is at the periastron at $t=0$).
Column 3: position of the shock front formation $z_{\mathrm{shock}}$ $[ 10^{-2} \, R_{*} ]$.
Column 4: star radius $R$ $[ 10^{-2} \, R_{*} ]$ at time $t$.
Column 5: position of the shock front formation relative to the star radius.
Column 6: time interval $\Delta t_{1}$ $[ \mathrm{s} ]$ between the formation of the shock wave and the maximum of central pressure. 
Column 7: time interval $\Delta t_{2}$ $[ \mathrm{s} ]$ between the formation of the shock wave and the end of the collapse of the whole stellar matter. 
Column 8: time interval $\Delta t_{3}$ $[ \mathrm{s} ]$ between the formation of the shock wave and its reflexion at the centre of the star. 
}
\label{P03Tab03}
\centering
\begin{tabular}{cccccccc}
\hline \hline
$\beta$ & $t$ & $z_{\mathrm{shock}}$ & $R$ & $z_{\mathrm{shock}}/R$ & $\Delta t_{1}$ & $\Delta t_{2}$ & $\Delta t_{3}$ \\
\hline
12  & 13.96 & 0.02 & 0.65 & 0.03 & 0.14 & 0.51 & 0.02 \\
13  & 11.68 & 0.10 & 0.69 & 0.15 & 0.24 & 0.53 & 0.16 \\
14  & 9.92  & 0.16 & 0.70 & 0.23 & 0.27 & 0.52 & 0.24 \\
15  & 8.52  & 0.22 & 0.73 & 0.30 & 0.29 & 0.52 & 0.28 \\
\hline
\end{tabular}
\end{table}
The evolution of the density and velocity profiles during the free fall and bounce-expansion phases in the case $\beta=15$ are also reproduced on Figs.~\ref{P03Fig11_A}-\ref{P03Fig11_B}. 
Actually, the shock wave forms further from the centre of the star as the penetration factor increases.
It therefore propagates on a longer distance before reflexion so that the shock front more increases.
We have noticed in these different star-BH encounters that the shock wave forms when the central pressure $\approx 10^{3}-10^{4} \, P_{*}$ (see Fig.~\ref{P03Fig09}).

The propagation velocity before and after reflexion of the shock wave and the compression ratios of the shocked matter are given in Table~\ref{P03Tab04}.  
\begin{table}[h!]
\caption{
Characteristic quantities of the shock wave for penetration factors $\beta=12$ and $15$ in the case $\gamma=5/3$.
Column 2: compression ratio $\sigma_{\mathrm{shock}}$ (defined as the ratio of density just upstream of the shock front to density just downstream of the shock front). 
Column 3: velocity of the shock front $v_{\mathrm{shock}}$ $[ 10^{3} \, \mathrm{km} \, \mathrm{s}^{-1} ]$. 
The values of time $t$ $[ \mathrm{s} ]$ correspond to those of Figs.~\ref{P03Fig08_A} and \ref{P03Fig10_A} for $\beta = 12$, and of Figs.~\ref{P03Fig11_A} and \ref{P03Fig11_B} for $\beta=15$.}
\label{P03Tab04}
\centering
\begin{tabular}{cccc}
\hline \hline
 & $t$ & $\sigma_{\mathrm{shock}}$ & $v_{\mathrm{shock}}$ \\
\hline
             & 13.99 & 1.2 & 3.3 \\
             & 14.04 & 1.3 & 5.9 \\
$\beta = 12$ & 14.11 & 1.5 & 2.4 \\
             & 14.18 & 1.7 & 2.3 \\
             & 14.29 & 2.3 & 2.2 \\
             & 14.41 & 3.3 & 2.1 \\
\hline
             & 8.71 & 1.6 & 5.9 \\
             & 8.77 & 1.7 & 5.6 \\
$\beta = 15$ & 8.82 & 1.7 & 3.5 \\
             & 8.89 & 2.2 & 2.7 \\
             & 8.94 & 2.6 & 2.5 \\
             & 8.98 & 3.2 & 2.9 \\
             & 9.02 & 3.7 & 3.8 \\
\hline
\end{tabular}
\end{table}
\\

It is interesting to notice that the tidal compression of stars presents a qualitative analogy with the core collapse of massive stars (i.e. type Ib/c and type II supernovae) in the sense that both processes involve homologous velocity profiles.
However they are significant physical differences.
In the hydrodynamical supernovae scenario, the collapse of the stellar core is due to the self-gravitational field and is spherically symmetric.
The core splits into a subsonically homologous collapsing inner part and a supersonically (non homologous) falling outer part (Yahil \& Lattimer \cite{Yah82}).
The shock wave development is due to the stiffening of the equation of state, i.e. the increase of the adiabatic index from a value somewhat below $4/3$ to a value $\approx 2.5-3$, when the central density reaches the nuclear matter density, which stops the collapse from the centre of the core and produces pressure waves moving outwards and accumulating near the sonic point where they steepen into a shock front (see e.g. M\"uller \cite{Mul98}). 
Moreover, after having bounced behind the shock wave, the inner core quickly settles into hydrostatic equilibrium, and the supersonic accretion continues for many dynamical timescales.
On the other hand, in the tidally compressed stars scenario, the collapse is triggered by the external (tidal) gravitational field of the BH and occurs vertically in the direction orthogonal to the orbital plane.
The core only collapses homologously, whereas the shock wave development is due to the sudden build-up of central pressure.
\subsubsection{Compression and heating factors}
With a view to subsequent consideration on the thermonuclear reactions that can be triggered within the tidally compressed star, particular interest attaches to the maximal values of the central density and temperature.

\begin{figure}[h!]
\includegraphics[width=8.5cm]{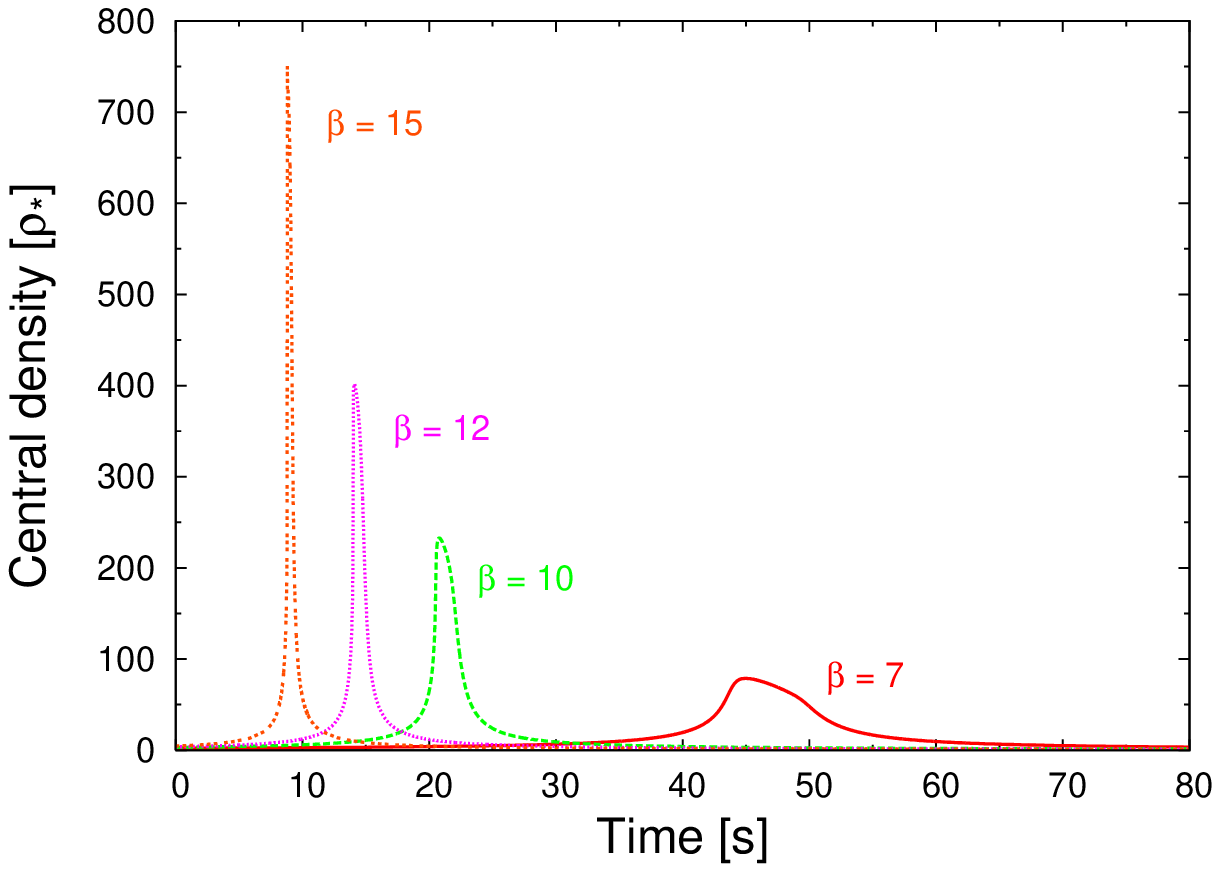} \\ \\
\includegraphics[width=8.5cm]{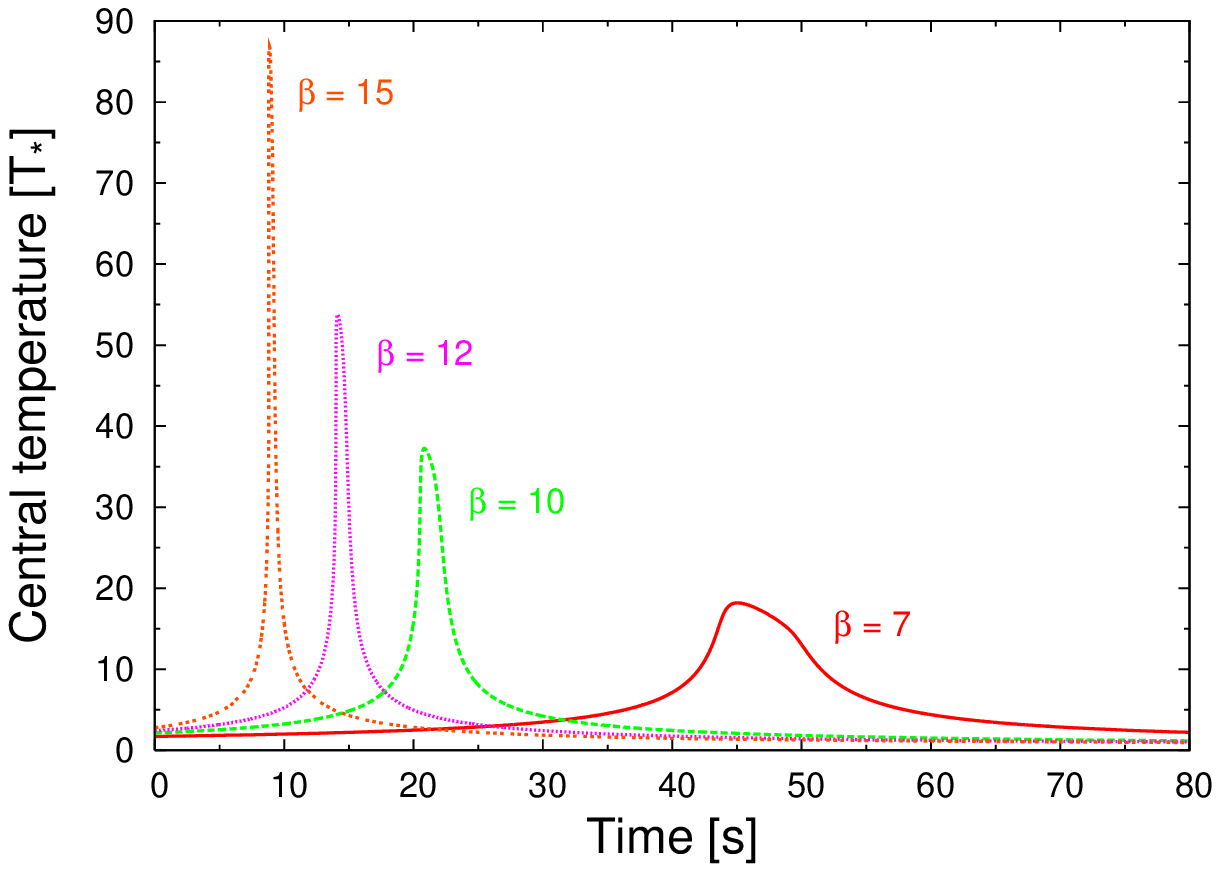}
\caption{Evolutions of the central density and temperature as function of time $t$ for different penetration factors $\beta$ in the case $\gamma=5/3$. 
After the passage of the star through the periastron at $t=0$, the stellar matter suddenly compresses at the centre of the star (see Fig.~\ref{P03Fig03} for $\beta=7$ and Fig.~\ref{P03Fig09} for $\beta=12$).
The increase of central density and temperature reaches its maximum at $t~\approx~44.97, 20.80, 14.10, 8.81$ for respectively $\beta=7,10,12,15$.}
\label{P03Fig12}
\end{figure}

The behaviour of the central density and temperature is reproduced on Fig.~\ref{P03Fig12} for the star-BH encounters previously considered.

\begin{figure}[h!]
\includegraphics[width=8.5cm]{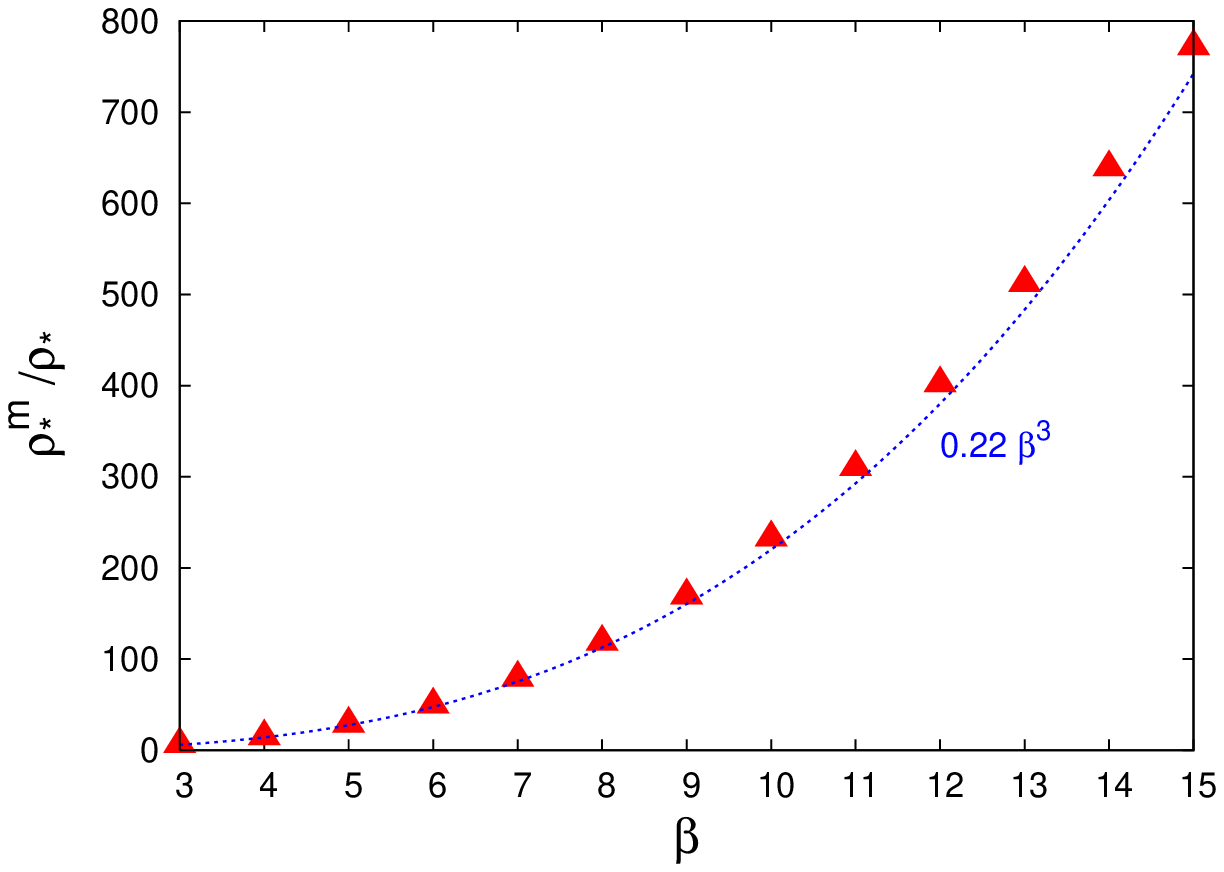} \\ \\
\includegraphics[width=8.5cm]{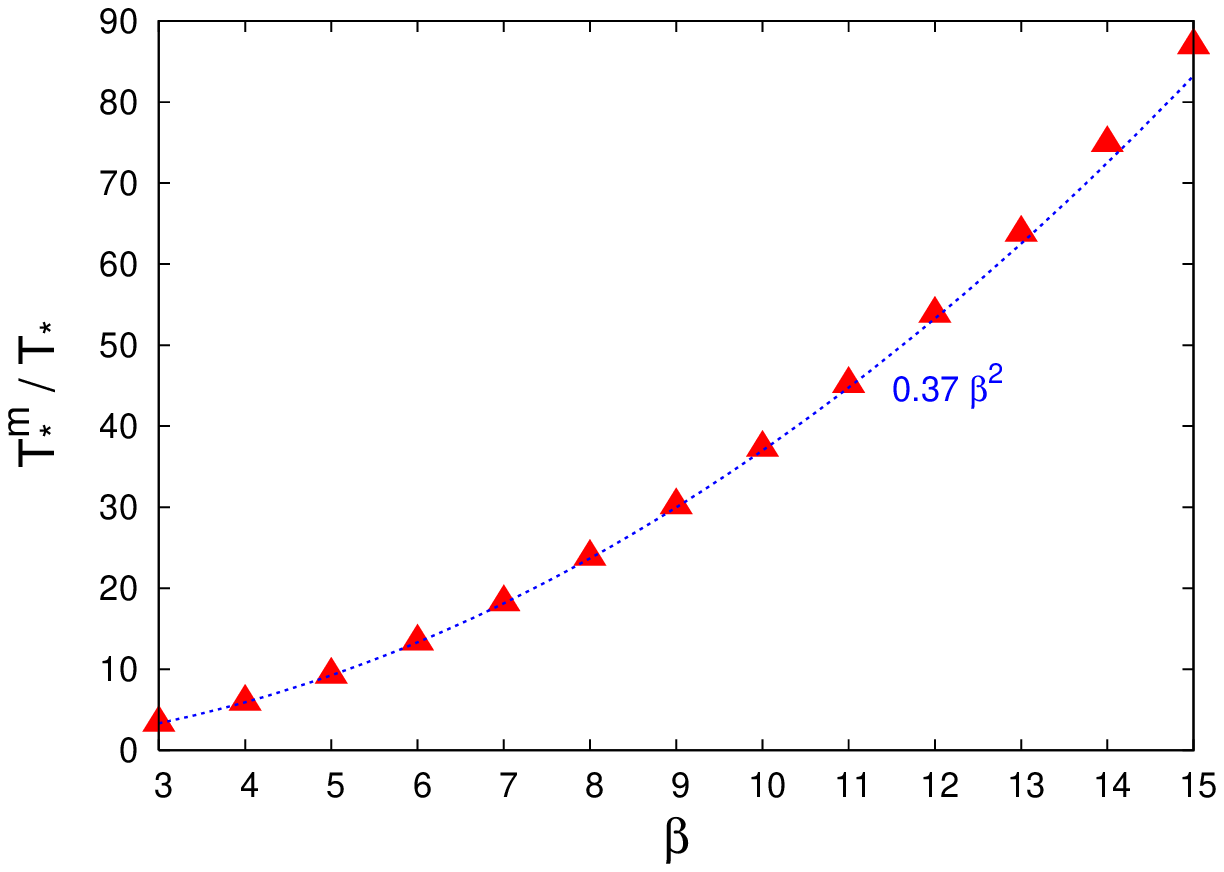} \\ \\
\includegraphics[width=8.5cm]{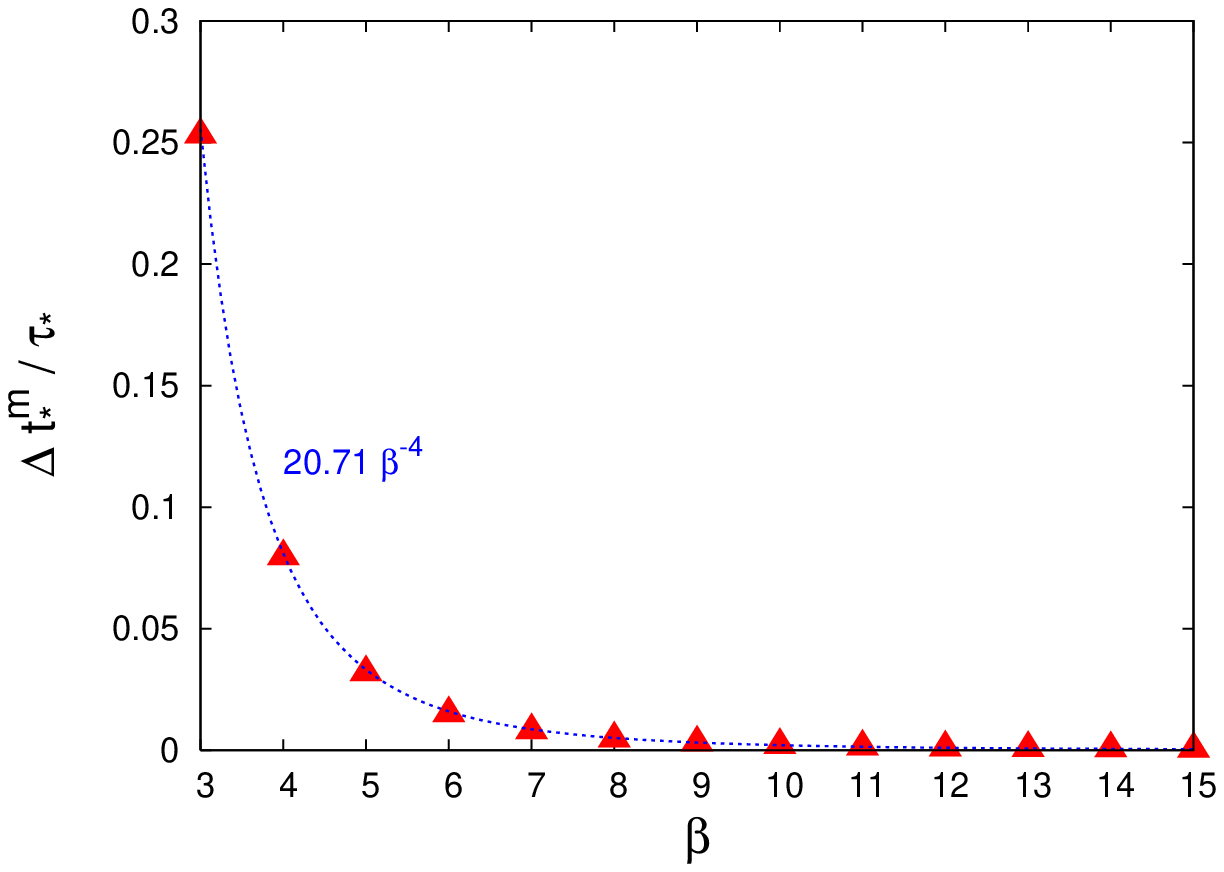}
\caption{Evolutions of the maximal central density, maximal central temperature, and duration during which the maximal values are maintained as function of the penetration factor $\beta$ in the case $\gamma=5/3$.
Triangle symbols: values computed by the hydrodynamical model (see Table~\ref{P03Tab05}). 
Dashed line: power law predicted by the AM (see Eqs.~(\ref{P03Eq04})-(\ref{P03Eq06})).
In bottom figure, the characteristic internal timescale of the star $\tau_{*} \equiv (1/G \rho_{*})^{1/2} \approx 1.33 \times 10^{3} \, \mathrm{s}$.} 
\label{P03Fig13}
\end{figure}

The evolution as function of the penetration factor, of the maxima $\rho_{*}^{\mathrm{m}}$ and $T_{*}^{\mathrm{m}}$ respectively reached by the central density and temperature at the instant of maximal compression, and of the duration $\Delta t_{*}^{\mathrm{m}}$ during which these values are maintained, is given in Table~\ref{P03Tab05} and is reproduced on Fig.~\ref{P03Fig13}.
\begin{table}[h!]
\caption{
Evolution of thermodynamical quantities at the centre of the star as function of the penetration factor $3 \le \beta \le 15$ in the case $\gamma=5/3$.
Column 2: time $t$ $[ \mathrm{s} ]$ when the central density and temperature reach their maximal value (the star is at the periastron at $t=0$).  
Column 3: maximal value $\rho_{*}^{\mathrm{m}}$ $[ \rho_{*} ]$ of the central density. 
Column 4: maximal value $T_{*}^{\mathrm{m}}$  $[ T_{*} ]$ of the central temperature.
Column 5: duration $\Delta t_{*}^{\mathrm{m}}$ $[ 10^{-4} \, \tau_{*} ]$, with $\tau_{*} \equiv \left( 1 / G\rho{*} \right)^{1/2} \approx 1.33 \times 10^{3} \, \mathrm{s}$, of the maximal central density and temperature. It is defined as the full width at half maximum of the peak of central temperature.
Column 6: star radius $R$ $[ 10^{-2} \, R_{*} ]$ at time $t$. 
Column 7: density $\rho_{\mathrm{R}}$ $[ 10^{-4} \, \rho_{*} ]$ at the star radius.
Column 8: temperature $T_{\mathrm{R}}$ $[ 10^{-2} \, T_{*} ]$ at the star radius.
}
\label{P03Tab05}
\centering
\begin{tabular}{cccccccc}
\hline \hline
$\beta$ & $t$ & $\rho_{*}^{\mathrm{m}}$ & $T_{*}^{\mathrm{m}}$ & $\Delta t_{*}^{\mathrm{m}}$ 
	& $R$ & $\rho_{\mathrm{R}}$ & $T_{\mathrm{R}}$ \\
\hline
3  & 299.55 & 6.03   & 3.30  & 2530.30 & 17.83 & 2.04 & 1.15 \\
4  & 156.18 & 14.39  & 5.88  & 795.25  & 8.14  & 5.06 & 1.39 \\
5  & 94.41  & 28.35  & 9.23  & 317.57  & 4.54  & 2.79 & 2.05 \\
6  & 63.09  & 49.33  & 13.33 & 149.09  & 2.76  & 8.12 & 2.55 \\
7  & 44.97  & 78.82  & 18.19 & 79.10   & 1.87  & 8.31 & 2.74 \\
8  & 33.63  & 118.26 & 23.80 & 45.97   & 1.35  & 3.69 & 1.83 \\
9  & 26.07  & 169.09 & 30.15 & 28.62   & 1.02  & 1.81 & 1.02 \\
10 & 20.80  & 232.73 & 37.25 & 18.47   & 0.79  & 1.47 & 2.52 \\
11 & 16.97  & 309.98 & 45.14 & 12.49   & 0.64  & 3.03 & 5.51 \\
12 & 14.10  & 401.94 & 53.79 & 8.74    & 0.54  & 7.79 & 6.34 \\
13 & 11.85  & 512.06 & 63.79 & 6.30    & 0.51  & 3.46 & 7.14 \\
14 & 10.16  & 638.83 & 74.84 & 4.69    & 0.41  & 4.19 & 4.01 \\
15 & 8.81   & 771.59 & 86.94 & 3.66    & 0.34  & 9.97 & 7.49 \\
\hline
\end{tabular}
\end{table}

Let us recall that the AM (Luminet \& Carter \cite{Lum86}) predicted for a polytropic gas $\gamma=5/3$ that the compression and heating factors of the stellar core were respectively given by 
\begin{eqnarray}
\left \{ \frac{\rho_{*}^{\mathrm{m}}}{\rho_{*}} \right \}_{\mathrm{AM}} & \approx & 0.22 \, \beta^{3}, \label{P03Eq04} \\
\left \{ \frac{T_{*}^{\mathrm{m}}}{T_{*}}       \right \}_{\mathrm{AM}} & \approx & 0.37 \, \beta^{2}, \label{P03Eq05} 
\end{eqnarray}
during the time interval
\begin{equation}
\left \{ \frac{\Delta t_{*}^{\mathrm{m}}}{\tau_{*}} \right \}_{\mathrm{AM}} \approx 20.71 \, \beta^{-4}, \label{P03Eq06}
\end{equation}
defining the characteristic internal timescale of the star 
\begin{equation}
\tau_{*} \equiv \left( \frac{1}{G \rho_{*}} \right)^{1/2}. \label{P03Eq07}
\end{equation}

For comparison, the fit of results of Table~\ref{P03Tab05} to power laws leads to the evolution laws
\begin{eqnarray}
\left \{ \frac{\rho_{*}^{\mathrm{m}}}{\rho_{*}}     \right \}_{\mathrm{hydro}} & \approx & 0.25  \, \beta^{2.97},  \label{P03Eq08} \\
\left \{ \frac{T_{*}^{\mathrm{m}}}{T_{*}}           \right \}_{\mathrm{hydro}} & \approx & 0.32  \, \beta^{2.06},  \label{P03Eq09} \\ 
\left \{ \frac{\Delta t_{*}^{\mathrm{m}}}{\tau_{*}} \right \}_{\mathrm{hydro}} & \approx & 21.61 \, \beta^{-4.05}. \label{P03Eq10}
\end{eqnarray}

There is actually an excellent agreement between both results (\ref{P03Eq04})-(\ref{P03Eq06}) and (\ref{P03Eq08})-(\ref{P03Eq10}) for the whole star-BH encounters $3 \le \beta \le 15$.
As recalled in Sect.~\ref{AMvsHydro}, the AM assumes that the vertical collapse occurs without shock wave development until the bounce of the stellar core.
We have clearly shown that this assumption is indeed verified for star-BH encounters with $\beta < 12$. 
On the other hand, for stronger encounters, the shock wave development as the whole stellar matter collapses has no particular effect on the compression and heating factors.
This is due to the fact that the shock waves form late enough during the sudden central compression, so that when they stop the process after collision and reflexion at the centre of the star, the central matter has already been compressed enough by the tidal field.

We have not taken into account the thermonuclear energy generation during the tidal compression process in the present study.
However, since we have confirmed the core thermodynamical quantities predicted by the AM at the instant of bounce, we can be confident that the corresponding results already obtained on the possible thermonuclear reactions remain valuable in the range $3 \le \beta \le 15$.
In this regard, coupling a nuclear network to the AM, Luminet \& Pichon (\cite{Lum89}) interested in the nucleosynthesis in a main-sequence stellar core approaching a BH of $10^{5}-10^{6} \, M_{\odot}$. 
They calculated that, starting from a typical population I chemical composition (i.e. a mixture of hydrogen, helium, carbon, nitrogen, and oxygen), the main nuclear reactions during the bounce phase will be accelerated proton captures on seed elements, that will be achieved for $\beta \gtrsim 10$.
An interesting point is that nucleosynthesis processes do not occur only during the maximal core compression when the density and temperature are very high, but will also continue during the cooling expansion phase, mainly due to weak decay of unstable isotopes on timescales of few tens of seconds.
The total thermonuclear energy release largely exceeds the gravitational binding energy of the star, so that the dynamics of the stellar debris will be modified in the sense that a larger fraction of the ejected gas could be unbound to the BH. 
In a forthcoming paper we shall examine in more detail how the hydrodynamical model with shock waves influences the nucleosynthesis processes. 
\subsection{Case $\gamma = 4/3$}
In order to see the influence of the compressibility of the stellar gas on the shock wave development, calculations have been also performed for the polytropic gas $\gamma=4/3$.
This case may give an interesting indication on what will happen with a more realistic equation of state.
Indeed for the strongest star-BH encounters, it is clear that the compressed stellar core will become more and more radiative owing to the excessively high values of density and temperature reached.
For this reason, Luminet \& Carter (\cite{Lum86}) already considered an initial mixture of a non-relativistic polytropic gas together with a black body photon gas within the AM.
They nevertheless found results, in particular the compression and heating factors of the stellar core, which were intermediate between those obtained with the pure polytropic gases $\gamma=5/3$ and $\gamma=4/3$.

The hydrodynamical simulations have shown that the tidal compression process occurs in the same way as the case $\gamma=5/3$, and that there exists a critical value of the penetration factor separating the shock wave development before or after the instant of maximal compression. 
Precisely in the case $\gamma=4/3$, shock waves form during the bounce-expansion phase of the stellar matter when $3 \le \beta < 6$, whereas they form during the preceeding free fall phase when $6 \le \beta \le 15$.

\begin{figure}[h!]
\includegraphics[width=8.5cm]{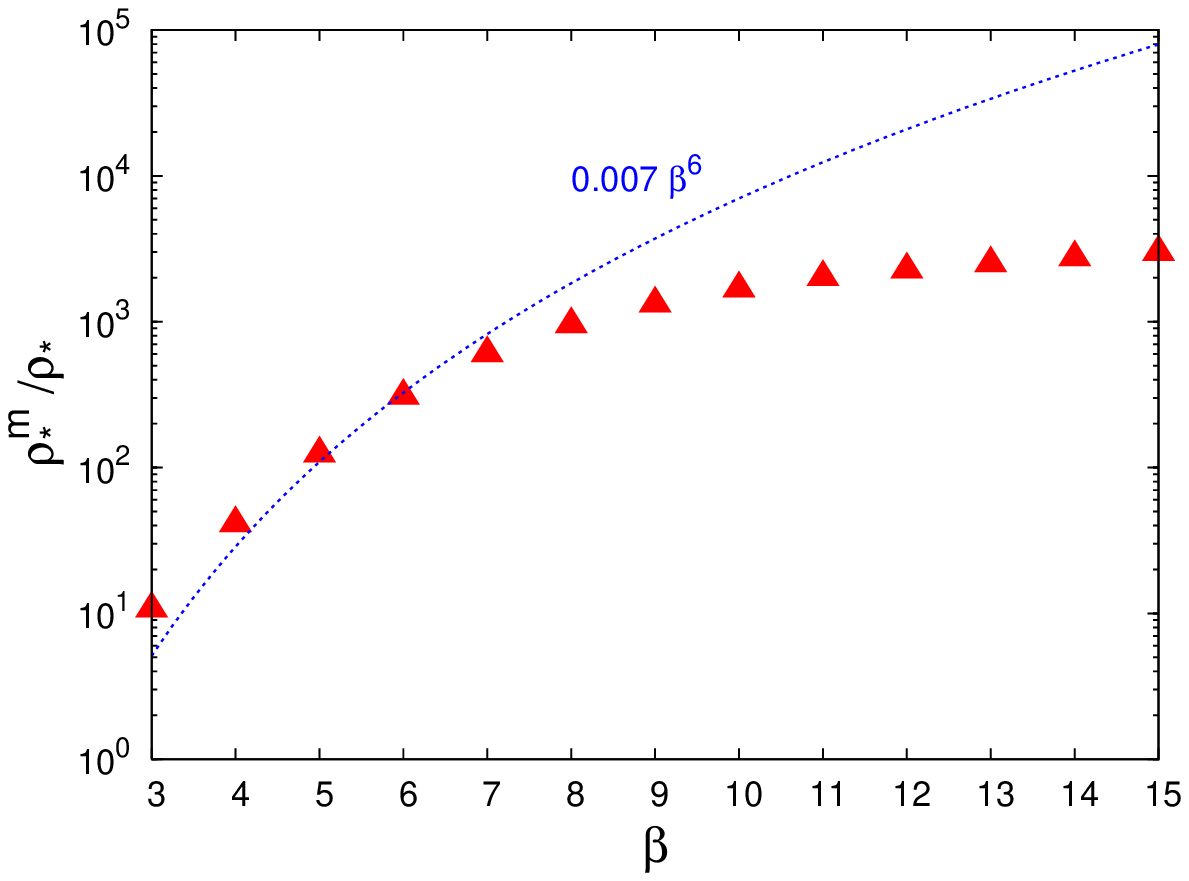} \\ \\
\includegraphics[width=8.5cm]{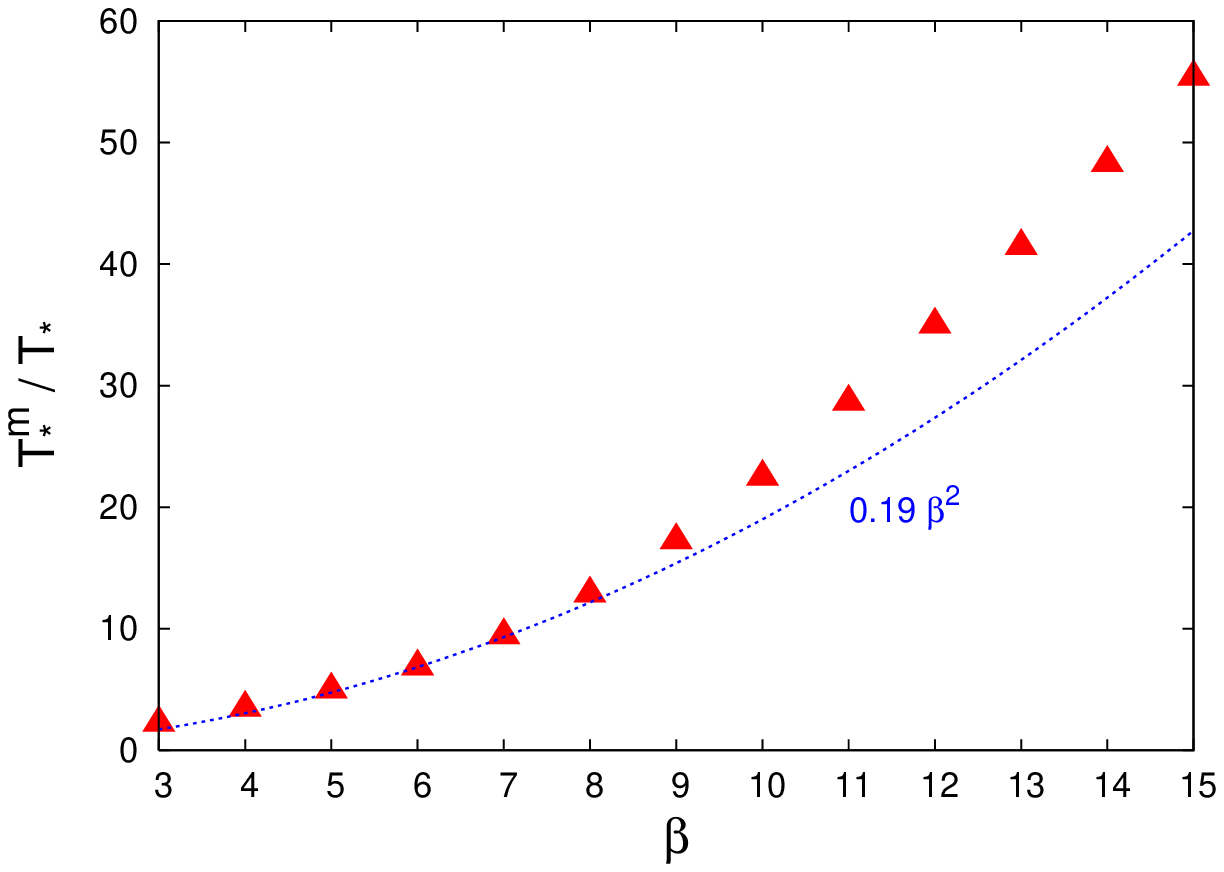}
\caption{Evolutions of the maximal central density and temperature as function of the penetration factor $\beta$ in the case $\gamma=4/3$.
Triangle symbols: values computed by the hydrodynamical model (see Table~\ref{P03Tab06}).  
Dashed line: power law predicted by the AM (see Eqs.~(\ref{P03Eq15})-(\ref{P03Eq16})).} 
\label{P03Fig14}
\end{figure}

The evolution of the compression and heating factors as function of the penetration factor are given in Table~\ref{P03Tab06} and are reproduced on Fig.~\ref{P03Fig14}.
\begin{table}[h!]
\caption{
Evolution of thermodynamical quantities at the centre of the star as function of the penetration factor $3 \le \beta \le 15$ in the case $\gamma=4/3$.
Column 2: time $t$ $[ \mathrm{s} ]$ when the central density and temperature reach their maximal value (the star is at the periastron at $t=0$).  
Column 3: maximal value $\rho_{*}^{\mathrm{m}}$ $[ \rho_{*} ]$ of the central density.
Column 4: maximal value $T_{*}^{\mathrm{m}}$  $[ T_{*} ]$ of the central temperature.
Column 5: duration $\Delta t_{*}^{\mathrm{m}}$ $[ 10^{-4} \, \tau_{*} ]$, with $\tau_{*} \equiv \left( 1 / G\rho{*} \right)^{1/2} \approx 1.33 \times 10^{3} \, \mathrm{s}$, of the maximal central density and temperature. It is defined as the full width at half maximum of the peak of central temperature.}
\label{P03Tab06}
\centering
\begin{tabular}{ccccc}
\hline \hline
$\beta$ & $t$ & $\rho_{*}^{\mathrm{m}}$ & $T_{*}^{\mathrm{m}}$ & $\Delta t_{*}^{\mathrm{m}}$ \\
\hline
3  & 318.43 & 10.67   & 2.20  & 4195.20 \\
4  & 159.05 & 41.12   & 3.44  & 774.50 \\
5  & 94.36  & 123.62  & 4.95  & 206.50 \\
6  & 62.09  & 306.94  & 6.84  & 66.89 \\
7  & 44.33  & 602.60  & 9.42  & 25.20 \\
8  & 33.20  & 953.05  & 12.85 & 13.37 \\
9  & 25.80  & 1317.68 & 17.23 & 8.65 \\
10 & 20.76  & 1675.42 & 22.46 & 5.94 \\
11 & 16.88  & 2000.36 & 28.63 & 4.44 \\
12 & 14.10  & 2260.15 & 34.97 & 3.57 \\
13 & 11.86  & 2488.20 & 41.45 & 2.99 \\
14 & 10.17  & 2732.89 & 48.27 & 2.49 \\
15 & 8.80   & 2974.74 & 55.36 & 2.12 \\
\hline
\end{tabular}
\end{table}
The fit of results to power laws correctly leads to the evolution laws
\begin{eqnarray}
\left \{ \frac{\rho_{*}^{\mathrm{m}}}{\rho_{*}} \right \}_{\mathrm{hydro}} & \approx & 0.042 \, \beta^{4.97}, \label{P03Eq11} \\
\left \{ \frac{T_{*}^{\mathrm{m}}}{T_{*}}       \right \}_{\mathrm{hydro}} & \approx & 0.34 \, \beta^{1.67}, \label{P03Eq12}
\end{eqnarray}
for $3 \le \beta < 6$, and   
\begin{eqnarray}
\left \{ \frac{\rho_{*}^{\mathrm{m}}}{\rho_{*}} \right \}_{\mathrm{hydro}} & \approx & 31.22 \, \beta^{1.70}, \label{P03Eq13} \\
\left \{ \frac{T_{*}^{\mathrm{m}}}{T_{*}}       \right \}_{\mathrm{hydro}} & \approx & 0.12 \, \beta^{2.26}, \label{P03Eq14}
\end{eqnarray}
for $6 \le \beta \le 15$.

The AM (Luminet \& Carter \cite{Lum86}) however predicted for the whole star-BH encounters that 
\begin{eqnarray}
\left \{ \frac{\rho_{*}^{\mathrm{m}}}{\rho_{*}} \right \}_{\mathrm{AM}} & \approx & 0.007 \, \beta^{6}, \label{P03Eq15} \\
\left \{ \frac{T_{*}^{\mathrm{m}}}{T_{*}}       \right \}_{\mathrm{AM}} & \approx & 0.19 \, \beta^{2}. 
\label{P03Eq16}
\end{eqnarray}

\begin{figure}[h!]
\includegraphics[width=8.5cm]{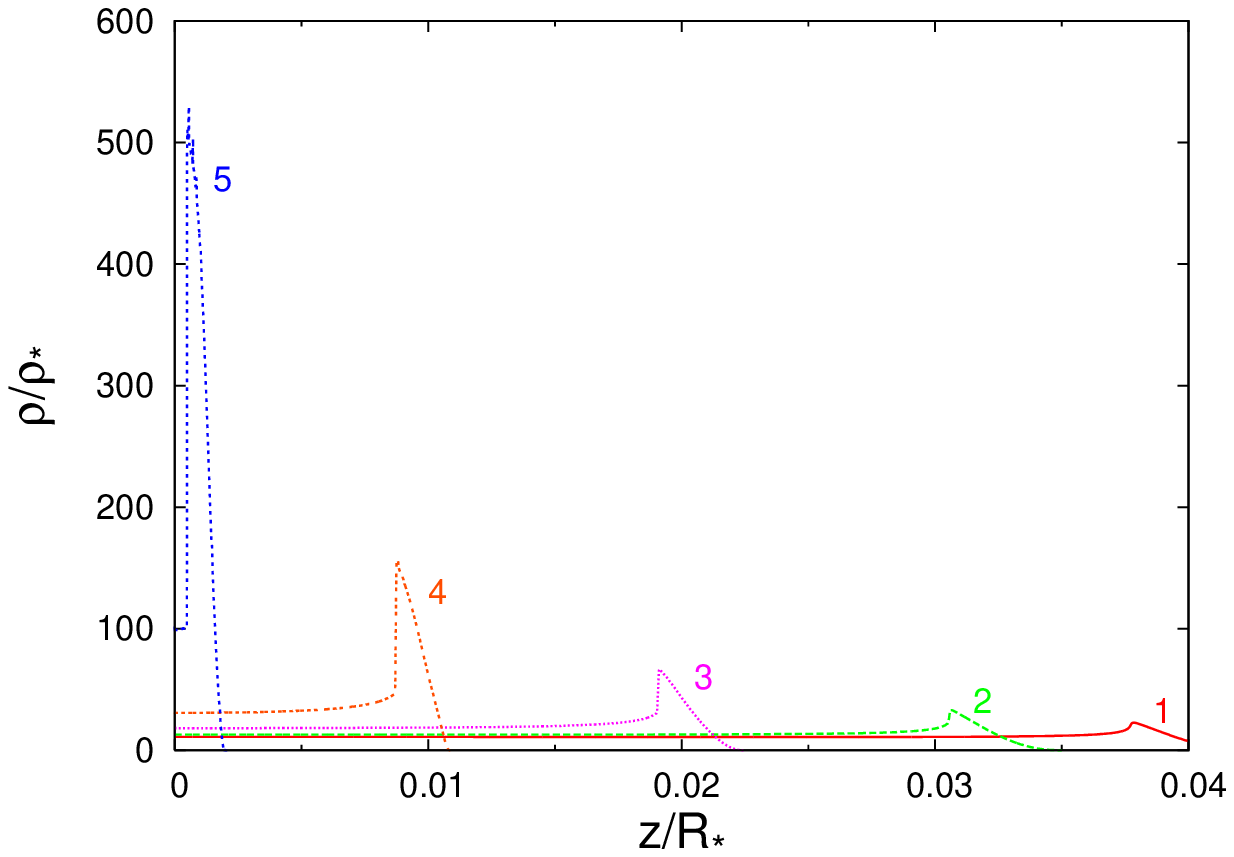} \\ \\
\includegraphics[width=8.5cm]{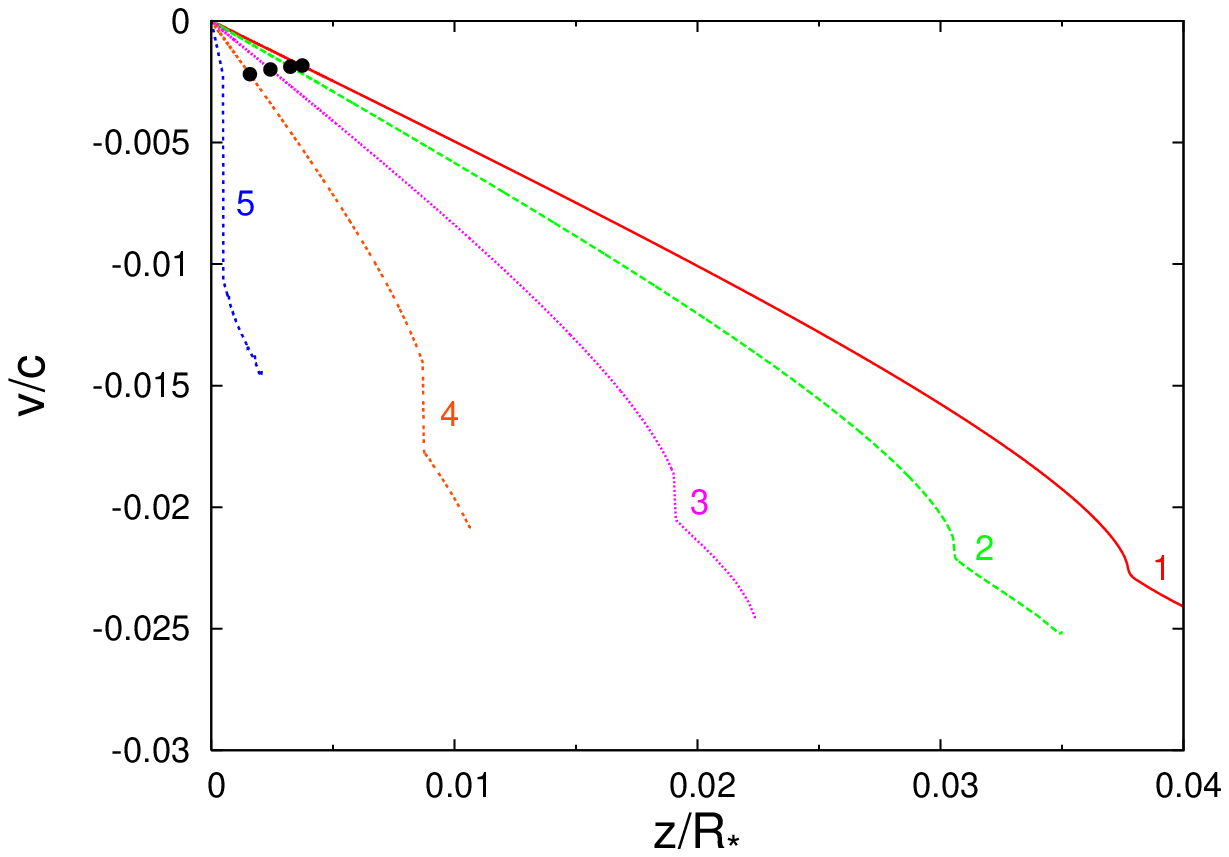}
\caption{ 
Density and velocity profiles along the positive vertical direction $z$ at different times $t$ during the free fall phase for $\beta=12$ and $\gamma=4/3$.
Labels stand for the following values of $t$ $[\mathrm{s}]$: (1)~9.76, (2)~10.46, (3)~11.63, (4)~12.78, (5)~13.93. 
The star is at the periastron at $t=0$. 
The position $z_{\mathrm{s}}$ of the sonic point where $\vert v(z_{\mathrm{s}},t) \vert = a(z_{\mathrm{s}},t)$, with $a(z,t)$ the local speed of sound given by Eq.~(\ref{P02Eq12}), is indicated by the black points. 
At left (resp. right) to the sonic point, the flow is subsonic (resp. supersonic) with $\vert v(z<z_{\mathrm{s}},t) \vert < a(z<z_{\mathrm{s}},t)$ (resp. $\vert v(z>z_{\mathrm{s}},t) \vert > a(z>z_{\mathrm{s}},t)$).
As the whole stellar matter collapses in free fall in the BH gravitational field, a shock wave forms and propagates up to the centre of the star.
Note the size of the shock front relative to the case $\gamma=5/3$ (see Fig.~\ref{P03Fig08_A}).}
\label{P03Fig15}
\end{figure}

\begin{figure}[h!]
\resizebox{\hsize}{!}{\includegraphics{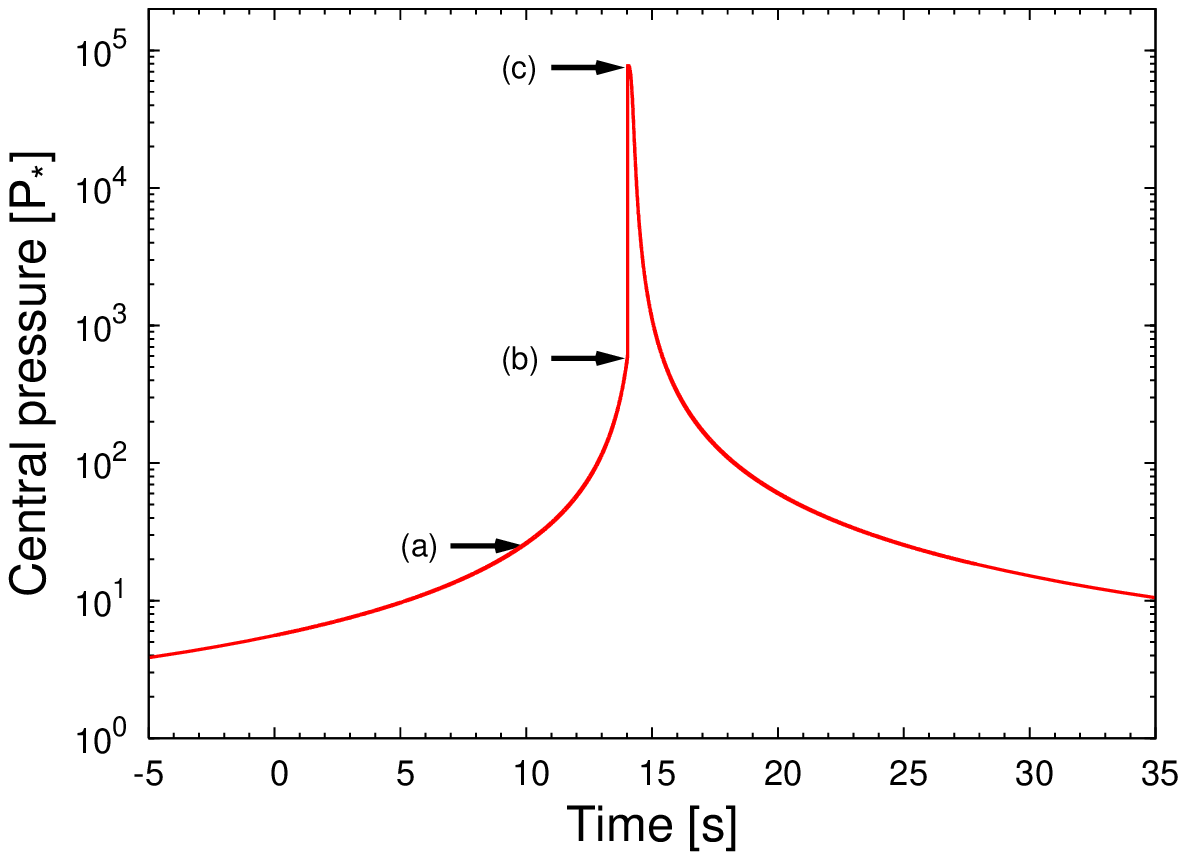}}
\caption{
Evolution of the central pressure as function of time $t$ for $\beta=12$ and $\gamma=4/3$.
After the passage of the star through the periastron at $t=0$, the stellar matter suddenly compresses at the centre of the star. 
During that phase, a shock wave forms at $t \approx 9.85$ (a) on both sides of the orbital plane (see Fig.~\ref{P03Fig15}).
Both symmetric shock waves propagate inwards until they collide at the centre of the star at $t \approx 14.01$ (b), which produces an additional (instantaneous) compression of the central matter (b) $\to$ (c).
The shock waves are then reflected outwards which stops the central compression at $t \approx 14.04$ (c).
Note that relative to the case $\gamma=5/3$ (see Fig.~\ref{P03Fig09}), the shock waves produce a superior compression at the instant of collision, because they form further to the centre (see Figs.~\ref{P03Fig08_A} and \ref{P03Fig15}) so that the shock fronts more increase before reaching the centre.}
\label{P03Fig16}
\end{figure}

In contrast with the polytropic gas $\gamma=5/3$ where the hydrodynamical and AM results are quite consistent, the shock wave development during the free fall phase (i.e. for $\beta \ge 6$) with $\gamma=4/3$ has for effect to substantially decrease the central density and to slightly increase the central temperature for the strongest encounters.
In such cases, it appears that shock waves indeed form earlier during the sudden central compression (Figs.~\ref{P03Fig15}-\ref{P03Fig16}), so that the central matter is less compressed by the tidal field when the shock waves stop the process after reflexion outwards from the centre of the star.
The amount of internal energy of compression lost is however balanced by a slightly superior amount of energy carried by the shock waves propagating inwards until reflexion.
%
%
%
%
\section{Summary and perspectives}
When a star deeply penetrates within the tidal radius of a massive BH without falling in the event horizon, the tidal effects produce a short-lived strong compression of the stellar core, at the end of which it becomes highly flattened according to a pancake-shape configuration.
The first investigations performed in the 1980's within the framework of the semi-analytical AM predicted that the compression could actually trigger some thermonuclear reactions within the stellar core which were indeed likely to lead to an explosion of the star. 
Afterwards, three-dimensional hydrodynamical simulations based on SPH carried out in the 1980-1990's came to the conclusion that the pancake stars phenomenon could be less spectacular than originally believed, finding lower estimates of the pancake quantities as to the compression and heating factors of the stellar core, and therefore questioned the explosive tidal disruption.    
The authors of these studies claimed that the development of shock waves during the tidal compression were mainly responsible for the discrepancy. 
However, the SPH results remained uncertain owing to potential issues of the numerical method to accurately simulate the compression phase, but also owing to a poor spatial resolution.

In this article, we have performed a detailed hydrodynamical study of the vertical tidal compression undergone by the star in the direction orthogonal to its orbital plane, through a one-dimensional hydrodynamical model based on the high-resolution shock-capturing Godunov-type approach to correctly deal with shock waves.
The main results are the following.

Firstly, we have established a more realistic scenario of the tidal compression process, and put in evidence two regimes depending on the strength of the star-BH encounter defined by the penetration factor $3 \le \beta \le 15$.
In the case of main-sequence stars evolving through a polytropic gas equation of state of adiabatic index $5/3$ and for encounters such as $\beta < 12$, the stellar matter collapses in free fall towards the orbital plane until the central pressure suddenly increases to revert the process. 
The stellar matter then bounces at the centre of the star before expanding. 
Shock waves develop during the expansion phase on both sides of the orbital plane and propagate outwards.
The description of the free fall phase until the instant of maximal compression is indeed in agreement with the initial assumption of the AM.
On the other hand, for stronger encounters, shock waves develop during the free fall phase on both sides of the orbital plane, while the central pressure suddenly increases, and propagate inwards until they collide at the centre of the star.
The shock waves are then reflected and propagate outwards, which reverts the free fall phase and causes the stellar matter to expand from the centre under the effect of the pressure.

Secondly, estimates of the compression and heating factors in both regimes are the same as those predicted by the semi-analytical model, thus showing that shock waves have no particular influence on the compression of the stellar core contrary to what concluded in the SPH studies.
Therefore we fully confirm the possible triggering of thermonuclear reactions within the stellar core, like they were already investigated on the basis of the semi-analytical model.

\begin{figure}[h!]
\resizebox{\hsize}{!}{\includegraphics{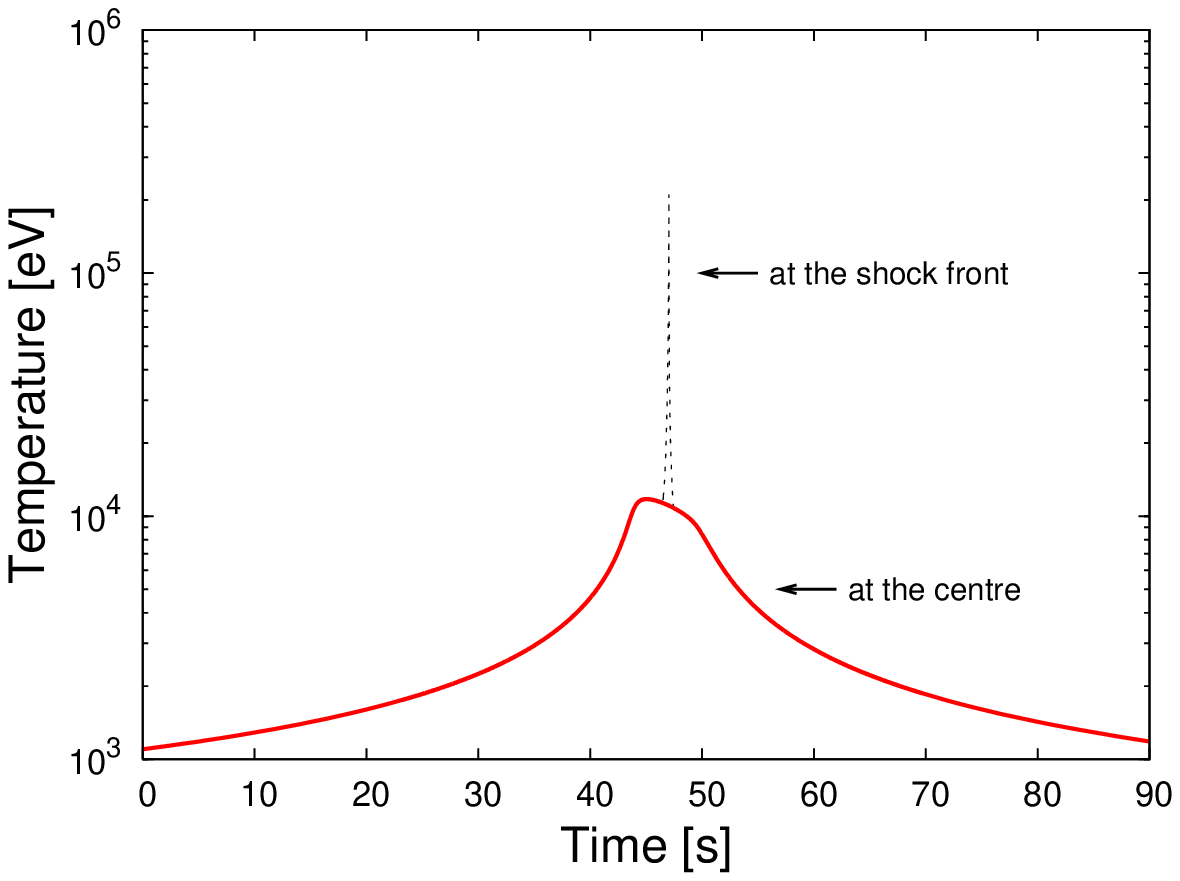}}
\caption{ 
Evolution of the temperature as function of time $t$ for $\beta=7$ and $\gamma=5/3$.
The star is at the periastron at $t=0$.
The solid line corresponds to the temperature at the centre of the star.
The dashed line corresponds to the increase of temperature produced by the shock wave as it propagates outwards within the stellar matter (see Fig.~\ref{P03Fig05_B} bottom).
The duration at half maximum of the peak of temperature at the shock front is $\approx 0.05 \, \mathrm{s}$.}
\label{P04Fig01}
\end{figure}

Thirdly, in both regimes of compression the shock waves carry outwards a brief $(< 0.1 \, \mathrm{s})$ but very high (above $100 \, \mathrm{keV}$) peak of temperature from the centre to the surface of the star (see Table~\ref{P04Tab01} and Fig.~\ref{P04Fig01}), and could provide an efficient mechanism to transfer the pancake energy into hard electromagnetic radiation. 
This last result is very promising since it could likely give rise to a new type of X- or $\gamma$-ray bursts, as already pointed out by Carter (\cite{Car92}) on the basis of qualitative arguments and recently considered by Kobayashi et al. (\cite{Kob04}) through hydrodynamical calculations. 
\begin{table}[h!]
\caption{
Evolution of characteristic quantities when the shock front reaches the star radius as function of the penetration factor $3 < \beta \le 15$ in the case $\gamma=5/3$. 
Column 2: time $t$ $[ \mathrm{s} ]$ when the shock front reaches the star radius (the star is at the periastron at $t=0$).
Column 3: star radius $R$ $[ 10^{-2} \, R_{*} ]$ at time $t$. It corresponds to the minimal value of the star radius reached during the compression.
Column 4: temperature $T_{\mathrm{R}}$ $[ T_{*} ]$ at the star radius. It corresponds to the maximal value of the temperature at the shock front. 
Column 5: velocity $v_{\mathrm{R}}$ $[ 10^{3} \, \mathrm{km} \, \mathrm{s}^{-1} ]$ at the star radius. It corresponds to the velocity of expansion when the shock front reaches the star radius.
Note that no shock wave forms for $\beta=3$.
}
\label{P04Tab01}
\centering
\begin{tabular}{ccccc}
\hline \hline
$\beta$ & $t$ & $R$ & $T_{\mathrm{R}}$ & $v_{\mathrm{R}}$ \\
\hline
4  & 174.50 & 4.42 & 21.85   & 1.86 \\
5  & 101.94 & 2.19 & 63.80   & 3.09 \\
6  & 66.73  & 1.29 & 111.72  & 4.03 \\
7  & 47.04  & 0.82 & 243.97  & 6.40 \\
8  & 34.92  & 0.56 & 298.17  & 6.95 \\
9  & 26.94  & 0.39 & 394.82  & 8.01 \\
10 & 21.42  & 0.29 & 467.76  & 8.85 \\
11 & 17.43  & 0.21 & 612.05  & 12.86 \\
12 & 14.47  & 0.17 & 868.26  & 16.34 \\
13 & 12.21  & 0.14 & 1117.86 & 18.85 \\
14 & 10.44  & 0.13 & 1340.36 & 21.67 \\
15 & 9.04   & 0.13 & 5751.44 & 42.97 \\
\hline
\end{tabular}
\end{table}

The present numerical simulations are a first step towards the resolution of the full problem of strong star-BH encounters. 
They have been performed in the case of a main-sequence star orbiting a massive BH along Newtonian orbits.
A more realistic situation should deal with the relativistic gravitational field of a static (Schwarzschild) or a rotating (Kerr) BH, for which it has already been shown that a deeply plunging star within the tidal radius undergoes several successive compression phases. 
This will require new calculations taking into account such relativistic effects, and one can expect that new hydrodynamical phenomena will arise due to the interaction of ingoing and outgoing shock waves. 
Including the self-gravity is also necessary in order to start the simulations well outside the tidal radius, and to better describe the post-pancake phase.  
Moreover, since our results confirm the possibility of thermonuclear explosion of the star, it will be necessary to couple a nuclear network to the hydrodynamics. 
The cases of more sophisticated equations of state for the stellar core (e.g. including radiative corrections) should be investigated, as well as the inclusion of non conservative terms (due to thermonuclear energy generation) in the energy equation. 
Eventually, in order to make more reliable predictions which could be compared to the available observations of high energy flares in galactic cores, the radiative transfer of energy through the disrupted star should be taken into account, as well as the aftermath of the stellar debris.

The subject of the tidal disruption of stars is called to develop in the near future. 
Indeed, the X-UV flares recently observed in the core of non-active galaxies can be interpreted as a consequence of such a phenomenon, thus providing new observational constraints for the models. 
A number of future (low- and high- energy) X-ray all-sky surveys are planned, the first ones to be launched in a few years such as XEUS and Constellation X missions, so predictions about the rate of these events and signals in the X- and $\gamma$-ray regimes are of strong interest. 
By providing a quick localization of such events, followed by the detection of the corresponding afterglows in the optical, infrared and radio bands, such missions could bring to the understanding of stellar disruptions as much as the Beppo-Sax and Swift telescopes have brought to the comprehension of $\gamma$-ray bursts. 
Some of the $\gamma$-ray bursts already observed by Swift and other telescopes could be tentatively interpreted as pancake stars.

\begin{figure}[h!]
\centering
\includegraphics[width=5cm]{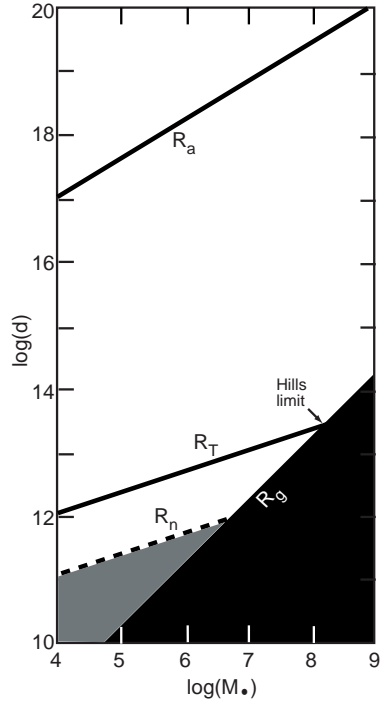}
\caption{Characteristic distances (in centimeters) for solar-type star / 
massive BH (in solar mass units) interactions. 
$R_{\mathrm{a}}$ is the gravitational influence radius (\ref{P04Eq02}), $R_{\mathrm{g}}$ is the gravitational radius (\ref{P04Eq04}), $R_{\mathrm{T}}$ is the tidal radius (\ref{P04Eq03}). 
Also represented is $R_{\mathrm{n}}$, the radius below which the star penetrates within $R_{\mathrm{T}}$ by a factor $\beta > 10$ and is likely to detonate thermonuclear reactions. 
The domain of astrophysical relevance for such a process is the grey area.} 
\label{P04Fig02}
\end{figure}

A massive BH imbedded in a dense cluster of stars strongly disturbs their dynamics within the gravitational influence radius (Frank \& Rees \cite{Fra76})
\begin{eqnarray}
R_{\mathrm{a}} & \equiv       & \frac{G M_{\bullet}}{v_{\infty}^{2}} \nonumber \\
               & \approx & 3 \times 10^{19} \, \mathrm{cm} \,
	                     \left( \frac{M_{\bullet}}{10^{8} M_{\odot}} \right)
			     \left( \frac{200 \, \mathrm{km} \, \mathrm{s}^{-1}}{v_{\infty}} \right)^{2}, \label{P04Eq01}
\end{eqnarray}
where $v_{\infty}$ is the stellar velocity dispersion in the galactic core. 
Currently, it is widely accepted the existence of an observational-based law connecting $M_{\bullet}$ and $v_{\infty}$ (Ferrarese \& Ford \cite{Fer05}) such that 
\begin{equation}
R_{\mathrm{a}} \approx 4 \times 10^{19} \, \mathrm{cm} \, 
                         \left( \frac{M_{\bullet}}{10^{8} M_{\odot}} \right)^{0.59}. \label{P04Eq02}
\end{equation}
If the star is diffused along an orbit within the tidal radius
\begin{eqnarray}
R_{\mathrm{T}} & \equiv       & R_{*} \left( \frac{M_{\bullet}}{M_{*}} \right)^{1/3} \nonumber \\ 
               & \approx & 3 \times 10^{13} \, \mathrm{cm} \, \left( \frac{R_{*}}{R_{\odot}} \right) 
	                                       \left( \frac{M_{\bullet}}{10^{8} M_{\odot}} \right)^{1/3}
	                                        \left( \frac{M_{\odot}}{M_{*}} \right)^{1/3}, \label{P04Eq03} 
\end{eqnarray}
it is ultimately disrupted by the tidal gravitational field.
However, since the BH gravitational radius 
\begin{eqnarray}
R_{\mathrm{g}} & \equiv       & \frac{G M_{\bullet}}{c^{2}} \nonumber \\ 
               & \approx & 10^{13} \, \mathrm{cm} \, \left( \frac{M_{\bullet}}{10^{8} M_{\odot}} \right) \label{P04Eq04}
\end{eqnarray}
grows faster than the tidal radius with the BH mass, the tidal disruption can only occur outside the horizon for BH mass below the Hills limit (Hills \cite{Hil75}), which is $\approx 10^{8} M_{\odot}$ for a star of solar mass and radius.
The different characteristic distances are reproduced on Fig.~\ref{P04Fig02}.
Our calculations require BHs in the range $10^5-10^6 \, M_{\odot}$, i.e. typically the massive BHs expected in ordinary galactic nuclei, such as our own Galactic Center.

Since the cross section for penetration of the star inside distance $R_{\mathrm{p}}$ is
\begin{equation}
\sigma (R_{\mathrm{p}}) \approx 
2 \pi \, G M_{\bullet} R_{\mathrm{p}} v_{\infty}^{-2},
\end{equation}
the rate of stars penetrating by a factor $\beta$ within the tidal radius is (Luminet \& Barbuy \cite{Lum90})
\begin{eqnarray}
\frac{dN (\beta)}{dt} 
&\approx& 
2 \pi \, G M_{\bullet} N_{\mathrm{c}} v_{\infty}^{-1} R_{\mathrm{R}} \, \beta^{-1} \\ \nonumber
&\approx& 
4 \times 10^{-4} \mathrm{yr}^{-1} \beta^{-1} 
\left( \frac{M_{\bullet}}{10^{6} M_{\odot}} \right)^{4/3}  
\left( \frac{N_{\mathrm{c}}}{10^{5} \, \mathrm{pc}^{-3}} \right)
\left( \frac{v_{\infty}}{60 \, \mathrm{km} \, \mathrm{s}^{-1}} \right)^{-1}, 
\end{eqnarray}
where $N_{\mathrm{c}}$ is the number density of stars at core  radius of the stellar cluster.
As mentioned in Sect.~\ref{Intro}, the coefficient $4 \times 10^{-4}$ could be considerably increased by the presence of a self-gravitating accretion disc around the BH.
The important thing is that the frequency of strong star-BH encounters (say with $\beta \gtrsim 3$) is by no means negligible, just one order of magnitude less than the frequency of tidal disruptions. 
Thus the rate of pancake stars should be about $10^{-5}$ per galaxy per year.
If the associated flares are generated in hard X- or $\gamma$-ray band, the full observable universe is transparent and several events of this kind per year would be expected.
%
%
%
%
\begin{acknowledgements}
We are grateful to J.~M.~Mart\'\i~who provided the sources of the RPPM code. \\
We thank B.~Carter, S.~Bonazzola, E.~M\"uller and J.~Cabrera for interesting discussions, and the referee for valuable comments to improve the present study. \\
\end{acknowledgements}
%
%
%
%
\begin{appendix}
\section{Derivation of the tidal gravitational acceleration} \label{A01}
We recall the general expression of the tidal gravitational acceleration from which (\ref{P02Eq05}) is derived.
As in the main text of the article, the BH gravitational field is considered in Newtonian dynamics.

\begin{figure}[h!]
\centering
\includegraphics[width=8cm]{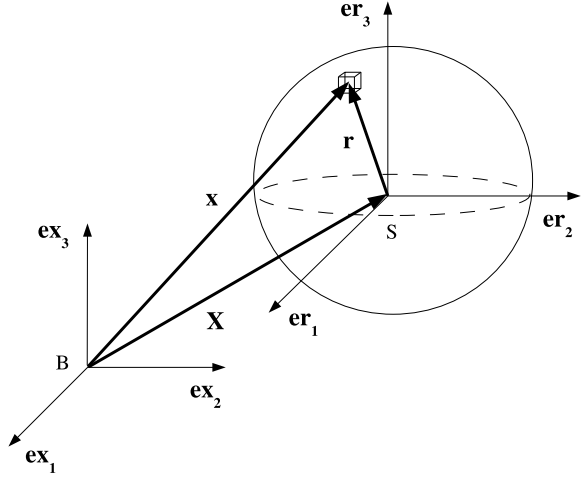}
\caption{Cartesian reference frame of the BH $(B,\vec{e{x}_{1}},\vec{e{x}_{2}},\vec{e{x}_{3}})$ and Cartesian reference frame of the star centre of mass $(S,\vec{e{r}_{1}},\vec{e{r}_{2}},\vec{e{r}_{3}})$. 
The star centre of mass is located by the position vector $\vec{X}(t)=X_{i}(t) \, \vec{e{x}_{i}}$.
A point-mass of the star is located by the position vector $\vec{x}(t)=x_{i}(t) \, \vec{e{x}_{i}}$ relative to the BH, and by the position vector $\vec{r}(t)=r_{i}(t) \, \vec{e{r}_{i}}$ relative to the star centre of mass with $\vec{r}(t)=\vec{x}(t)-\vec{X}(t)$.} 
\label{A01Fig01}
\end{figure}

Let us introduce (see Fig.~\ref{A01Fig01}) a first Cartesian frame $(B,\vec{e{x}_{1}},\vec{e{x}_{2}},\vec{e{x}_{3}})$ with origin $B$ at the BH position (reference frame $\mathcal{B}$ of the BH), and a second Cartesian frame $(S,\vec{e{r}_{1}},\vec{e{r}_{2}},\vec{e{r}_{3}})$ with origin $S$ at the star centre of mass (reference frame $\mathcal{S}$ of the star centre of mass, which relative to the inertial reference frame $\mathcal{B}$ is parallely propagated along the star motion). 
At time $t$ in reference frame $\mathcal{B}$, the star centre of mass is at position $\vec{X}(t)=X_{i}(t) \, \vec{e{x}_{i}}$, and a point-mass of the star at position $\vec{x}(t)=x_{i}(t) \, \vec{e{x}_{i}}$, where hereafter the summation convention over repeated indices is used and indices vary from $1$ to $3$. 
In reference frame $\mathcal{S}$, the same stellar point-mass is at position $\vec{r}(t)=r_{i}(t) \, \vec{e{r}_{i}}$ with $r_{i}(t)=x_{i}(t) - X_{i}(t)$. 

Let us place in reference frame $\mathcal{B}$, and let us compute the (gravitational) acceleration at position $\vec{x}(t)$ in the star due to the instantaneous action of the BH gravitational field.
At distance $x \equiv (x_{i}x_{i})^{1/2}$ from the BH, the stellar point-mass is exposed to the Newtonian gravitational potential 
\begin{equation}
\Phi(\vec{x}) = - \frac{GM_{\bullet}}{x}. 
\label{A01Eq01}
\end{equation}

Since in the present case, the star radius is very small compared to the distance $X \equiv (X_{i}X_{i})^{1/2}$ between the star centre of mass and the BH, $r \ll X$ with $r \equiv (r_{i}r_{i})^{1/2}$. 
Therefore, the gravitational potential (\ref{A01Eq01}) can be expanded on the neighbourhood of the star centre of mass $\vec{X}$:
\begin{eqnarray}
\Phi(\vec{x}) & = & 
\Phi(\vec{X}+\vec{r}) \nonumber \\
              & = & 
\Phi(\vec{X}) + (x_{i}-X_{i}) \left[ \partial_{x_{i}} \Phi(\vec{x}) \right]_{\vec{x}=\vec{X}} \nonumber \\
              & + & 
\frac{1}{2} (x_{i}-X_{i}) (x_{j}-X_{j}) \left[ \partial_{x_{i}} \partial_{x_{j}} \Phi(\vec{x}) \right]_{\vec{x}=\vec{X}} \nonumber \\
              & + & 
\frac{1}{6} (x_{i}-X_{i}) (x_{j}-X_{j}) (x_{k}-X_{k}) \left[ \partial_{x_{i}} \partial_{x_{j}} \partial_{x_{k}} \Phi(\vec{x}) \right]_{\vec{x}=\vec{X}} \nonumber \\
              & + & 
O(\vert \vec{x} \vert^{4}). 
\label{A01Eq02}
\end{eqnarray}

The expansion (\ref{A01Eq02}) is rewritten in the form
\begin{equation}
\Phi(\vec{x}) = 
\Phi(\vec{X}) + (x_{i}-X_{i}) \left[ \partial_{x_{i}} \Phi(\vec{x}) \right]_{\vec{x}=\vec{X}} + \Phi_{\mathrm{T}}(\vec{x}), 
\label{A01Eq03}
\end{equation}
defining the \textit{tidal gravitational potential}
\begin{eqnarray}
\Phi_{\mathrm{T}}(\vec{x}) \equiv 
& - & \frac{1}{2} (x_{i}-X_{i}) (x_{j}-X_{j}) C_{ij}(\vec{X}) \nonumber \\
& - & \frac{1}{6} (x_{i}-X_{i}) (x_{j}-X_{j}) (x_{k}-X_{k}) D_{ijk}(\vec{X}) \nonumber \\
& + & O(\vert \vec{x} \vert^{4}), 
\label{A01Eq04}
\end{eqnarray}
along with the second-order \textit{tidal gravitational tensor} 
\begin{equation}
C_{ij}(\vec{X})  \equiv 
- \left[ \partial_{x_{i}} \partial_{x_{j}} \Phi(\vec{x}) \right]_{\vec{x}=\vec{X}}, 
\label{A01Eq05} 
\end{equation}
and the third-order \textit{deviation gravitational tensor}
\begin{equation}
D_{ijk}(\vec{X}) \equiv 
- \left[ \partial_{x_{i}} \partial_{x_{j}} \partial_{x_{k}} \Phi(\vec{x}) \right]_{\vec{x}=\vec{X}}. 
\label{A01Eq06}
\end{equation}

The deviation tensor (\ref{A01Eq06}) measures, at the lowest order, the deviation of the motion of the star centre of mass from the motion that it would have if all the mass $M_{*}$ of the star was concentrated at this point. 
One can indeed show that the equation of motion of the star centre of mass is rigorously given by 
\begin{equation}
M_{*} \frac{d^{2}X_{i}}{dt^{2}} = 
- M_{*} \partial_{X_{i}} \Phi(\vec{X}) + \frac{1}{2} D_{ijk} \int r_{j} r_{k} \, \rho \, d^{3}\vec{r} 
+ O(\vert \vec{r} \vert^{3}) 
\label{A01Eq07}
\end{equation}
with 
\begin{equation}
\partial_{X_{i}} \Phi(\vec{X}) = \frac{GM_{\bullet}}{X^{3}} X_{i}, 
\label{A01Eq08}
\end{equation}
and where the integration is performed over the volume of the star of density $\rho$. 
However since $R_{*} \ll X$, the star appears as a point-mass relative to the BH so that the deviation effect is weak enough for the deviation tensor to be completely neglected. 
The motion of the star centre of mass can therefore be considered as strictly equivalent to the motion of a particle of mass $M_{*}$ in free fall in the external gravitational potential $\Phi(\vec{X})$, and (\ref{A01Eq07}) then limits to the first term given by (\ref{A01Eq08}). 
In the same way, the tidal potential (\ref{A01Eq04}) limits to the tidal tensor term (\ref{A01Eq05}).

The first and second-order partial derivatives of the gravitational potential (\ref{A01Eq01}) are equal to
\begin{eqnarray}
\partial_{x_{i}} \Phi(\vec{x})                  & = & 
\frac{GM_{\bullet}}{x^{3}} x_{i}, 
\label{A01Eq09} \\
\partial_{x_{i}} \partial_{x_{j}} \Phi(\vec{x}) & = & 
\frac{GM_{\bullet}}{x^{3}} \left( \delta_{ij} - \frac{3x_{i}x_{j}}{x^{2}} \right), 
\label{A01Eq10}
\end{eqnarray}
where $\delta_{ij}$ denotes the usual Kronecker delta:
\begin{displaymath}
\delta_{ij} = 
\left \{ 
\begin{array}{ll}
1 & \quad i  =  j \\
0 & \quad i \ne j.
\end{array} 
\right.
\end{displaymath}

Using (\ref{A01Eq01}), (\ref{A01Eq09}), and (\ref{A01Eq10}) applied at the star centre of mass $\vec{X}$, one can express with (\ref{A01Eq04}) and (\ref{A01Eq05}) the gravitational potential (\ref{A01Eq03}) at position $\vec{x}$. From that result, relative to reference frame $\mathcal{B}$ the acceleration at position $\vec{x}$ deduces directly:
\begin{eqnarray}
\frac{d^{2}x_{i}}{dt^{2}} & = & 
- \partial_{x_{i}} \Phi(\vec{x}) 
\label{A01Eq11} \\
                          & = & 
- \left[ \partial_{x_{i}} \Phi(\vec{x}) \right]_{\vec{x}=\vec{X}}
+ (x_{j}-X_{j}) \, C_{ij}(\vec{X}) 
\label{A01Eq12}
\end{eqnarray}
with
\begin{equation}
\left[ \partial_{x_{i}} \Phi(\vec{x}) \right]_{\vec{x}=\vec{X}} = 
\frac{GM_{\bullet}}{X^{3}} X_{i}, 
\label{A01Eq13} 
\end{equation}
and
\begin{equation}
C_{ij}(\vec{X}) = 
\frac{GM_{\bullet}}{X^{3}} \left( - \delta_{ij} + \frac{3X_{i}X_{j}}{X^{2}} \right). 
\label{A01Eq14}
\end{equation}

The first term of (\ref{A01Eq12}) given by (\ref{A01Eq13}) is identical to the acceleration (\ref{A01Eq08}) of the star centre of mass in the BH gravitational field. 
The tidal gravitational effects, which are by nature differential effects, are given by the second term of (\ref{A01Eq12}) referred to as the \textit{tidal gravitational acceleration}.
Therefore in the reference frame $\mathcal{S}$ of the star centre of mass, the acceleration at position $\vec{r}=\vec{x}-\vec{X}$ is equal to the tidal acceleration:   
\begin{eqnarray}
\frac{d^{2}r_{i}}{dt^{2}} & = & 
\frac{d^{2}x_{i}}{dt^{2}} - \frac{d^{2}X_{i}}{dt^{2}} 
\label{A01Eq15} \\ 
                          & = & 
r_{j} \, C_{ij}(\vec{X}) 
\label{A01Eq16}
\end{eqnarray}
using (\ref{A01Eq12}) with (\ref{A01Eq13}), and (\ref{A01Eq07}) with (\ref{A01Eq08}).
\section{Properties of the tidal gravitational field} \label{A02}
Following the assumptions mentioned in appendix \ref{A01}, we recall the general properties of the tidal gravitational field. 
As made explicit by the expression (\ref{A01Eq16}) of the tidal acceleration, the behaviour of the tidal field is imposed, actually at the lowest order, by the tidal tensor (\ref{A01Eq14}).

The expression of the tidal tensor can be simplified by choosing the orientation of the coordinate axes $(\vec{e{x}_{1}},\vec{e{x}_{2}},\vec{e{x}_{3}})$ such that the motion of the star centre of mass remains in the $(\vec{e{x}_{1}},\vec{e{x}_{2}})$ plane during the encounter with the BH. This choice, which is of course possible since the Newtonian orbit of the star is planar, implies that $X_{3}(t)=0$ at each time $t$. The expression (\ref{A01Eq14}) of the tidal tensor then gives
\begin{equation}
\vec{C}(\vec{X}) = 
\frac{GM_{\bullet}}{X^{3}}
\left[
\begin{array}{ccc}
-1 + \displaystyle{\frac{3X_{1}^{2}}{X^{2}}} & \displaystyle{\frac{3X_{1}X_{2}}{X^{2}}}     & 0 \\
\displaystyle{\frac{3X_{1}X_{2}}{X^{2}}}     & -1 + \displaystyle{\frac{3X_{2}^{2}}{X^{2}}} & 0 \\
0                                            & 0                                            & -1
\end{array} 
\right]. 
\label{A02Eq01}
\end{equation}

Moreover, in order to easily show the induced effects by the tidal field on the star, let us place in the principal directions frame of the tidal tensor. 
The latter being of real values, and by definition symmetric, can always be diagonalized. 
The eigenvalues of (\ref{A02Eq01}) are
\begin{eqnarray}
\lambda_{1} & = &   \frac{2GM_{\bullet}}{X^{3}}, \label{A02Eq02} \\
\lambda_{2} & = & - \frac{GM_{\bullet}}{X^{3}},  \label{A02Eq03} \\
\lambda_{3} & = &   \lambda_{2},                     \label{A02Eq04} 
\end{eqnarray}
and the normalized (right) eigenvectors associated with the eigenvalues (\ref{A02Eq02})-(\ref{A02Eq04}) are
\begin{eqnarray}
\vec{e{u}_{1}} & = & 
\frac{\vert X_{2} \vert}{X}  
\left( \frac{X_{1}}{X_{2}}\vec{e{x}_{1}} + \vec{e{x}_{2}} \right), 
\label{A02Eq05} \\
\vec{e{u}_{2}} & = & 
\frac{\vert X_{1} \vert}{X}  
\left( -\frac{X_{2}}{X_{1}}\vec{e{x}_{1}} + \vec{e{x}_{2}} \right), 
\label{A02Eq06} \\
\vec{e{u}_{3}} & = & 
\vec{e{x}_{3}}.
\label{A02Eq07}  
\end{eqnarray}

In the principal directions frame $(S,\vec{e{u}_{1}},\vec{e{u}_{2}},\vec{e{u}_{3}})$, the tidal tensor is diagonal:
\begin{equation}
C_{ij} = 
\left \{ 
\begin{array}{ll}
\lambda_{i} & \quad i  =  j \\
0           & \quad i \ne j.
\end{array} 
\right.
\label{A02Eq08}
\end{equation} 
From (\ref{A01Eq16}) the tidal acceleration at position $\vec{u}(t)=u_{i}(t) \, \vec{e{u}_{i}}$ in the star is equal to
\begin{equation}
\frac{d^{2}u_{i}}{dt^{2}} = u_{j} \, C_{ij}(\vec{X}), 
\label{A02Eq09}
\end{equation}
which with (\ref{A02Eq08}) simplifies to
\begin{equation}
\frac{d^{2}u_{i}}{dt^{2}} = \lambda_{j} \, u_{j} \, \delta_{ij}. 
\label{A02Eq10}   
\end{equation}

\begin{figure}[h!]
\centering
\includegraphics[width=12cm]{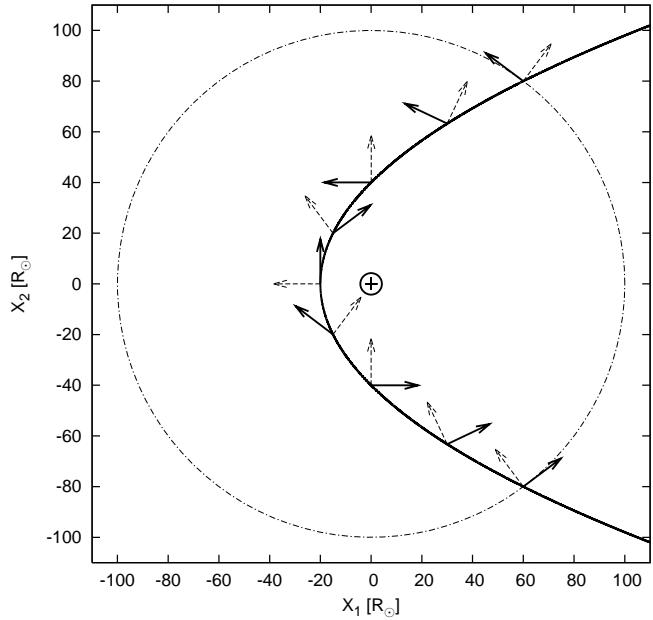}
\caption{Representation of the eigenvectors (\ref{A02Eq05}) and (\ref{A02Eq06}) of the tidal gravitational tensor (\ref{A02Eq01}) at different positions along the star orbit.
The eigenvectors $\vec{e{u}_{1}}$ and $\vec{e{u}_{2}}$ indicate the principal directions of the tidal tensor and are contained in the plane of the star orbit. 
$\vec{e{u}_{1}}$ (dashed arrow) is a stretching principal direction whereas $\vec{e{u}_{2}}$ (solid arrow) is a compressive principal direction, and both continually change as the star moves. 
The orbit of the star centre of mass (solid line), whose coordinates relative to the BH (cross) are $(X_{1},X_{2})$, is assumed parabolic and is computed with (\ref{P02Eq01})-(\ref{P02Eq02}). 
In this example, $M_{\bullet} = 10^{6} M_\mathrm{\odot}$, $M_{*} = M_\mathrm{\odot}$, and $R_{*} = R_\mathrm{\odot}$. 
The dot-dashed circle represents the BH tidal radius (\ref{P04Eq03}) and the solid circle the BH gravitational radius (\ref{P04Eq04}). 
The star enters in the tidal radius from the right bottom corner with a penetration factor $\beta=5$.} 
\label{A02Fig01}
\end{figure}

From (\ref{A02Eq10}), it is directly noticeable that one of the effects of the tidal field is to \textit{stretch} the stellar matter along the direction $\vec{e{u}_{1}}$, since by (\ref{A02Eq02}) the eigenvalue $\lambda_{1}$ is positive.
On the contrary, the eigenvalues $\lambda_{2}$ and $\lambda_{3}$ given by (\ref{A02Eq03}) and (\ref{A02Eq04}) being negative, a second effect of the tidal field is to \textit{compress} the stellar matter along both directions $\vec{e{u}_{2}}$ and $\vec{e{u}_{3}}$.
Actually, as can be seen from (\ref{A02Eq05}) and (\ref{A02Eq06}), both principal directions $\vec{e{u}_{1}}$ and $\vec{e{u}_{2}}$ are contained in the plane of the star orbit $(\vec{e{x}_{1}},\vec{e{x}_{2}})$. 
Because these directions depend on the star centre of mass coordinates $X_{i}(t)$, they continually change with time $t$ (see Fig.~\ref{A02Fig01}). 
It is different for the third principal direction $\vec{e{u}_{3}}$ which from (\ref{A02Eq07}) is orthogonal to the orbital plane, but also constant so that this direction remains always fixed during the motion of the star.
\end{appendix}
%
%
%
%

%
%
\end{document}